\documentclass[a4paper,11pt]{article}
 \pdfoutput=1
 \usepackage{lscape}
 \usepackage[all]{xy} 
 \usepackage{longtable} % to make Mass_hierarchy table
 \usepackage{jheppub} 
 \usepackage{graphicx}
 \usepackage{amssymb}
 \usepackage{amsmath}
\usepackage{mathtools}
\usepackage{listings}
\usepackage{dcolumn}% Align table columns on decimal point
\usepackage{bm}% bold math 
\usepackage{color}
\usepackage{multirow}
\usepackage{upquote} 
 
\newcommand{\met}{{/\!\!\! E_T}} 
 
\allowdisplaybreaks

\usepackage[table]{xcolor}

\definecolor{mag}{RGB}{255,0,255}
\definecolor{cy}{RGB}{0,255,255}
\definecolor{blu}{RGB}{51,51,255}

\newcommand{\lsim}{{\;\raise0.3ex\hbox{$<$\kern-0.75em\raise-1.1ex\hbox{$\sim$}}\;}}
\newcommand{\gsim}{{\;\raise0.3ex\hbox{$>$\kern-0.75em\raise-1.1ex\hbox{$\sim$}}\;}}
\newcommand{\beq}{\begin{equation}}
\newcommand{\eeq}{\end{equation}}
\newcommand{\bea}{\begin{eqnarray}}
\newcommand{\eea}{\end{eqnarray}}
\mathchardef\minus="002D

\title{\boldmath Detecting kinematic boundary surfaces in phase space: 
particle mass measurements in SUSY-like events}
\author[a]{Dipsikha Debnath,}
\author[b]{James~S.~Gainer,}
\author[c]{Can Kilic,}
\author[a,d]{Doojin Kim{${}^1$}\note{Corresponding author: {\tt immworry@gmail.com}},}
\author[a]{Konstantin~T.~Matchev,}
\author[c]{and Yuan-Pao Yang}

\affiliation[a]{Physics Department, University of Florida, Gainesville, FL
  32611, USA.}
\affiliation[b]{Department of Physics and Astronomy, University of Hawaii,
  Honolulu, HI 96822, USA}
\affiliation[c]{Theory Group, Department of Physics and Texas Cosmology Center,
  The University of Texas at Austin, Austin, TX 78712} 
\affiliation[d]{Theory Division, CERN, CH-1211 Geneva 23, Switzerland} 

\abstract{We critically examine the classic endpoint method for particle mass determination,
focusing on difficult corners of parameter space, where some of the measurements are not 
independent, while others are adversely affected by the experimental resolution. In such scenarios,
mass differences can be measured relatively well, but  the overall mass scale remains poorly constrained.
Using the example of the standard SUSY decay chain $\tilde q\to \tilde\chi^0_2\to \tilde \ell \to \tilde \chi^0_1$, 
we demonstrate that sensitivity to the remaining mass scale parameter can be recovered by 
measuring the two-dimensional kinematical boundary in the relevant three-dimensional phase space of
invariant masses squared. We develop an algorithm for detecting this boundary, which uses 
the geometric properties of the Voronoi tessellation of the data, and in particular, the relative 
standard deviation (RSD) of the volumes of the neighbors for each Voronoi cell in the tessellation.
We propose a new observable, $\bar\Sigma$, which is the average RSD per unit area, calculated over the hypothesized 
boundary. We show that the location of the $\bar\Sigma$ maximum correlates very well with the true 
values of the new particle masses. Our approach represents the natural extension of the 
one-dimensional kinematic endpoint method to the relevant three dimensions of invariant mass phase space.}

\date{October 28, 2016}
\preprint{
\begin{flushleft} 
UTTG-21-16\\
CERN-TH-2016-235\\
UH511-1268-2016
\end{flushleft} 
}

\begin{document} 
\maketitle
\flushbottom

\section{Introduction}
\label{sec:introduction}

The dark matter problem is currently our best experimental evidence for the existence of new particles 
and interactions beyond the Standard Model (BSM). A great number of ongoing experiments are 
trying to discover dark matter particles through direct \cite{Cushman:2013zza} or indirect detection \cite{Buckley:2013bha}. 
In principle, dark matter particles could also be produced in high energy collisions at the Large Hadron Collider (LHC) at CERN,
providing a complementary discovery probe in a controlled experimental environment \cite{Arrenberg:2013rzp}.

\begin{figure}[t]
\centering
\includegraphics[width=0.7\textwidth]{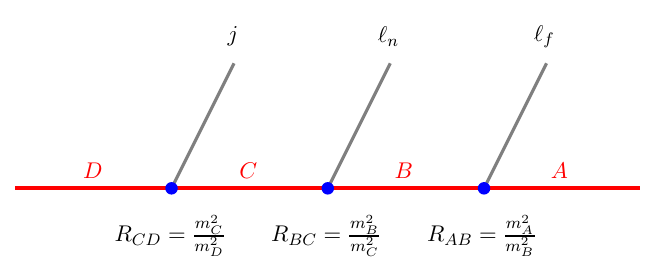}
\caption{The generic decay chain under consideration in this paper: $D\to jC\to j\ell_n B\to j\ell_n\ell_f A$, 
where $A$, $B$, $C$ and $D$ are new BSM particles, while the SM decay products consist of one jet $j$ and 
two leptons, labelled ``near" $\ell_n$ and ``far" $\ell_f$. In the SUSY case, $D$ represents a squark $\tilde q$,
$C$ is a heavier neutralino $\tilde\chi^0_2$, $B$ is a charged slepton $\tilde \ell$ and $A$ is the lightest neutralino
$\tilde\chi^0_1$, which escapes undetected. The masses of the BSM particles are denoted by 
$m_D$, $m_C$, $m_B$ and $m_A$. The corresponding ratios of squared masses $R_{CD}$, $R_{BC}$ and $R_{AB}$
are introduced for convenience in writing the kinematic endpoint formulas (\ref{lldef}-\ref{jllthetadef})
and delineating the relevant regions in the mass parameter space (\ref{massparspace}) 
(see also eq.~(\ref{Rparspace}) and Fig.~\ref{fig:regions} below). 
\label{fig:decay} 
}
\end{figure}

Since the dark matter particles must be stable on cosmological timescales, in many popular BSM models they carry some conserved 
quantum number. The simplest choice is a ${\mathbb Z}_2$ parity, which is known as 
$R$-parity in models with low energy supersymmetry (SUSY) \cite{Martin:1997ns},
Kaluza-Klein (KK) parity in models with universal extra dimensions (UED) \cite{Appelquist:2000nn}, 
$T$-parity in Little Higgs models \cite{Cheng:2003ju}, etc. As a result, the dark matter particles are necessarily 
produced in pairs: either directly, or in the cascade decays of other, heavier BSM particles \cite{Abercrombie:2015wmb}.
The prototypical such cascade decay is shown in Fig.~\ref{fig:decay}, in which a new particle $D$ undergoes a series of 
two-body decays, terminating in the dark matter candidate $A$, which is neutral and stable, and thus escapes undetected.
Under those circumstances, measuring the set of four masses 
\beq
\left\{ m_D, m_C, m_B, m_A\right\}
\label{massparspace}
\eeq
is a difficult problem, which has been attracting a lot of attention over the last 20 years (for a review, see \cite{Barr:2010zj}).
The main challenge stems from the fact that the momentum of particle $A$ is not measured, so that the standard technique
of directly reconstructing the new particles as invariant mass resonances does not apply. Instead, one has to somehow infer the 
new masses (\ref{massparspace}) from the measured kinematic distributions of the {\em visible} SM decay products.

In the decay chain of Fig.~\ref{fig:decay}, the SM decay products are taken to be a quark jet $j$ and 
two leptons, labelled ``near" $\ell_n$ and ``far" $\ell_f$. This choice is motivated by the following arguments:
\begin{itemize}
\item At a hadron collider like the LHC, strong production dominates, thus particle $D$ is very likely to be colored.
At the same time, the dark matter candidate $A$ is neutral, therefore the color must be shed somewhere along the decay chain
in the form of a QCD jet. Here we assume that this ``color-shedding" occurs in the $D\to C$ transition\footnote{We note that 
in principle one can test this assumption experimentally, e.g. by constructing suitably defined on-shell constrained $M_2$ variables 
corresponding to the competing event topologies \cite{Cho:2014naa}, or by studying the shapes and the correlations for the
invariant mass variables considered below \cite{Cho:2012er}. Such an exercise is useful, but beyond the scope of this paper.}, 
since one expects the strong decays of particle $D$ to be the dominant ones. 
\item The presence of leptons among the SM decay products in Fig.~\ref{fig:decay} is theoretically not guaranteed, 
but is nevertheless experimentally motivated. First, leptonic signatures have significantly lower SM backgrounds and thus 
represent clean discovery channels. Second, the momentum of a lepton is measured much better than that of a jet, 
therefore the masses (\ref{massparspace}) will be measured with a better precision in a leptonic channel 
(as opposed to a purely jetty channel).
Finally, if the SM decay products in Fig.~\ref{fig:decay} were all jets, in light of the arising combinatorial 
problem \cite{MP,Matsumoto:2006ws,Rajaraman:2010hy,Bai:2010hd,Baringer:2011nh,Choi:2011ys}, 
we would have to resort to sorted invariant mass variables \cite{Kim:2015bnd,Klimek:2016axq}, whose kinematic endpoints are less pronounced 
and thus more difficult to measure over the SM backgrounds.
\item From a historical perspective, the best motivation for considering the decay chain of Fig.~\ref{fig:decay}
is that it is ubiquitous in SUSY, where $D$ represents a squark $\tilde q$,
$C$ is a heavier neutralino $\tilde\chi^0_2$, $B$ is a charged slepton $\tilde \ell$ and $A$ is the lightest neutralino
$\tilde\chi^0_1$, which escapes the detector and leads to missing transverse energy $\met$. In
the two most popular frameworks of SUSY breaking, gravity-mediated and gauge-mediated, the combination of (a) specific high scale 
boundary conditions, and (b) renormalization group evolution of the soft SUSY parameters down to the weak scale,
leads to just the right mass hierarchy for the decay chain of Fig.~\ref{fig:decay} to occur. In the late 1990's and early 2000's, this
prompted a flurry of activity on the topic of mass determination in such ``SUSY-like" missing energy events. 
Soon afterwards, it was also realized that the decay chain of Fig.~\ref{fig:decay} is not exclusive to supersymmetry, 
but the same final state signature also appears in other models, e.g. minimal UED \cite{Cheng:2002ab} and littlest Higgs \cite {Freitas:2006vy}.
\end{itemize}

To date, a large variety of mass measurement techniques for SUSY-like events have been developed. Roughly speaking, they all can be 
divided into two categories. 
\begin{itemize}
\item {\em Exclusive methods.} In this case, one takes advantage of the presence of two decay chains in the event 
(they are often assumed identical) {\em and} the available $\met$ measurement. Several approaches are then possible. 
For example, in the so-called ``polynomial methods'' one attempts to solve explicitly for the momenta of the invisible 
particles in a given event, possibly using additional information from prior measurements of kinematic 
endpoints~\cite{Hinchliffe:1998ys,Nojiri:2003tu,Kawagoe:2004rz,Cheng:2007xv,Nojiri:2008ir, Cheng:2008mg, Cheng:2009fw,
Webber:2009vm, Autermann:2009js, Kang:2009sk, Nojiri:2010dk, Kang:2010hb, Hubisz:2010ur, Cheng:2010yy, Gripaios:2011jm}.\footnote{For 
long enough decay chains, the polynomial methods are able to solve for the invisible momenta, even without additional experimental 
input and without a second decay chain in the event. If the decay chain of Fig.~\ref{fig:decay} contained an additional two-body decay to a visible particle,
just 5 events are sufficient for solving the event kinematics \cite{Nojiri:2003tu,Cheng:2009fw}. }
Alternatively, utilizing information from both branches, one could introduce suitable transverse\footnote{Transversality is not strictly necessary,
in fact it may even be beneficial to work with $3+1$-dimensional variants of those variables 
\cite{Ross:2007rm,Barr:2008ba,Barr:2008hv,Konar:2008ei,Konar:2010ma,Barr:2011xt,Robens:2011zm,Mahbubani:2012kx,Bai:2013ema,Cho:2014yma,Cho:2015laa,Konar:2015hea}. }
variables whose distributions exhibit an upper kinematic endpoint indicative of the parent particle mass
\cite{Lester:1999tx, Barr:2003rg, Baumgart:2006pa,
Lester:2007fq, Tovey:2008ui, Serna:2008zk, Nojiri:2008vq, Cho:2008tj, 
Cheng:2008hk, Burns:2008va, Choi:2009hn, Matchev:2009ad,Polesello:2009rn, 
Konar:2009wn, Cho:2009ve, Nojiri:2010mk}. 
In the latter case, one still retains a residual dependence on the unknown dark matter particle mass $m_A$,
which must be fixed by some other means, e.g. via the kink method
\cite{Cho:2007qv, Gripaios:2007is, Barr:2007hy, Cho:2007dh, Nojiri:2008hy, 
Barr:2009jv,Matchev:2009fh,Konar:2009qr} or by performing a sufficient number of independent measurements
\cite{Serna:2008zk,Burns:2008va}. While they could be potentially quite sensitive, these exclusive methods are 
also less robust, since they rely on the correct identification of all objects in the event, and are thus prone 
to combinatorial ambiguities, the effects from $\met$ resolution, initial and final state radiation, underlying event and pileup, etc. 
\item {\em Inclusive methods.} In this case, one focuses on the decay chain from Fig.~\ref{fig:decay} itself,
disregarding what else is going on in the event. Using only the measured momenta of the visible 
SM decay products, i.e., the jet and the two leptons, one could form all possible invariant mass combinations\footnote{In general,
one is not limited to Lorentz-invariant variables only, e.g., recently it was suggested to study the peak of the {\em energy}
distribution as a measure of the mass scale \cite{Agashe:2012bn,Agashe:2013eba,Agashe:2015wwa,Agashe:2015ike}.}, 
namely $m_{\ell\ell}$, $m_{j\ell_n}$, $m_{j\ell_f}$, and $m_{j\ell\ell}$, measure their respective 
upper kinematic endpoints
\beq
\left\{m^{max}_{\ell\ell}, m^{max}_{j\ell_n}, m^{max}_{j\ell_f}, m^{max}_{j\ell\ell} \right\},
\label{4endpoints}
\eeq
and use them to solve for the four input parameters (\ref{massparspace}).
As just described, this approach is too naive, as it overlooks the 
remaining combinatorial problem involving the two leptons $\ell_n$ and $\ell_f$. 
Since ``near" and ``far" cannot be distinguished on an event by event basis, the variables
$m_{j\ell_n}$ and $m_{j\ell_f}$ are ill defined. This is why it has become customary to redefine the two
jet-lepton invariant mass combinations as\footnote{A more recent alternative approach is to
introduce new invariant mass variables which are symmetric functions of $m_{j\ell_n}$
and $m_{j\ell_f}$, thus avoiding the need to distinguish $\ell_n$ from $\ell_f$ on an event per event basis
\cite{Matchev:2009iw}.} 
\begin{eqnarray}
m_{jl(lo)}
&\equiv& \min \left\{m_{jl_n}, m_{jl_f} \right\}  , \label{mjllodef} \\ [2mm]
m_{jl(hi)}
&\equiv& \max \left\{m_{jl_n}, m_{jl_f} \right\}  . \label{mjlhidef}
\end{eqnarray}
The distributions of the newly defined quantities (\ref{mjllodef}) and (\ref{mjlhidef}) also exhibit upper 
kinematic endpoints, $m^{max}_{jl(lo)}$ and $m^{max}_{jl(hi)}$, respectively. Then, instead of (\ref{4endpoints}),
one can use the new well-defined set of measurements
\begin{equation}
\left\{m_{ll}^{max}, m_{jll}^{max}, m_{jl(lo)}^{max}, m_{jl(hi)}^{max}\right\}
\label{4measurements}
\end{equation}
to invert and solve for the input mass parameters (\ref{massparspace}).
This procedure constitutes the classic kinematic endpoint method for mass measurements, which has 
been successfully tested for several SUSY benchmark points~\cite{Hinchliffe:1996iu, Bachacou:1999zb, Allanach:2000kt,
Lester:2001zx, Gjelsten:2004ki, Gjelsten:2005aw, Lester:2005je, Miller:2005zp}.
\end{itemize}

However, despite its robustness and simplicity, the kinematic endpoint method still has a couple of weaknesses.
As we show below, taken together, they essentially lead to an almost flat direction in the solution space, thus jeopardizing the uniqueness of the mass determination.
The first of these two problems is purely theoretical --- it is well known that in certain regions of the parameter space (\ref{massparspace})
the four measurements (\ref{4measurements}) are not independent, but obey the relation \cite{Gjelsten:2004ki}
\begin{equation}
\left( m_{jll}^{max}\right)^2 = \left( m_{jl(hi)}^{max}\right)^2 + \left( m_{ll}^{max}\right)^2.
\label{mjllcorrelation}
\end{equation}
In practice, this means that the measurements (\ref{4measurements}) fix only three out of the four mass parameters 
(\ref{massparspace}), leaving one degree of freedom undetermined. In what follows, we shall choose to parametrize this
``flat direction" with the mass $m_A$ of the lightest among the four new particles $D$, $C$, $B$ and $A$. 
One can then use, e.g., the first three of the measurements in (\ref{4measurements}) and solve uniquely for the 
three heavier masses $m_D$, $m_C$, and $m_B$, leaving $m_A$ as a free parameter. We list the relevant inversion 
formulas in Appendix \ref{sec:inversion}. The obtained one-parameter family of mass spectra 
\beq
\left\{ 
\begin{array}{l}
m_D = m_D(m_A; m_{ll}^{max}, m_{jll}^{max}, m_{jl(lo)}^{max}), \\
m_C = m_C(m_A; m_{ll}^{max}, m_{jll}^{max}, m_{jl(lo)}^{max}), \\
m_B = m_B(m_A; m_{ll}^{max}, m_{jll}^{max}, m_{jl(lo)}^{max}), \\
m_A
\end{array}
\right.
\label{massfamily}
\eeq
will satisfy the three measured kinematic endpoints $m_{ll}^{max}$, $m_{jll}^{max}$, and $m_{jl(lo)}^{max}$ by construction.
What is more, in parameter space regions where eq.~(\ref{mjllcorrelation}) holds, the family (\ref{massfamily}) will also obey the fourth 
measurement of $m_{jl(hi)}^{max}$, so that the four measurements (\ref{4measurements}) will be insufficient to lift the $m_A$ 
degeneracy in (\ref{massfamily}).

These considerations beg the following two questions, which will be addressed in this paper.
\begin{enumerate}
\item In the remaining part of the parameter space, where (\ref{mjllcorrelation}) does {\em not} hold and
$m_{jl(hi)}^{max}$ provides an independent fourth measurement, how well is the $m_A$ degeneracy lifted after all?
With the explicit examples of Sections~\ref{sec:case31} and \ref{sec:case32} below, we shall show that 
although in theory the additional measurement of $m_{jl(hi)}^{max}$ determines the value of $m_A$, 
in practice this may be difficult to achieve, since the effect is very small and will be swamped by the experimental resolution.
\item In the region of parameter space in which (\ref{mjllcorrelation}) holds, what additional measurement 
should be used, and how well does it lift the degeneracy? In the existing literature, the standard approach is 
to consider the constrained\footnote{The distribution $m_{jll(\theta>\frac{\pi}{2})}$ is nothing but the
usual $m_{jll}$ distribution taken over a subset of the original events, namely those which satisfy
the additional dilepton mass constraint 
\begin{equation}
\frac{m_{ll}^{max}}{\sqrt{2}} < m_{ll} < m_{ll}^{max}\, . \nonumber
\end{equation}
In the rest frame of particle $B$, this cut implies the following 
restriction on the opening angle $\theta$ between the two leptons \cite{Nojiri:2000wq}
\begin{equation}
\theta > \frac{\pi}{2}\, , \nonumber
\end{equation}
thus justifying the notation for $m_{jll(\theta>\frac{\pi}{2})}$.
} distribution $m_{jll(\theta>\frac{\pi}{2})}$, which exhibits
a useful {\em lower} kinematic endpoint $m_{jll(\theta>\frac{\pi}{2})}^{min}$ \cite{ATLAS:1999vwa,Nojiri:2000wq}.
In what follows, we shall therefore always supplement the 
original set of 4 measurements (\ref{4measurements}) with the additional
measurement of $m_{jll(\theta>\frac{\pi}{2})}^{min}$ to obtain the extended set
\begin{equation}
\left\{m_{ll}^{max}, m_{jll}^{max}, m_{jl(lo)}^{max}, m_{jl(hi)}^{max}, m_{jll(\theta>\frac{\pi}{2})}^{min}\right\},
\label{5measurements}
\end{equation}
so that in principle there is sufficient information to determine 
the four unknown masses. Even then, we shall show that the sensitivity of the 
additional experimental input $m_{jll(\theta>\frac{\pi}{2})}^{min}$ to the previously found flat direction
(\ref{massfamily}) is very low. First of all, it is already well appreciated that the measurement
of $m_{jll(\theta>\frac{\pi}{2})}^{min}$ is very challenging, since in the vicinity of its lower endpoint, the
shape of the signal distribution is concave downward, which makes it difficult to extract the endpoint with 
simple linear fitting, and one has to use the whole shape of the $m_{jll(\theta>\frac{\pi}{2})}$ distribution \cite{Lester:2006yw}.
Secondly, as we shall show in the examples below, the variation of the value of $m_{jll(\theta>\frac{\pi}{2})}^{min}$
along the flat direction (\ref{massfamily}) can be numerically quite small, and therefore the sensitivity of the added fifth measurement
along the flat direction (\ref{massfamily}) is not that great.
\end{enumerate}

Either way, we see that the known methods for lifting the degeneracy of the flat direction (\ref{massfamily}) will
face severe limitations once we take into account the experimental resolution, finite statistics, backgrounds, etc. 
\cite{Gjelsten:2005sv,Burns:2009zi} Thus the first goal of this paper will be to illustrate the severity of the problem, 
i.e. to quantify the ``flatness" of the family of solutions (\ref{massfamily}). For this purpose,
we shall reuse the study points from Ref.~\cite{Burns:2009zi}, which at the time were meant to illustrate {\em discrete}
ambiguities, i.e. cases where two distinct points in mass parameter space (\ref{massparspace}) accidentally 
happen to give mathematically identical values for all five measurements (\ref{5measurements}).
Here we shall extend those study points to a family of mass spectra (\ref{massfamily}) which give mathematically identical 
values for the first three\footnote{And sometimes four, if we are in parts of parameter space where 
(\ref{mjllcorrelation}) holds.} of the measurements (\ref{5measurements}),
and numerically {\em very similar} values for the remaining measurements.

Having identified the problem, the second goal of the paper is to propose a novel solution to it and investigate its viability. 
Our starting point is the observation that the signal events from the decay chain in Fig.~\ref{fig:decay}
populate the interior of a compact region in the $(m_{j\ell_n}, m_{\ell\ell}, m_{j\ell_f})$
space, whose boundary is given by the surface ${\cal S}$ defined 
by the constraint \cite{Costanzo:2009mq,CLTASI,Kim:2015bnd}
\beq
{\cal S}:\quad
\hat{m}^2_{j\ell_f} = 
\left[ \sqrt{ \hat{m}^2_{\ell\ell} \left(1- \hat{m}^2_{j\ell_n}\right) }
\pm  \frac{m_B}{m_C} 
\sqrt{ \hat{m}^2_{j\ell_n} \left(1-\hat{m}^2_{\ell\ell}\right)} \right]^2 ,
\label{eq:samosa2}                               
\eeq
which, for convenience, is written in terms of the unit-normalized variables 
\beq
\hat{m}_{j\ell_n} =  \frac{m_{j\ell_n}}{ m^{max}_{j\ell_n}}, \quad
\hat{m}_{\ell\ell} =  \frac{m_{\ell\ell}}{ m^{max}_{\ell\ell}}, \quad 
\hat{m}_{j\ell_f} =  \frac{m_{j\ell_f}}{ m^{max}_{j\ell_f}}.  
\eeq
We note that both the shape and the size of the surface ${\cal S}$ depend on the input mass spectrum (\ref{massparspace}),
i.e., ${\cal S}(m_A,m_B,m_C,m_D)$, and this dependence is precisely what we will be targeting with our method to be described below.

In its traditional implementation, the kinematic endpoint method is essentially\footnote{The fact that 
one has to use $m_{jl(lo)}$ and $m_{jl(hi)}$ in place of $m_{j\ell_n}$ and $m_{j\ell_f}$
does not change the gist of the argument.} using the kinematic endpoints (\ref{4endpoints}) 
of the one-dimensional projections of the signal population onto each of the three axes $m_{j\ell_n}$, $m_{\ell\ell}$ and $m_{j\ell_f}$, 
as well as onto the ``radial" direction $m_{j\ell\ell} = \sqrt{m^2_{j\ell_n} + m^2_{\ell\ell} + m^2_{j\ell_f} }$. 
This approach is suboptimal because it ignores correlations and misses endpoint features along the other possible projections. 
The only way to guarantee that we are using the full available information in the data is to fit to the three-dimensional boundary 
(\ref{eq:samosa2}) itself \cite{Agrawal:2013uka,Debnath:2016mwb}, which will be the approach advocated here.
As previously observed in \cite{Agrawal:2013uka} (and extended to a broader class of event topologies in \cite{Altunkaynak:2016bqe}), 
most of the signal events are populated near the phase space boundary 
(\ref{eq:samosa2}), on which the signal number density $\rho_s$ formally becomes singular. This fact is rather fortuitous, since
it implies a relatively sharp change in the local number density as we move across the phase space boundary, 
even in the presence of SM backgrounds (with some number density $\rho_b$, which is expected to be a relatively smooth function). 
Thus, we need to develop a suitable method for identifying regions in phase space where the gradient of the total number 
density $\rho\equiv \rho_b+\rho_s$ is relatively large, and then fit to them the analytical parametrization (\ref{eq:samosa2}) 
in order to obtain the best fit values for the four new particle masses (\ref{massparspace}).

The first step of this program was already accomplished in our earlier paper \cite{Debnath:2016mwb}, building on 
the idea originally proposed in \cite{Debnath:2015wra} for finding ``edges" in two-dimensional stochastic distributions of point data. 
Ref.~\cite{Debnath:2015wra} suggested that interesting features in the data, e.g., edge discontinuities, kinks, and so on, can be
identified by analyzing the geometric properties of the Voronoi tessellation \cite{Voronoi} of the data.\footnote{We 
note the existence of efficient codes for finding Voronoi tessellations in the form of the {\tt qHULL} algorithms~\cite{qhull}.
Wrappers that allow the use of these algorithms in many frameworks also exist, and in this work we use a private {\tt Python} code
to compute the geometric attributes of the Voronoi cells.} 
The volume $v_i$ of a given Voronoi cell generated by a data point at some location $\vec{r}_i$ provides an estimate of the functional value of the number 
density $\rho$ at that location,
\beq
\rho (\vec{r}_i) \sim \frac{1}{v_i}.
\eeq
Therefore, in order to obtain an estimate of $|\vec {\nabla} \rho(\vec{r})|$, we can construct variables which compare the properties of the 
Voronoi cell and its direct neighbors. Among the different options investigated in Refs.~\cite{Debnath:2015wra,Debnath:2015hva}, the relative standard 
deviation (RSD), $\bar\sigma_i$, of the volumes of neighboring cells, was identified as the most promising tagger of edge cells.
The RSD was defined as follows.  Let $N_i$ be the set of neighbors of the $i$-th Voronoi cell $C_i$, 
with volumes, $\{v_j\}$, for $j\in N_i$.  The RSD, $\bar\sigma_i$, is now defined by
\begin{equation}
\bar{\sigma}_i \equiv 
\frac{1}{\langle {v}(N_i)\rangle}\, \sqrt{\sum_{j\in N_i} \frac{\left(v_j-\langle {v}(N_i)\rangle \right)^2}{|N_i|-1}},
\label{defvar}
\end{equation}
where we have normalized by the average volume of the set of neighbors, $N_i$, of the $i$-th cell
\begin{equation}
\langle {v}(N_i)\rangle \equiv \frac{1}{|N_i|}\sum_{j\in N_i} v_j.
\label{ave-vol-neighbors}
\end{equation}
Subsequently, in \cite{Debnath:2016mwb} we showed that this procedure for tagging edge cells can be readily extended to three-dimensional 
point data, as is the case here. The end result of the method was a set of Voronoi cells which have been tagged as ``edge cell candidates"
since their values of $\bar\sigma_i$ were above the chosen threshold \cite{Debnath:2016mwb}. With the thus obtained set of edge cells in hand, 
it appears that we are in a perfect position to perform a mass measurement, simply by finding the set of values for $\{m_A,m_B,m_C,m_D\}$
which maximize the overlap between our tagged edge cells and the hypothesized surface ${\cal S}$. We have checked that 
this approach indeed works and gives a reasonable estimate of the true mass spectrum. However, here we prefer to suggest 
a slightly modified alternative, which accomplishes the same goal, but with somewhat better precision.

The problem with fitting to a subset of the original data set (namely the set of Voronoi cells which happened to pass the $\bar\sigma_i$ cut)
is that we are still throwing away useful information, e.g., the Voronoi cells which barely failed the cut. In spite of formally failing, 
those cells are nevertheless still quite likely to be edge cells. Thus, in order to retain the full amount of information in our data, 
we prefer to abandon this ``cut and fit" approach, and instead design a global variable which is calculated over the full data set.
The only requirement is that the variable is maximized (or minimized, as the case may be) for the true values of the masses $\{m_A,m_B,m_C,m_D\}$.

In order to motivate such a variable, consider for a moment the case when the function $\rho(\vec{r})$ is known analytically, then
let us investigate the (normalized) surface integral 
\beq
\frac{\int_{\tilde {\cal S}(\tilde m_A,\tilde m_B,\tilde m_C,\tilde m_D)} da\, |\vec {\nabla} \rho(\vec{r})|}
{\int_{\tilde {\cal S}(\tilde m_A,\tilde m_B,\tilde m_C,\tilde m_D)} da}
\label{eq:integralovergradient}
\eeq
for some arbitrary trial\footnote{From here on trial values for the masses will carry a tilde to distinguish from the true values of the 
masses which will have no tilde. Correspondingly, $\tilde {\cal S}$ stands for a hypothesized ``trial" boundary surface (\ref{eq:samosa2})
obtained with trial values of the mass parameters. } values $(\tilde m_A, \tilde m_B, \tilde m_C, \tilde m_D)$ of the unknown masses (\ref{massparspace}).
The meaning of the quantity (\ref{eq:integralovergradient}) is very simple: it is the average gradient of $\rho(\vec{r})$ over 
the chosen surface $\tilde {\cal S}$.
We expect the dominant contributions to the integral to come from regions where the gradient is large, and we know that the gradient is 
largest on the true phase space boundary ${\cal S}(m_A,m_B,m_C,m_D)$, defined in terms of the {\em true} values of the 
particle masses. However, if our choice for $(\tilde m_A, \tilde m_B, \tilde m_C, \tilde m_D)$ is wrong, the integration surface 
$\tilde {\cal S}$ will be far from the true phase space boundary ${\cal S}$, and those large contributions will be missed.
The only way to capture {\em all} of the large contributions to the integral is to have $\tilde {\cal S}$ coincide with the true 
${\cal S}$, and this is only possible if in turn the trial masses are exactly equal to the true particle masses. This suggests a method of
mass measurement whereby the true mass spectrum is obtained as the result of an optimization problem involving the 
quantity (\ref{eq:integralovergradient}).

Of course, in our case the analytical form of the integrand $|\vec {\nabla} \rho(\vec{r})|$ is unknown, but we can obtain
a closely related quantity using the Voronoi tessellation of the data. Following \cite{Debnath:2015wra,Debnath:2016mwb}, we shall utilize
the RSD $\bar{\sigma}_i $ defined in (\ref{defvar}), which has been shown to be a good indicator of edge cells, and replace 
the integrand in (\ref{eq:integralovergradient}) as 
\beq
|\vec {\nabla} \rho(\vec{r})| \longrightarrow g(\vec{r}) \equiv \bar{\sigma}_i\ {\rm for}\  \vec{r}\in C_i.
\label{eq:replacement}
\eeq
In other words, the gradient estimator\footnote{Note that $g(\vec{r})$ is not supposed to be an approximation for $|\vec {\nabla} \rho(\vec{r})|$,
the crucial property for us is that the two functions peak in the same location.} function $g(\vec{r})$ is defined so that it is equal to the RSD $ \bar{\sigma}_i$ 
of the Voronoi cell $C_i$ in which the point $\vec{r}$ happens to be. Eqs.~(\ref{eq:integralovergradient}) and (\ref{eq:replacement}) 
suggest that the variable which we should be maximizing is
\beq
\bar\Sigma(\tilde m_A,\tilde m_B,\tilde m_C,\tilde m_D) \equiv 
\frac{\int_{\tilde {\cal S}(\tilde m_A,\tilde m_B,\tilde m_C,\tilde m_D)} da\, g(\vec{r})}
{\int_{\tilde {\cal S}(\tilde m_A,\tilde m_B,\tilde m_C,\tilde m_D)} da}.
\label{Sigmabardef}
\eeq
It obviously depends on our choice of trial masses $(\tilde m_A,\tilde m_B,\tilde m_C,\tilde m_D)$, and as argued above, 
we expect the maximum of $\bar\Sigma$ to occur for the correct choice $(m_A,m_B,m_C,m_D)$, i.e.
\beq
\max_{\tilde m_A,\tilde m_B,\tilde m_C,\tilde m_D} \bar\Sigma(\tilde m_A,\tilde m_B,\tilde m_C,\tilde m_D) \simeq
\bar\Sigma(m_A,m_B,m_C,m_D).
\label{hypothesis}
\eeq
This hypothesis will be tested and validated with explicit examples below in Sections~\ref{sec:case31} and \ref{sec:case32}.

The paper is organized as follows. In the next Section~\ref{sec:endpoints} we shall first review
the well known formulas for the one-dimensional kinematic endpoints (\ref{5measurements})
and introduce the corresponding relevant partitioning of the mass parameter space into domain regions.
In the next two sections we shall concentrate on the two most troublesome regions, $(3,2)$ and $(3,1)$,
where the problematic relationship (\ref{mjllcorrelation}) holds. We shall pick one study point in each region, 
then study how well our conjecture (\ref{hypothesis}) is able to determine the true mass spectrum.
In principle, (\ref{hypothesis}) involves optimization over 4 continuous variables, which is very time consuming
(additionally, we have to perform the integration in the numerator of (\ref{Sigmabardef}) by Monte Carlo).
This is why for simplicity we choose to illustrate the power of our method with a one-dimensional 
toy study along the problematic flat direction (\ref{massfamily}). In particular, for each of our two study points we shall 
assume that the first three kinematic endpoints $m_{ll}^{max}$, $m_{jll}^{max}$, and $m_{jl(lo)}^{max}$
are already measured, leaving us only the task of determining the remaining degree of freedom $m_A$ along the
flat direction defined in (\ref{massfamily}). Correspondingly, we shall consider the whole family of mass 
spectra (\ref{massfamily}) which passes through a given study point. This family will eventually take us into the 
neighboring parameter space regions, including the third potentially problematic region, namely $(2,3)$, 
in which (\ref{mjllcorrelation}) is satisfied. For each family, we shall perform the following investigations
\begin{itemize}
\item As a warm up, we shall first illustrate that for each of the three distributions,
$m_{ll}$, $m_{jll}$, and $m_{jl(lo)}$, the endpoint along the flat direction is the same 
(as expected by construction).
\item We shall then investigate the variation of the kinematic endpoint of the $m_{jl(hi)}$ distribution
along the flat direction (\ref{massfamily}). The endpoint value $m_{jl(hi)}^{max}$  is expected to be constant in regions $(3,2)$, $(3,1)$ and $(2,3)$,
so the main question will be, how much does it vary in the remaining parameter space regions.
\item We shall similarly investigate the variation of the lower kinematic endpoint $m_{jll(\theta>\frac{\pi}{2})}^{min}$
along the flat direction (\ref{massfamily}). Together with the previous item, this will serve as an illustration 
of the main weakness of the classic kinematic endpoint method for mass measurements.  
\item Then we shall illustrate the distortion of the kinematic boundary surface (\ref{eq:samosa2}) 
along the flat direction (\ref{massfamily}). The size of the distortion will be indicative of the precision with which 
one can hope to perform the mass measurement (\ref{hypothesis}) using the kinematic boundary {\em surface} in phase space.
\item Finally, we shall perform the fitting (\ref{hypothesis}) along the flat direction parameterized by $\tilde m_A$.
We shall show results in two cases: (a) when the background events are distributed uniformly in $m^2$ phase space, 
and (b) when the background is coming from dilepton $t\bar{t}$ events.
\end{itemize}
We shall summarize and conclude in Section~\ref{sec:conclusions}.
Appendix~\ref{sec:inversion} contains the inversion formulas needed to define the flat direction (\ref{massfamily}).

\section{Endpoint formulas and partitioning of parameter space}
\label{sec:endpoints}

\subsection{Notation and conventions}

Following \cite{Burns:2009zi}, we introduce for convenience some shorthand notation for the mass squared ratios
\begin{equation}
R_{ij} \equiv \frac{m_i^2}{m_j^2}
\ ,
\label{RABdef}
\end{equation}
where $i,j\in\left\{A,B,C,D\right\}$. Note that in (\ref{RABdef}) there are only three independent quantities, 
which can be taken to be the set $\{R_{AB}, R_{BC}, R_{CD}\}$. To save writing, we will also introduce
convenient shorthand notation for the five kinematic endpoints as follows
\beq
a = \left(m_{ll}^{max}\right)^2, \quad
b = \left(m_{jll}^{max}\right)^2, \quad
c = \left(m_{jl(lo)}^{max}\right)^2, \quad
d = \left( m_{jl(hi)}^{max}\right)^2, \quad
e = \left( m_{jll(\theta>\frac{\pi}{2})}^{min}\right)^2. 
\label{abcde}
\eeq
Note that these represent the kinematic endpoints of the mass {\em squared} distributions\footnote{Contrast to
the notation of Ref.~\cite{Gjelsten:2004ki}, which uses $a,b,c,d$ to label the same endpoints, but for the {\em linear} masses.}.

In the next two sections we shall use the three endpoint measurements $m_{ll}^{max}$, $m_{jll}^{max}$, and $m_{jl(lo)}^{max}$
to fix $m_D$, $m_C$ and $m_B$, leaving $m_A$ as a free parameter. Another way to think about this procedure is to note that 
the parameter space (\ref{massparspace}) can be equivalently parametrized as
\beq
\left\{ R_{CD}, R_{BC}, R_{AB}, m_A \right\}.
\label{Rparspace}
\eeq
Then, the endpoint measurements of $m_{ll}^{max}$, $m_{jll}^{max}$, and $m_{jl(lo)}^{max}$ can be used to fix the ratios
$R_{CD}$, $R_{BC}$ and $R_{AB}$ (see Appendix~\ref{sec:inversion}), leaving the overall mass scale undetermined and parametrized by $m_A$.

\subsection{Endpoint formulas}

The kinematical endpoints are given by the following formulas:
\bea
a &\equiv& \left(m_{ll}^{max}\right)^2 = m_D^2\, R_{CD}\, (1-R_{BC})\, (1-R_{AB}); 
\label{lldef}
\\ [4mm]
b &\equiv& \left(m_{jll}^{max}\right)^2 = 
\left\{ 
\begin{array}{lll}
m_D^2 (1-R_{CD})(1-R_{AC}),       & ~{\rm for}\ R_{CD}<R_{AC},       & {\rm case}\ (1,-),~~ \\[4mm] 
m_D^2 (1-R_{BC})(1-R_{AB}R_{CD}), & ~{\rm for}\ R_{BC}<R_{AB}R_{CD}, & {\rm case}\ (2,-),~~ \\[4mm] 
m_D^2 (1-R_{AB})(1-R_{BD}),       & ~{\rm for}\ R_{AB}<R_{BD},       & {\rm case}\ (3,-),~~ \\[4mm] 
m_D^2\left(1-\sqrt{R_{AD}}\,\right)^2,   & ~{\rm otherwise}, & {\rm case}\ (4,-);~~
\end{array}
\right.
\label{jlldef}
\\ [4mm]
c &\equiv& \left(m_{jl(lo)}^{max}\right)^2=\left\{ 
\begin{array}{lll}
\left(m_{jl_n}^{max}\right)^2,   & ~{\rm for}\ (2-R_{AB})^{-1} < R_{BC} < 1,   & {\rm case}\ (-,1), \\[4mm] 
\left(m_{jl(eq)}^{max}\right)^2, & ~{\rm for}\ R_{AB}< R_{BC}<(2-R_{AB})^{-1}, & {\rm case}\ (-,2),\\[4mm] 
\left(m_{jl(eq)}^{max}\right)^2, & ~{\rm for}\ 0< R_{BC}<R_{AB},               & {\rm case}\ (-,3);
\end{array}%
\right .
\label{jllodef}
\\ [4mm]
d &\equiv& \left( m_{jl(hi)}^{max}\right)^2 =\left\{ 
\begin{array}{lll}
\left(m_{jl_f}^{max}\right)^2, & ~{\rm for}\ (2-R_{AB})^{-1} < R_{BC} < 1,   & {\rm case}\ (-,1), \\[4mm] 
\left(m_{jl_f}^{max}\right)^2, & ~{\rm for}\ R_{AB}< R_{BC}<(2-R_{AB})^{-1}, & {\rm case}\ (-,2), \\[4mm] 
\left(m_{jl_n}^{max}\right)^2, & ~{\rm for}\ 0< R_{BC}<R_{AB},               & {\rm case}\ (-,3);
\end{array}%
\right .
\label{jlhidef}
\eea
where
\bea
\left(m_{jl_n}^{max}\right)^2   &=& m_D^2\, (1-R_{CD})\, (1-R_{BC})\, , \label{mjlnmax}\\
\left(m_{jl_f}^{max}\right)^2   &=& m_D^2\, (1-R_{CD})\, (1-R_{AB})\, , \label{mjlfmax}\\
\left(m_{jl(eq)}^{max}\right)^2 &=& m_D^2\, (1-R_{CD})\, (1-R_{AB})\, (2-R_{AB})^{-1} \, . \label{mjleqmax}
\eea
Finally, the endpoint $m_{jll(\theta>\frac{\pi}{2})}^{min}$ 
introduced earlier in the Introduction, is given by 
\begin{eqnarray}
e &\equiv& \left( m_{jll(\theta>\frac{\pi}{2})}^{min}\right)^2 = 
\frac{1}{4}m_D^2 \Biggl\{ (1-R_{AB})(1-R_{BC})(1+R_{CD})  \label{jllthetadef} 
\\ \nonumber
&+& 2\, (1-R_{AC})(1-R_{CD})
-(1-R_{CD})\sqrt{(1+R_{AB})^2 (1+R_{BC})^2-16 R_{AC}}\Biggr\} .
\end{eqnarray}

\subsection{Partitioning of the mass parameter space}

\begin{figure}[t]
\centering
\includegraphics[width=0.7\textwidth]{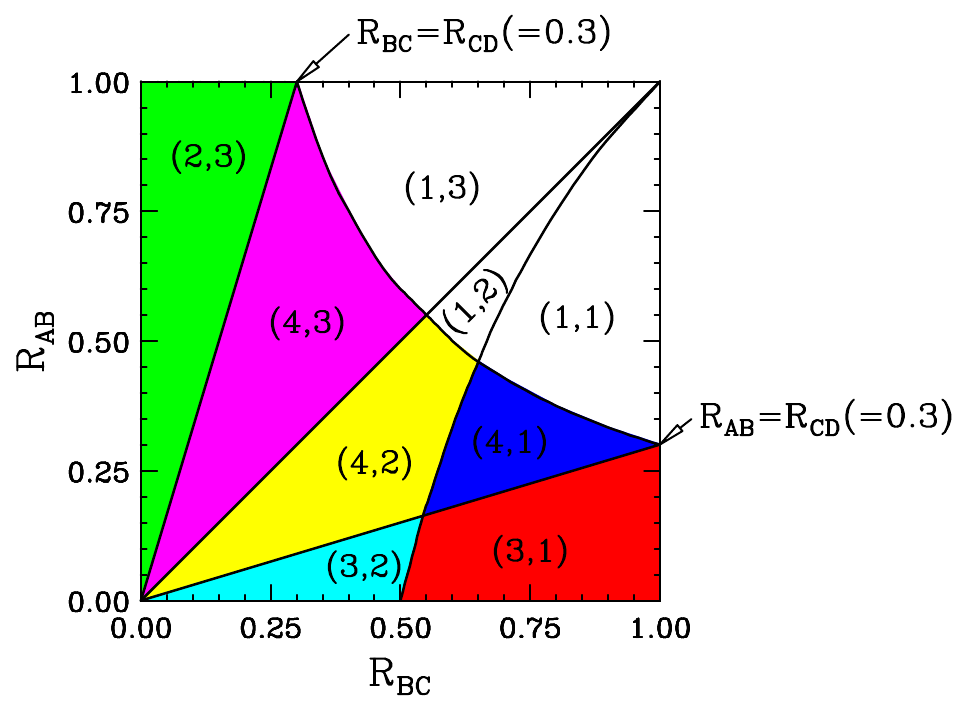}
\caption{ A slice through the $\left\{ R_{CD}, R_{BC}, R_{AB} \right\}$ 
parameter space at a fixed $R_{CD}=0.3$. The $(R_{BC}, R_{AB})$ plane
exhibits the nine definition domains $(N_{jll},N_{jl})$ of the set of equations (\ref{jlldef}-\ref{jlhidef}).
For the purposes of this paper, only six of those regions will be in play, and we have color-coded 
them as follows: 
region $(3,1)$ in red,
region $(4,1)$ in blue,
region $(3,2)$ in cyan,
region $(4,2)$ in yellow,
region $(4,3)$ in magenta, and
region $(2,3)$ in green.
\label{fig:regions} 
}
\end{figure}

One can see that the formulas (\ref{jlldef}-\ref{jlhidef}) are piecewise-defined:
they are given in terms of different expressions, 
depending on the parameter range for $R_{CD}$, $R_{BC}$ and $R_{AB}$.
This divides the $\{ R_{CD},R_{BC},R_{AB}\}$ parameter subspace from (\ref{Rparspace})
into several distinct regions, illustrated in Fig.~\ref{fig:regions}.
Following \cite{Gjelsten:2004ki}, we label those by a pair of integers $(N_{jll},N_{jl})$.
As already indicated in eqs.~(\ref{jlldef}-\ref{jlhidef}), the first integer $N_{jll}$ 
identifies the relevant case for $m_{jll}^{max}$, while the second integer $N_{jl}$ 
identifies the corresponding case for $(m_{jl(lo)}^{max},m_{jl(hi)}^{max})$.
One can show that only 9 out of the 12 pairings
$(N_{jll},N_{jl})$ are physical, and they are all exhibited within the
unit square of Fig.~\ref{fig:regions}. In what follows, an individual study point
within a given region $(N_{jll},N_{jl})$ will be marked with corresponding subscripts as $P_{N_{jll}N_{jl}}$.

Using (\ref{lldef}), (\ref{jlldef}) and (\ref{jlhidef}),
it is easy to check that the ``bad" relation (\ref{mjllcorrelation}), which
can be equivalently rewritten in the new notation as
\beq
b = a + d,
\label{bad}
\eeq
is identically satisfied in regions (3,1), (3,2) and (2,3) of Fig.~\ref{fig:regions}.
Therefore, as already discussed, in these regions one would necessarily have to rely on the additional 
information provided by the measurement of the $e$ endpoint (\ref{jllthetadef}).

Before concluding this rather short preliminary section, we direct the reader's attention to the 
color-coding in Fig.~\ref{fig:regions}, where we have shaded in color six of the parameter space regions: 
region $(3,1)$ in red, region $(4,1)$ in blue, region $(3,2)$ in cyan, 
region $(4,2)$ in yellow, region $(4,3)$ in magenta, and region $(2,3)$ in green.
It will turn out that the two families of mass spectra considered in the next two sections will visit the
six color-shaded regions. For the benefit of the reader, in the remainder of the paper we shall 
strictly adhere to this color scheme --- for example, results obtained for a study point from a particular 
region will always be plotted with the color of the respective region: study points in region $(3,1)$ are \textcolor{red}{red},
study points in region $(4,1)$ are \textcolor{blue}{blue}, etc.

\section{A case study in region $(3,1)$}
\label{sec:case31}

\subsection{Kinematical properties along the flat direction}
\label{sec:flat31}

In this section we shall study the flat direction (\ref{massfamily}) in mass parameter space 
which is generated by a study point $P_{31}$ from region $(3,1)$ (the same study point was used in
\cite{Burns:2009zi} for a slightly different purpose). 
\begin{table}[t]
\centering
\begin{tabular}{|l|c||r|r||r|r||}
%\hline
\cline{3-6}
\multicolumn{2}{c||}{}
         &  \multicolumn{2}{c||}{true branch}
         &  \multicolumn{2}{c||} {auxilliary branch}    \\ [0.5mm]
%\cline{3-6}
\hline
\multicolumn{2}{|c||}{Region }
         & \cellcolor{red} $(3,1)$ & \cellcolor{blue}$(4,1)$   & \cellcolor{mag}$(4,3)$  & \cellcolor{green}$(2,3)$     \\  [0.5mm] 
%\cline{3-6}
\hline
\multicolumn{2}{|c||}{Study point}
         & \cellcolor{red} $P_{31}$& \cellcolor{blue}$P_{41}$ & \cellcolor{mag}$P_{43}$ & \cellcolor{green}$P_{23}$     \\  [0.5mm] \hline \hline
\multicolumn{2}{|c||}{$m_A$ (GeV)}
         & \cellcolor{red} 236.64 & \cellcolor{blue}5000.00  & \cellcolor{mag}2,000.00 &  \cellcolor{green}100.00    \\ [0.5mm] \hline 
\multicolumn{2}{|c||}{$m_B$ (GeV)}
         & \cellcolor{red} 374.16& \cellcolor{blue}5126.02  & \cellcolor{mag}2040.56 & \cellcolor{green}124.78    \\ [0.5mm] \hline 
\multicolumn{2}{|c||}{$m_C$ (GeV)}
         & \cellcolor{red} 418.33& \cellcolor{blue}5168.03  & \cellcolor{mag}2167.36 & \cellcolor{green}272.54    \\ [0.5mm] \hline 
\multicolumn{2}{|c||}{$m_D$ (GeV)}
         & \cellcolor{red} 500.00 & \cellcolor{blue}5256.90  & \cellcolor{mag}2256.90  & \cellcolor{green}362.23    \\ [0.5mm] \hline 
\multicolumn{2}{|c||}{$R_{AB}$}
         & \cellcolor{red} 0.400 &  \cellcolor{blue}0.951   &   \cellcolor{mag}0.960 & \cellcolor{green} 0.642     \\ [0.5mm] \hline 
\multicolumn{2}{|c||}{$R_{BC}$}
         & \cellcolor{red} 0.800 & \cellcolor{blue} 0.984   &   \cellcolor{mag}0.886 &  \cellcolor{green}0.210    \\ [0.5mm] \hline 
\multicolumn{2}{|c||}{$R_{CD}$}
         & \cellcolor{red}  0.700 &  \cellcolor{blue}0.966   &   \cellcolor{mag}0.922 & \cellcolor{green} 0.566     \\ [0.5mm] \hline 
\hline
$m_{ll}^{max}$ (GeV)  & $\sqrt{a}$
         & \multicolumn{4}{c||}{144.91}  \\ [0.5mm] \hline
$m_{jll}^{max}$ (GeV) & $\sqrt{b}$
         & \multicolumn{4}{c||}{256.90 }   \\ [0.5mm] \hline
$m_{jl(lo)}^{max}$ (GeV)  & $\sqrt{c}$
         & \multicolumn{4}{c||}{122.47 }  \\ [0.5mm] \hline
$m_{jl(hi)}^{max}$ (GeV)  & $\sqrt{d}$
         & \cellcolor{red}  212.13  &  \cellcolor{blue}212.12    & \cellcolor{mag}212.13     & \cellcolor{green}212.13 \\ [0.5mm] \hline
$m_{jll(\theta>\frac{\pi}{2})}^{min}$ (GeV)  & $\sqrt{e}$
         &\cellcolor{red} 132.10   &  \cellcolor{blue}129.73    & \cellcolor{mag}130.79     &  \cellcolor{green}141.78   \\ [0.5mm] \hline
\hline
%$m_{jl_f}^{max}$ (GeV)  & $\sqrt{f}$
%         &212     & 127      & 200     & 183                   \\ [0.5mm] \hline
%$m_{jl_f}^{(p)}$ (GeV)  & $\sqrt{p}$
%         &190     & 112      & 126     & 115                   \\ [0.5mm] \hline
%$m_{jl_n}^{max}$ (GeV)  & $\sqrt{n}$
%         &122     & 212      & 173     & 200                   \\ [0.5mm] \hline
%$m_{jl(eq)}^{max}$ (GeV)& $\sqrt{q}$
%         &  NA   %167.705 no equal point in region 1 
%                  & 122      & 149     & 149                   \\ [0.5mm] \hline
%\hline
\end{tabular}
\caption{\label{tab:31}
Mass spectrum and expected kinematic endpoints for the study point $P_{31}$ from region $(3,1)$
which was discussed in Ref.~\cite{Burns:2009zi}, together with three additional study points
illustrating the different regions from Fig.~\ref{fig:regions} encountered by the
parameter space trajectories from Fig.~\ref{fig:trajectory}. By construction,
all study points give identical values for the
kinematic endpoints $m_{ll}^{max}$, $m_{jll}^{max}$ and $m_{jl(lo)}^{max}$.
Furthermore, in accordance with (\ref{mjllcorrelation}), the two study points $P_{31}$ and $P_{23}$ 
from regions $(3,1)$ and $(2,3)$ have identical values of $m_{jl(hi)}^{max}$.
The remaining two study points $P_{41}$ and $P_{43}$, representing regions $(4,1)$ and $(4,3)$,
have essentially the same value for $m_{jl(hi)}^{max}$ as well.
The last row lists the predicted values for $m_{jll(\theta>\frac{\pi}{2})}^{min}$, which 
are slightly different, and allow discriminating between the four endpoints in theory, but not in practice.}
\end{table}
Table~\ref{tab:31} lists some relevant information for the study point $P_{31}$: the input mass spectrum (\ref{massparspace}),
the corresponding mass squared ratios (\ref{RABdef}), and the predicted kinematic endpoints (\ref{5measurements}), 
also reminding the reader of the alternative shorthand notation (\ref{abcde}). As discussed in the Introduction, 
starting from the point $P_{31}$, we can follow a one-dimensional trajectory (\ref{massfamily})
through the parameter space (\ref{Rparspace}) so that everywhere along the trajectory the 
prediction for the three endpoints $a$, $b$ and $c$ is unchanged (see Fig.~\ref{fig:3dist_31} below). 
This trajectory is illustrated in Fig.~\ref{fig:trajectory},
where we show its projections onto the three planes $(R_{BC}, R_{AB})$ (left panel),
$(R_{AB}, R_{CD})$ (middle panel) and $(R_{BC}, R_{CD})$ (right panel).
\begin{figure}[t]
\centering
\includegraphics[width=0.32\textwidth]{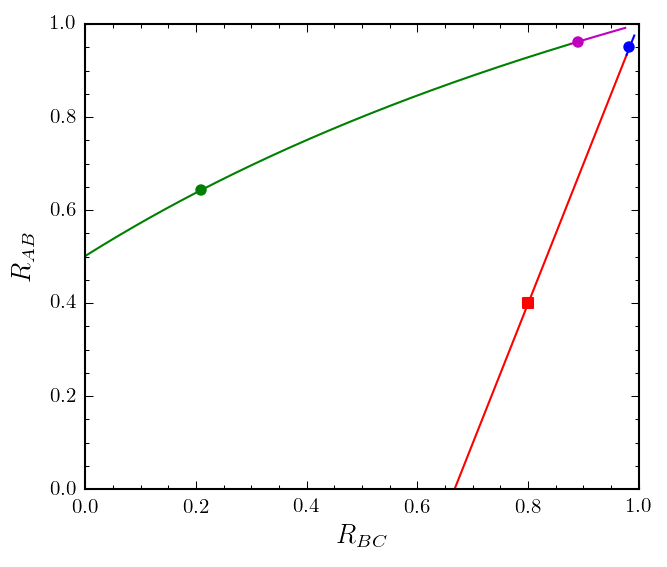}
\includegraphics[width=0.32\textwidth]{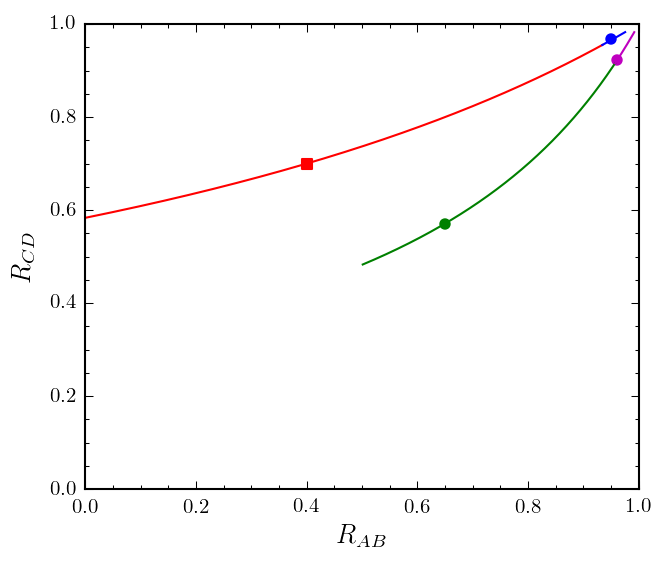}  
\includegraphics[width=0.32\textwidth]{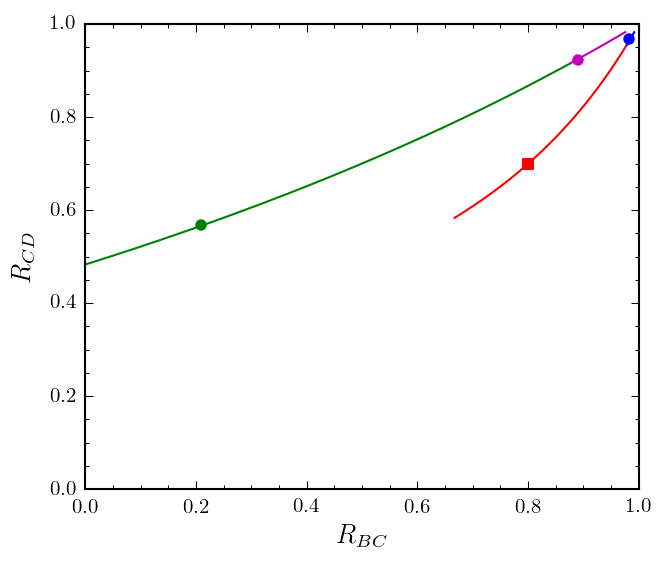}
\caption{The two trajectories in mass parameter space leading to the same endpoints $a$, $b$ and $c$. 
The lines are colored in accordance with the coloring convention for the regions depicted in Fig.~\ref{fig:regions}.
The red square marks the original study point $P_{31}$ from Table~\ref{tab:31}, while the circles 
denote the other three study points from Table~\ref{tab:31}: 
$P_{41}$ in region $(4,1)$ (blue circle), 
$P_{43}$ in region $(4,3)$ (magenta circle), and
$P_{23}$ in region $(2,3)$ (green circle).  \label{fig:trajectory} 
}
\end{figure}
The lines in Fig.~\ref{fig:trajectory} are parametrized by the continuous test mass parameter $\tilde m_A$.
For any given fixed value of $\tilde m_A$, the trajectory in Fig.~\ref{fig:trajectory} predicts the test values 
for the other three mass parameters, namely $\tilde m_B$, $\tilde m_C$ and $\tilde m_D$.
This is shown more explicitly in Fig.~\ref{fig:mass_diff_31}, where we plot the mass differences
$\tilde m_B-\tilde m_A$ (solid lines), $\tilde m_C-\tilde m_A$ (dashed lines), and
$\tilde m_D-\tilde m_A$ (dotted lines), as a function of $\tilde m_A$.
\begin{figure}[t]
\centering
\includegraphics[width=0.5\textwidth]{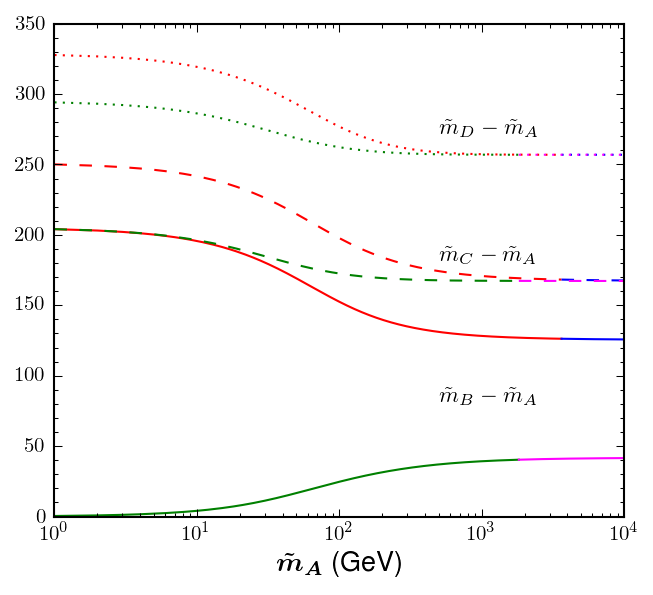}
\caption{Mass spectra along the flat direction specified by the study point $P_{31}$.
As a function of $\tilde m_A$, we plot the mass differences $\tilde m_B-\tilde m_A$ (solid lines), 
$\tilde m_C-\tilde m_A$ (dashed lines), and $\tilde m_D-\tilde m_A$ (dotted lines), 
which would preserve the values for the three kinematic endpoints $a$, $b$ and $c$.
\label{fig:mass_diff_31} 
}
\end{figure}
All lines in Figs.~\ref{fig:trajectory} and \ref{fig:mass_diff_31}
are color-coded using the same color conventions as for the parameter space regions in Fig.~\ref{fig:regions}. 
Initially, as we move away from point $P_{31}$ (marked with the red square in Fig.~\ref{fig:trajectory}), 
we are still within the red region $(3,1)$, and the trajectory is therefore colored in red and 
parametrically given by eqs.~(\ref{reg31mB}-\ref{reg31mD}).
As the value of $\tilde m_A$ is reduced from its nominal value (236.6 GeV) at the point $P_{31}$,
the mass spectrum gets lighter and eventually we reach $\tilde m_A=0$, where (the red portion of) the 
trajectory terminates at $R_{AB}=0$, $R_{BC}\simeq0.67$ and $R_{CD}\simeq0.58$.
If, on the other hand, we start increasing $\tilde m_A$ from its nominal $P_{31}$ value,
the spectrum gets heavier, and we start approaching the neighboring region $(4,1)$.
Eventually, at around $\tilde m_A\sim 3600$ GeV, the trajectory crosses into 
region $(4,1)$ and thus changes its color to blue. This transition is illustrated 
in the left panel of Fig.~\ref{fig:mass_ratios_31}, where we plot the mass squared ratios
$R_{AB}$, $R_{BC}$ and $R_{CD}$ (solid lines), together with some other relevant quantities
(dotted lines). In particular, the boundary between regions $(3,1)$ and $(4,1)$ is given by the 
relation $R_{AB}=R_{BD}$, see (\ref{reg31R_AB}) and (\ref{reg41R_AB}).
We can see that crossover more clearly in the insert in the left panel of Fig.~\ref{fig:mass_ratios_31},
where the line color changes from red to blue as soon as the $R_{BD}$ (dotted) line crosses the $R_{AB}$ (solid) line.
 
\begin{figure}[t]
\centering
\includegraphics[width=0.4\textwidth]{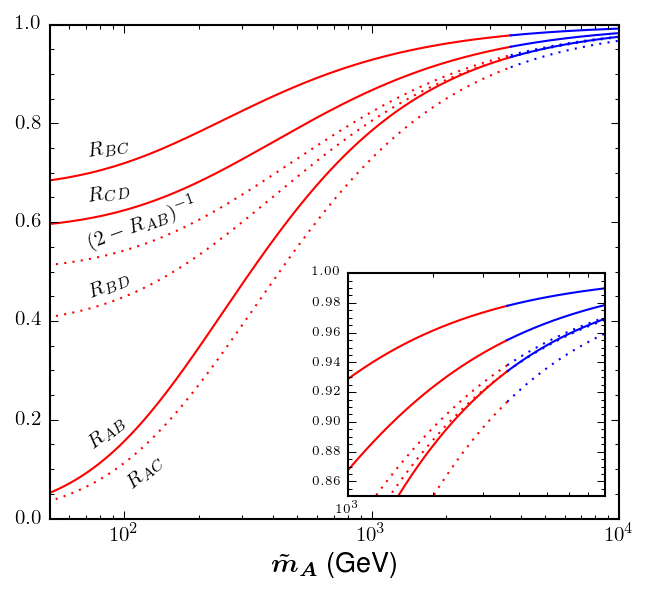}
\includegraphics[width=0.4\textwidth]{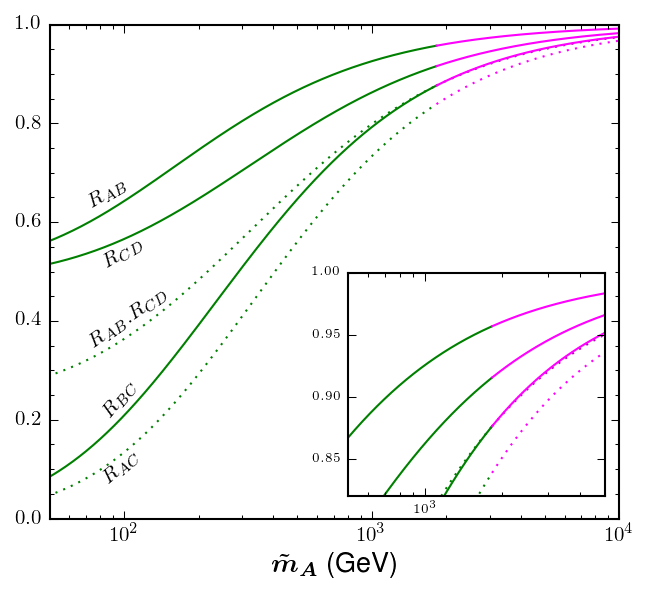}
\caption{The equivalent representation of Fig.~\ref{fig:mass_diff_31} in terms of the 
mass squared ratios $R_{AB}$, $R_{BC}$ and $R_{CD}$ (solid lines).
The dotted lines depict various quantities of interest which are used to delineate
the regions in Fig.~\ref{fig:regions}. 
The left panel shows the true branch passing through regions $(3,1)$ (red) and $(4,1)$ (blue),
while the right panel shows the auxiliary branch through regions $(2,3)$ (green) and $(4,3)$ (magenta). 
The left insert zooms in on the transition between regions $(3,1)$ and $(4,1)$ near $\tilde m_A=3600$ GeV, while the
right insert focuses on the transition between regions $(2,3)$ and $(4,3)$ near $\tilde m_A=1800$ GeV.
\label{fig:mass_ratios_31} 
}
\end{figure}

Once we are in region $(4,1)$, we follow the blue portion of the trajectory in Fig.~\ref{fig:trajectory},
which is parametrically defined by eqs.~(\ref{reg41mB}-\ref{reg41mD}). We 
choose a representative study point for region $(4,1)$ as well ---
it is denoted by $P_{41}$ and listed in the third (blue shaded) column of Table~\ref{tab:31}.
The corresponding mass spectrum is clearly very heavy, but is nevertheless perfectly consistent with the 
three measured endpoints $a$, $b$ and $c$, as shown in Fig.~\ref{fig:3dist_31}. 
As seen in Fig.~\ref{fig:trajectory}, the blue portion of the mass trajectory appears headed for the point
$(R_{AB}, R_{BC}, R_{CD})=(1,1,1)$, which is indeed reached in the limit of $\tilde m_A\to \infty$,
without ever entering into the neighboring region $(1,1)$\footnote{Note that as the value of $R_{CD}$ increases, 
the $(1,1)$ region shrinks and for $R_{CD}=1$ it disappears altogether.}.

Fig.~\ref{fig:trajectory} reveals that the mass family (\ref{massfamily}) through our study point $P_{31}$ 
includes a segment which starts at $(R_{AB}, R_{BC}, R_{CD})=(0,0.67,0.58)$ and ends at $(R_{AB}, R_{BC}, R_{CD})=(1,1,1)$,
visiting regions $(3,1)$ and $(4,1)$. Since the actual study point $P_{31}$ belongs to this segment, in what follows 
we shall refer to it as ``the true branch". However, Fig.~\ref{fig:trajectory} also shows that there is an additional 
disconnected segment of the mass trajectory through the green region $(2,3)$ and the magenta region $(4,3)$. 
In the following, we shall refer to this additional segment as ``the auxiliary branch". Note that this terminology is 
introduced only for clarity and should not be taken too literally --- as far as the measured endpoints $a$, $b$ and $c$
are concerned, all points on the true and auxiliary branches are on the same footing, since the experimenter 
would have no way of knowing {\em a priori} which is the true branch and which is the auxiliary branch.
This is why we have to seriously consider points on the auxiliary branch as well. We choose two representative study points, 
which are listed in the last two columns of Table~\ref{tab:31}: point $P_{43}$ belongs to the magenta region $(4,3)$,
while point $P_{23}$ is in the green region $(2,3)$. As shown in Fig.~\ref{fig:trajectory}, the auxiliary branch starts at 
$(R_{AB}, R_{BC}, R_{CD})=(0.5,0,0.48)$ and asymptotically meets the true branch at the corner point 
$(R_{AB}, R_{BC}, R_{CD})=(1,1,1)$. The transition between the two regions $(2,3)$ and $(4,3)$ along the 
auxiliary branch is illustrated in the right panel of Fig.~\ref{fig:mass_ratios_31}. According to 
(\ref{reg23R_BC}) and (\ref{reg43R_BC}), the boundary between regions $(2,3)$ and $(4,3)$ is defined by the relation 
$R_{BC}=R_{AB}R_{CD}$. The right panel of Fig.~\ref{fig:mass_ratios_31} confirms this: the color of the auxiliary 
branch in Figs.~\ref{fig:trajectory} and \ref{fig:mass_diff_31} changes from green to magenta as soon as the dotted 
line representing the product $R_{AB}R_{CD}$ crosses the solid line for $R_{BC}$.

\begin{figure}[t]
\centering
\includegraphics[width=0.32\textwidth]{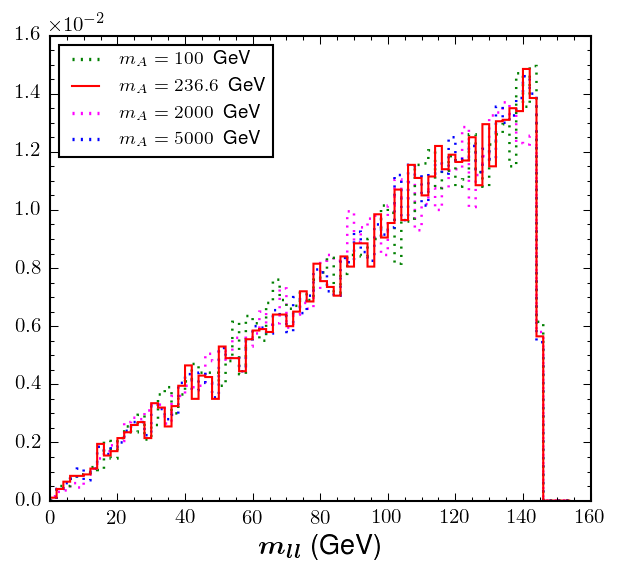}
\includegraphics[width=0.32\textwidth]{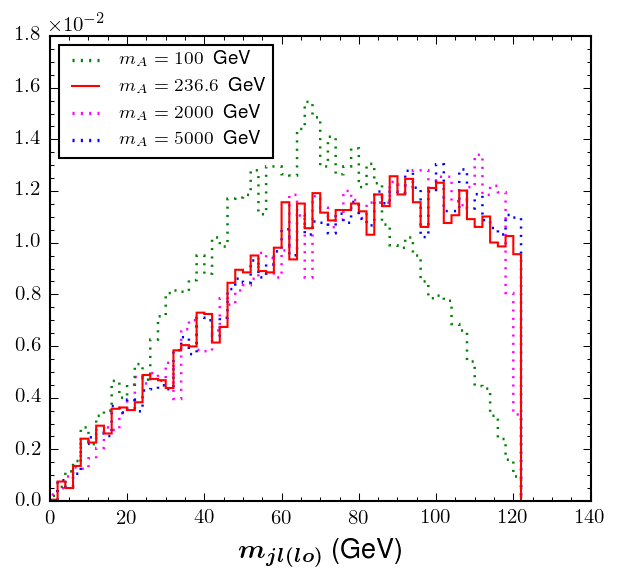}  
\includegraphics[width=0.32\textwidth]{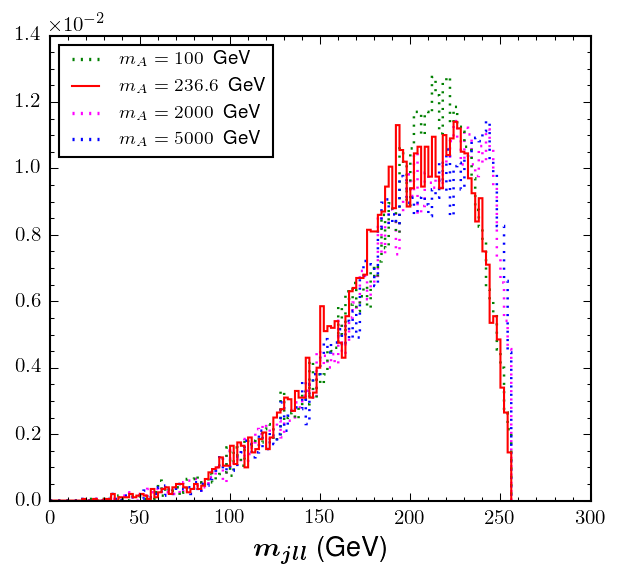}
\caption{ Unit-normalized invariant mass distributions for the four study points from Table~\ref{tab:31}:
the distribution of $m_{\ell\ell}$ (left panel), $m_{j\ell(lo)}$ (middle panel), and $m_{j\ell\ell}$ (right panel).
The lines are color coded according to our conventions from Fig.~\ref{fig:regions} and Table~\ref{tab:31}:
red for $P_{31}$, blue for $P_{41}$, magenta for $P_{43}$ and green for $P_{23}$. \label{fig:3dist_31} 
}
\end{figure}

To summarize our discussion so far, we have imposed the three endpoint measurements $a$, $b$ and $c$ 
on the four-dimensional parameter space (\ref{massparspace}), reducing it to the one-dimensional parameter curve 
depicted in Figs.~\ref{fig:trajectory} and \ref{fig:mass_diff_31}. The curve consists of two branches which
visit four of the colored regions in Fig.~\ref{fig:regions}, and we have chosen one study point in each region.
The four study points are listed in Table~\ref{tab:31}, and their predicted invariant mass distributions
from the {\sc ROOT} phase space generator \cite{root}
are shown in Fig.~\ref{fig:3dist_31}: $m_{\ell\ell}$ in the left panel, $m_{j\ell(lo)}$ in the middle panel 
and $m_{j\ell\ell}$ in the right panel. By construction, for any points along the mass trajectory (\ref{massfamily}),
and in particular for the four study points from Table~\ref{tab:31}, these distributions share 
common kinematic endpoints. Furthermore, as Fig.~\ref{fig:3dist_31} reveals, the shapes of most 
distributions are also very similar, which makes it difficult to pinpoint our exact location along the 
mass trajectory (\ref{massfamily}). This is why in the remainder of this section, we shall focus on 
the question, what additional measurements may allows us to discriminate experimentally points 
along the two branches in Figs.~\ref{fig:trajectory} and \ref{fig:mass_diff_31}, 
and in particular distinguish between the four study points in Table~\ref{tab:31}. 

\begin{figure}[t]
\centering
\includegraphics[width=0.49\textwidth]{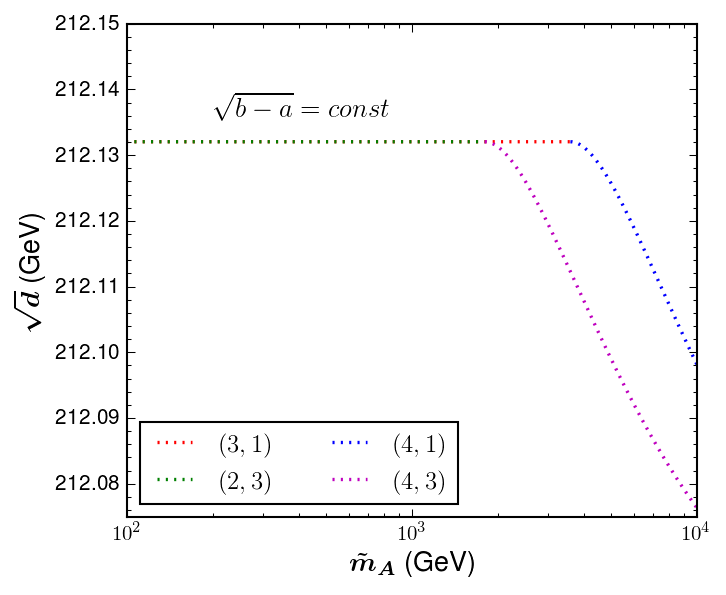}
\includegraphics[width=0.45\textwidth]{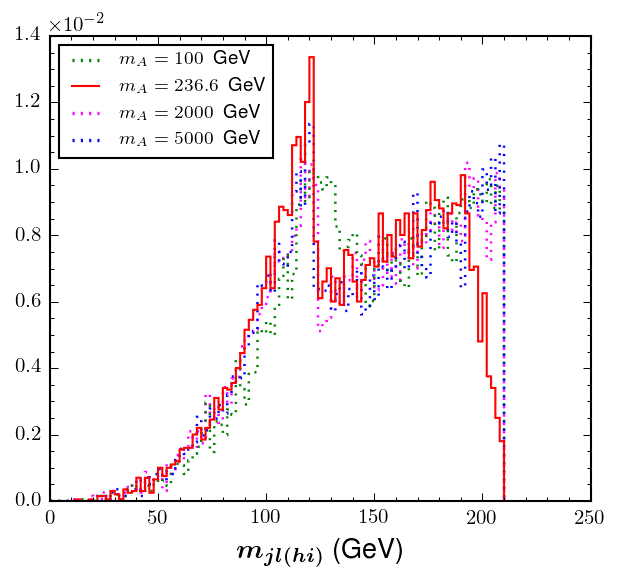}  
\caption{ Left: The prediction for the kinematic endpoint $\sqrt d$ along the flat direction (\ref{massfamily}) generated by $P_{31}$, 
as a function of the trial value of the parameter $\tilde m_A$. Right: The same as Fig.~\ref{fig:3dist_31}, but for the distribution $m_{j\ell(hi)}$. 
\label{fig:d_31} 
}
\end{figure}

\begin{figure}[t]
\centering
\includegraphics[width=0.49\textwidth]{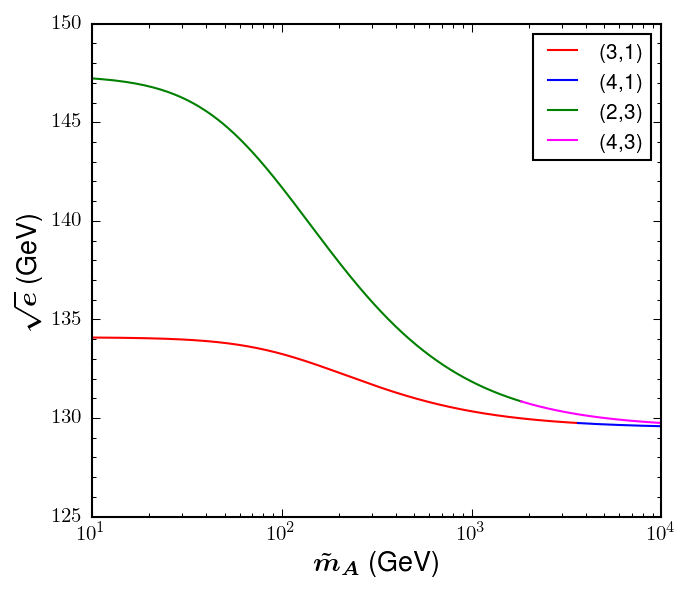}
\includegraphics[width=0.45\textwidth]{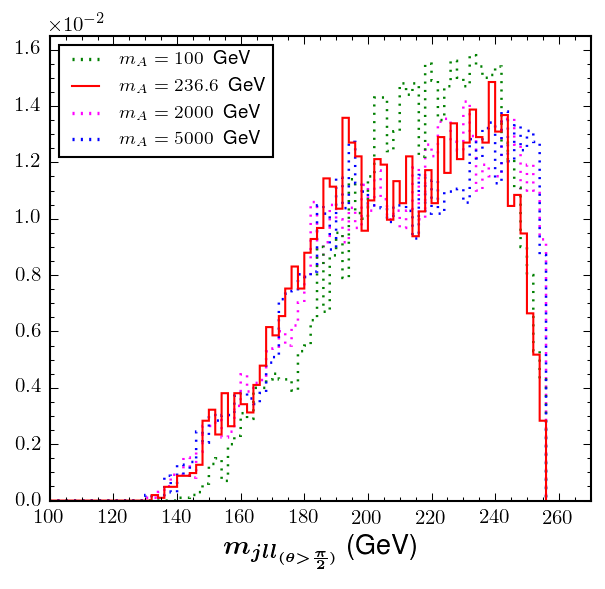}  
\caption{ The same as Fig.~\ref{fig:d_31}, but for the endpoint $\sqrt e$ and the 
corresponding distribution $m_{jll(\theta>\frac{\pi}{2})}$. 
\label{fig:e_31} 
}
\end{figure}

One obvious possibility is to investigate the remaining kinematic endpoints $d$ and $e$,
which are analyzed in Figs.~\ref{fig:d_31} and \ref{fig:e_31}, respectively.
The left panels show the theoretical predictions for the kinematic endpoints $\sqrt{d}=m_{j\ell(hi)}^{max}$ and 
$\sqrt{e}=m_{jll(\theta>\frac{\pi}{2})}^{min}$ along the flat direction (\ref{massfamily}) as a function of $\tilde m_A$,
while the right panels exhibit the corresponding invariant mass distributions for each of our four study points from Table~\ref{tab:31}.

Let us first focus on Fig.~\ref{fig:d_31} which illustrates the $\tilde m_A$ dependence of the $m_{j\ell(hi)}$ distribution 
and its kinematic endpoint $\sqrt{d}$. As we have already discussed, in regions $(3,1)$ and $(2,3)$ the additional measurement
of $\sqrt{d}$ is not useful, since it is not independent --- the value of $d$ is predicted by the relation (\ref{bad}), as confirmed by the 
left panel in Fig.~\ref{fig:d_31}, where the red and green dotted lines representing those two regions are perfectly flat 
and insensitive to $\tilde m_A$. However, this still leaves open the possibility that in the remaining two regions, namely
$(4,1)$ and $(4,3)$, the measurement of the $d$ endpoint will be able to lift the degeneracy and determine the value of $m_A$,
since, at least in theory, $d$ is a non-trivial function of $\tilde m_A$, see (\ref{reg41dformula}) and (\ref{reg43dformula}).
Unfortunately, Fig.~\ref{fig:d_31} demonstrates that this is not the case in practice --- the $\tilde m_A$ dependence is 
extremely weak, and the endpoint value for $\sqrt{d}$ only changes by a few tens of MeV as $\tilde m_A$ is varied over
a range of several TeV! This lack of sensitivity is the reason why we have been referring to the family of mass spectra (\ref{massfamily})
as a ``flat direction" in mass parameter space. Clearly, due to the finite experimental resolution, an endpoint measurement 
with a precision of tens of MeV is not feasible, the anticipated experimental errors at the LHC are significantly higher, 
on the order of a few GeV \cite{Chatrchyan:2013boa}.

It is instructive to understand this lack of sensitivity analytically, by studying, e.g.~the mathematical expression (\ref{reg41dformula}) for $d$
which is relevant for region $(4,1)$. Figs.~\ref{fig:mass_diff_31} and \ref{fig:d_31} already showed that region $(4,1)$ occurs at 
large values of $\tilde m_A$, where the spectrum is relatively heavy --- on the order of several TeV. At the same time, 
the measured parameter inputs into (\ref{reg41dformula}), namely the endpoints $a$, $b$ and $c$, are all on the order of several hundred GeV. 
This suggests an expansion in terms of $1/\tilde m_A$ as
\beq
d(a,b,c,\tilde m_A) \equiv K_0 + \frac{K_1}{\tilde m_A} + {\cal O} \left(\frac{1}{\tilde m_A^2}\right).
\label{dexpansion41}
\eeq
Using (\ref{reg41dformula}), we get the expansion coefficients to be
\bea
K_0& =& \frac{ac}{(a+c)^2}\left(\sqrt{b}+\sqrt{b-a-c}\right)^2,  \label{K0}\\[2mm]
K_1 &=& \frac{ac\left[\left(\sqrt{b}+\sqrt{b-a-c}\right)(a^2+ac-2ab+2bc)+(a^2-c^2)\sqrt{b}\right]}{(a+c)^3}.\label{K1}
\eea
Interestingly, the numerical value of $K_0$ is extremely close to $b-a$:
\beq
K_0\equiv \lim_{\tilde m_A\to \infty} d = (212.047\ {\rm GeV})^2  \leftrightarrow b-a = (212.132\ {\rm GeV})^2.
\label{closeness}
\eeq
Since $K_0$ is the leading order prediction for $d$, (\ref{closeness}) implies that even in region $(4,1)$,
the relation (\ref{bad}) will still hold to a very good approximation --- any deviations from it will be $1/\tilde m_A$ suppressed.
We can formalize this observation by introducing the value $m_A^{(b)}$ which the parameter $\tilde m_A$ takes when
the mass trajectory (\ref{massfamily}) crosses the boundary between regions $(3,1)$ and $(4,1)$.
Using the continuity of the function $d(a,b,c,\tilde m_A)$, we can write
\beq
b-a = K_0 + \frac{K_1}{\tilde m_A^{(b)}} + {\cal O} \left(\frac{1}{\left(\tilde m_A^{(b)}\right)^2}\right),
\label{matching}
\eeq
where the left-hand side is the value of $d$ in region $(3,1)$ which is given by (\ref{reg31dformula}), while the
right-hand side is the value of $d$ as predicted by the Taylor expansion (\ref{dexpansion41}) in region $(4,1)$.
Eliminating $K_0$ from (\ref{matching}), we can rewrite the expansion (\ref{dexpansion41}) in the form
\begin{subequations}
\bea
d(a,b,c,\tilde m_A) &\equiv& b-a + \frac{K_1}{\tilde m_A} - \frac{K_1}{m_A^{(b)}} + {\cal O} \left(\frac{1}{\tilde m_A^2}\right) \label{d41expansion1}\\
&=& b-a - K_1\frac{\tilde m_A-m_A^{(b)}}{\tilde m_Am_A^{(b)}} + {\cal O} \left(\frac{1}{\tilde m_A^2}\right),   \label{d41expansion2}
\eea
\end{subequations}
which manifestly shows that the deviations from the relation (\ref{bad}) are $1/m_A$ suppressed.
One can check that the sign of the $K_1$ coefficient (\ref{K1}) is positive, then (\ref{d41expansion2}) 
explains why $d$ is a {\em decreasing} function of $\tilde m_A$ in region $(4,1)$, as observed in the left panel of 
Fig.~\ref{fig:d_31}.

Starting from (\ref{reg43dformula}), one can repeat the same analysis for the magenta portion of the auxiliary branch 
which is located in region $(4,3)$. As the left panel of Fig.~\ref{fig:d_31} shows, the conclusions will be the same --- 
the $d$ endpoint is still given approximately by the ``bad" relation (\ref{bad}), and the corrections to it are tiny
and $1/m_A$ suppressed. The right panel in Fig.~\ref{fig:d_31} explicitly demonstrates that the variation of the 
$d$ endpoint along the flat direction is unnoticeable by eye even with perfect resolution, large statistics and no background.
The shapes of the $m_{j\ell(hi)}$ distributions are also very similar. As a result, we anticipate that the additional
measurement of the $d$ kinematic endpoint and the analysis of the associated $m_{j\ell(hi)}$ distribution
will not help much in lifting the degeneracy of the flat direction (\ref{massfamily}).

We now turn to the discussion of the fifth and final kinematic endpoint, $e$, illustrated in Fig.~\ref{fig:e_31}.
The left panel now shows a more promising result --- the variation along the flat direction is much larger than 
what we saw previously in Fig.~\ref{fig:d_31}. This is especially noticeable for the auxiliary branch, where 
the prediction for $\sqrt{e}$ can vary by as much as 17 GeV, suggesting that one might be able to at least 
rule out some portions of it. At the same time, the variation of $\sqrt{e}$ along the true branch is only 4 GeV,
once again making it rather difficult to pinpoint an exact location along the true branch. Unfortunately, these
theoretical considerations are dwarfed by the experimental challenges in measuring the $e$ endpoint, as suggested
by the right panel of Fig.~\ref{fig:e_31}. Unlike the other four kinematic endpoints, $e$ is a lower endpoint (a.k.a. ``threshold"), 
which places it in a region where one expects more background. More importantly, the signal distribution is very poorly populated 
near its lower endpoint - the vast majority of signal events appear sufficiently far away from the threshold, and the measurement 
will suffer from a large statistical uncertainty. This casts significant doubts on the feasibility of this measurement --- 
in previous studies, the $\sqrt{e}$ endpoint was either the measurement with the largest experimental error from the fit 
(on the order of 10 GeV \cite{Allanach:2000kt}), or one could not obtain a measurement for it at all \cite{Gjelsten:2004ki}.
One could hope to improve on the precision by utilizing shape information \cite{Gjelsten:2006tg}, but this introduces
additional systematic uncertainty, since the background shape and the shape distortion due to cuts has to be modeled 
with Monte Carlo.

Being mindful of the challenges involved with the measurement of the $e$ endpoint, in this paper we shall 
look for an alternative method for lifting the degeneracy along the flat direction. Our proposal is to study the 
shape of the kinematic boundary (\ref{eq:samosa2}), which is a two-dimensional surface in the three-dimensional 
space of observables 
\beq
\left\{ m_{j\ell(lo)}^2, m_{j\ell(hi)}^2, m_{\ell\ell}^2 \right\}.
\label{3diminvmassspace}
\eeq

\begin{figure}[t]
\centering
\includegraphics[width=0.342\textwidth]{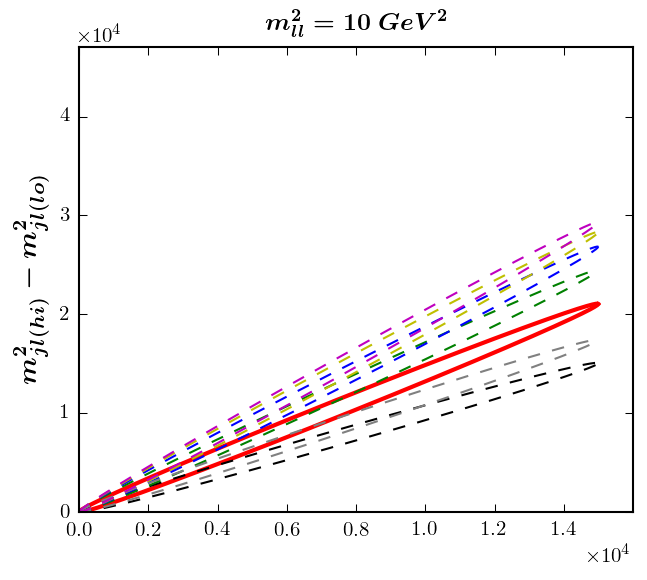}
\includegraphics[width=0.32\textwidth]{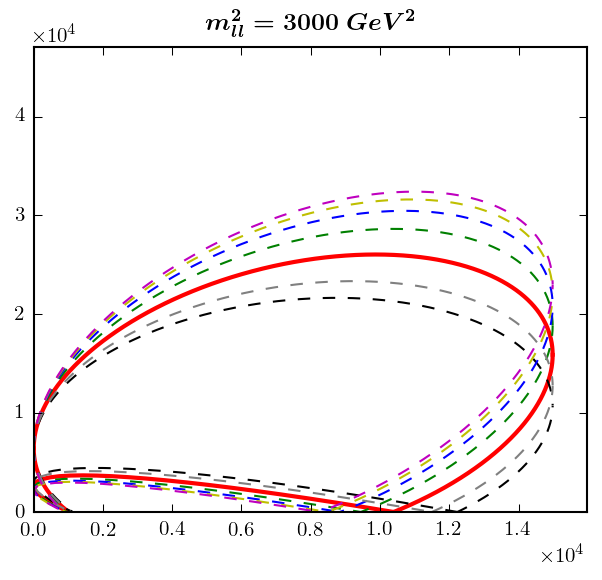}  
\includegraphics[width=0.32\textwidth]{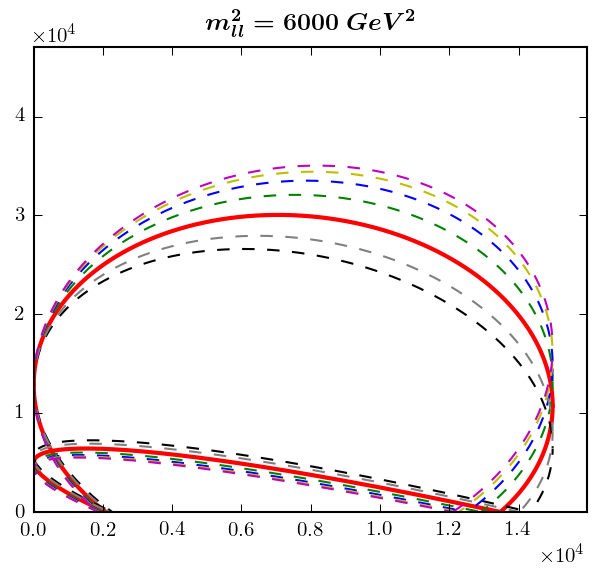}\\
\includegraphics[width=0.342\textwidth]{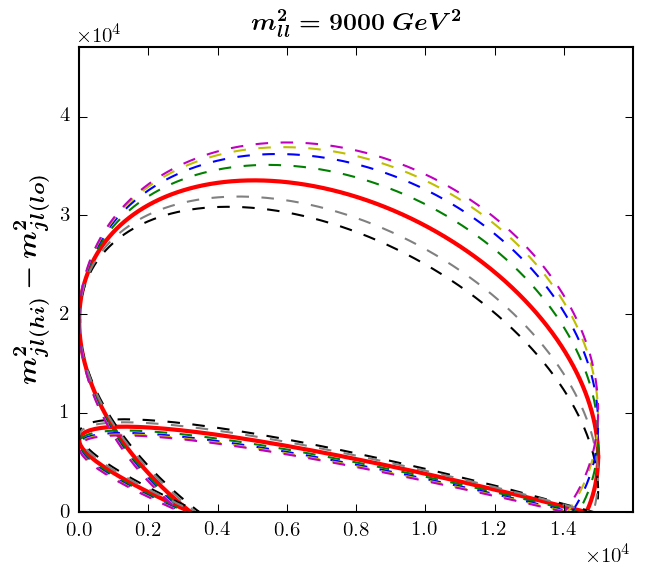}
\includegraphics[width=0.32\textwidth]{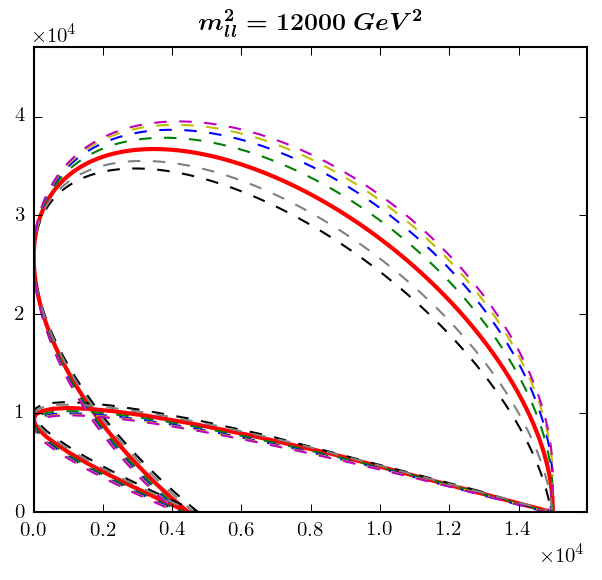}  
\includegraphics[width=0.32\textwidth]{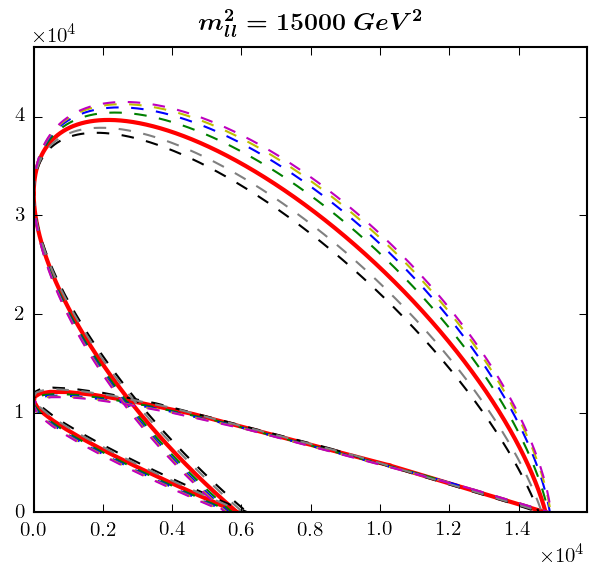}\\
\includegraphics[width=0.342\textwidth]{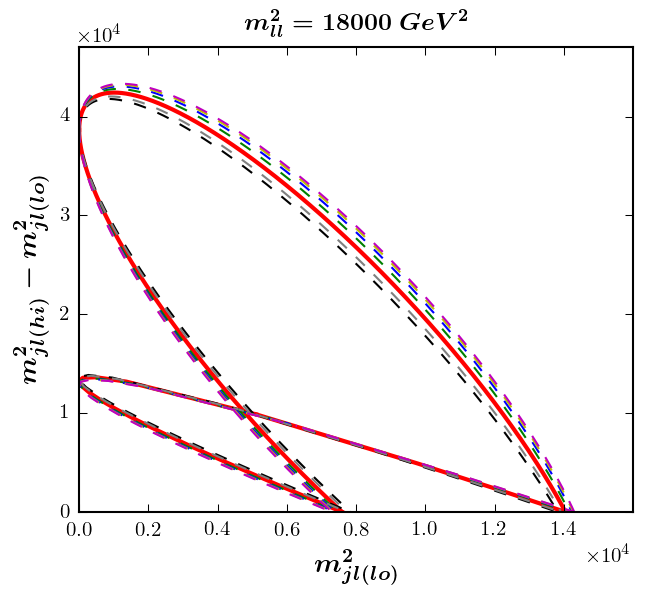}
\includegraphics[width=0.32\textwidth]{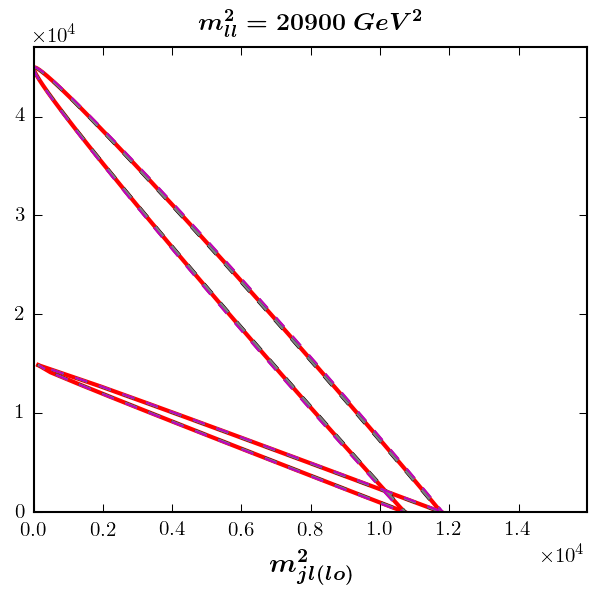}  
\includegraphics[width=0.32\textwidth]{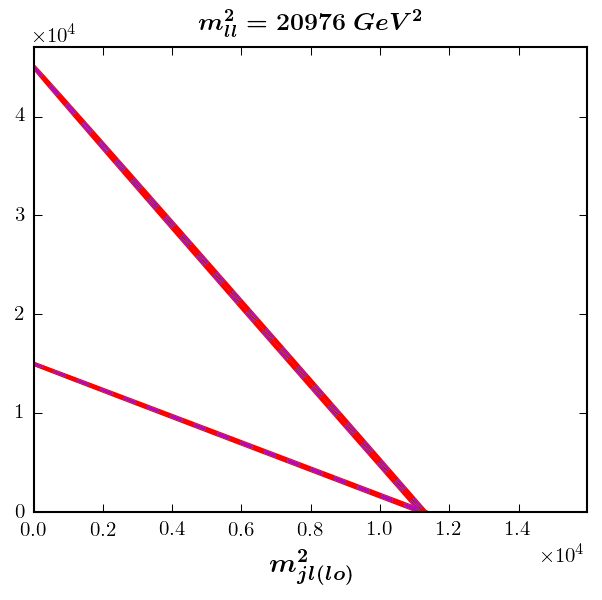}
\caption{ Signal kinematic boundaries in the $(m_{j\ell(lo)}^2, m_{j\ell(hi)}^2-m_{j\ell(lo)}^2)$ plane,
at nine fixed values of $m_{\ell\ell}^2$. Results are shown for several points along the true branch 
in regions $(3,1)$ and $(4,1)$. The red solid line represents the case of the $P_{31}$ study point
with $\tilde m_A=236.6$ GeV, while the dashed lines correspond to other values of $\tilde m_A$ along the true branch:
$\tilde m_A=0$ (black),
$\tilde m_A=100$ GeV (gray),
$\tilde m_A=500$ GeV (green),
$\tilde m_A=1000$ GeV (blue),
$\tilde m_A=2000$ GeV (yellow) and
$\tilde m_A=5000$ GeV (magenta).  
\label{fig:contours_31_true} 
}
\end{figure}

\begin{figure}[t]
\centering
\includegraphics[width=0.342\textwidth]{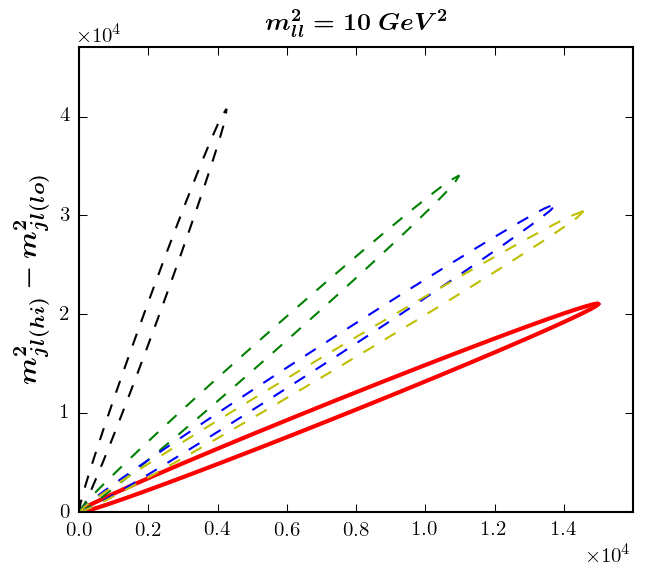}
\includegraphics[width=0.32\textwidth]{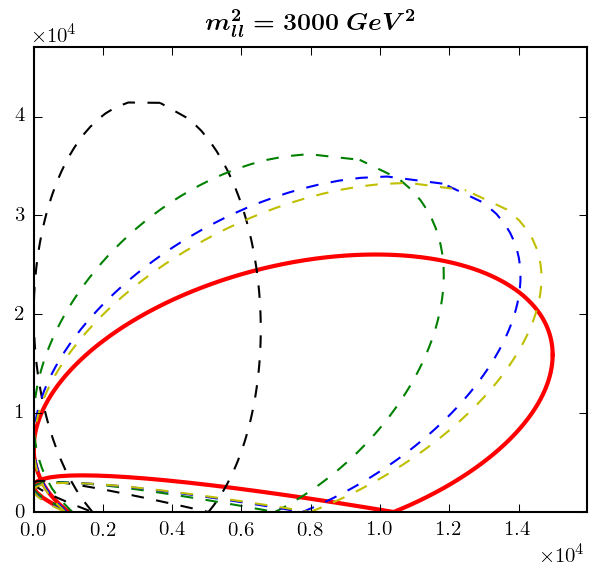}  
\includegraphics[width=0.32\textwidth]{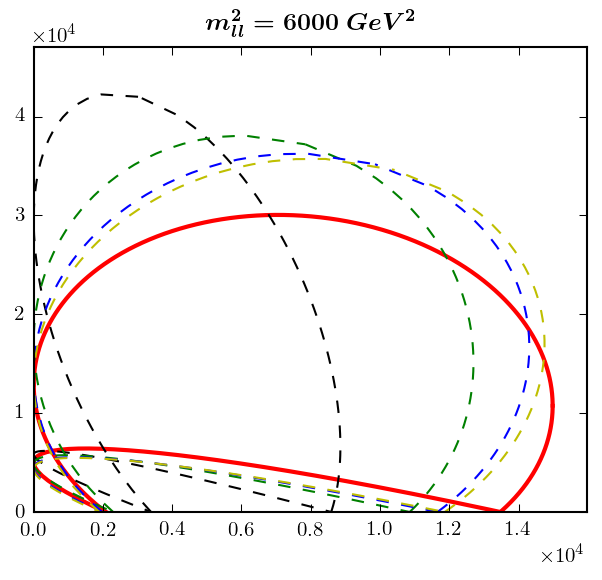}\\
\includegraphics[width=0.342\textwidth]{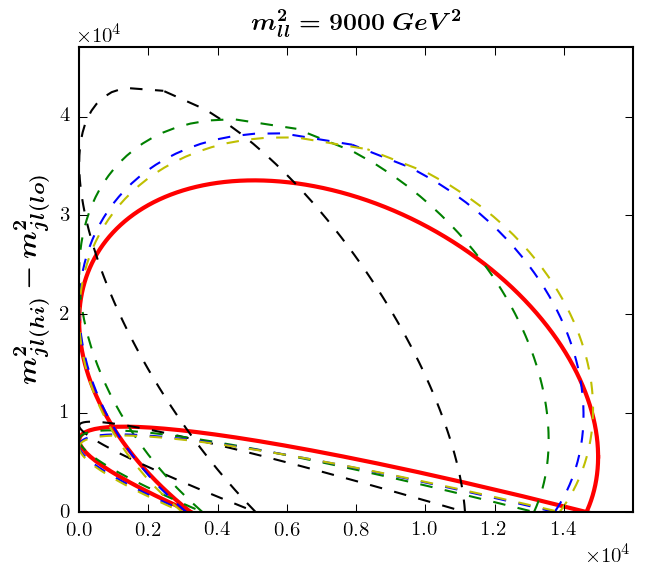}
\includegraphics[width=0.32\textwidth]{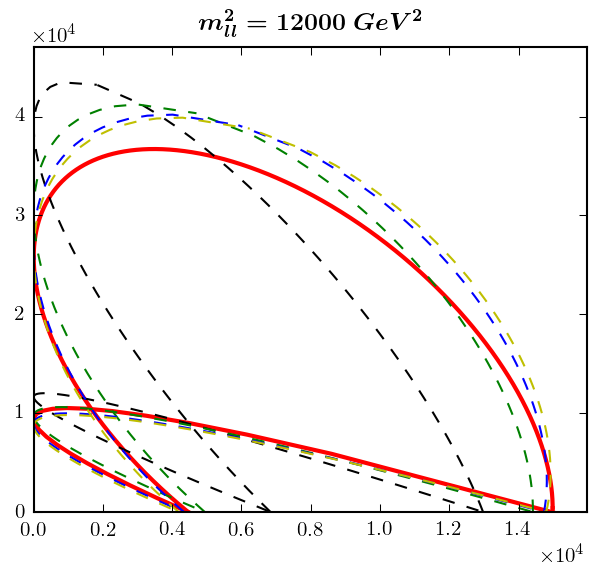}  
\includegraphics[width=0.32\textwidth]{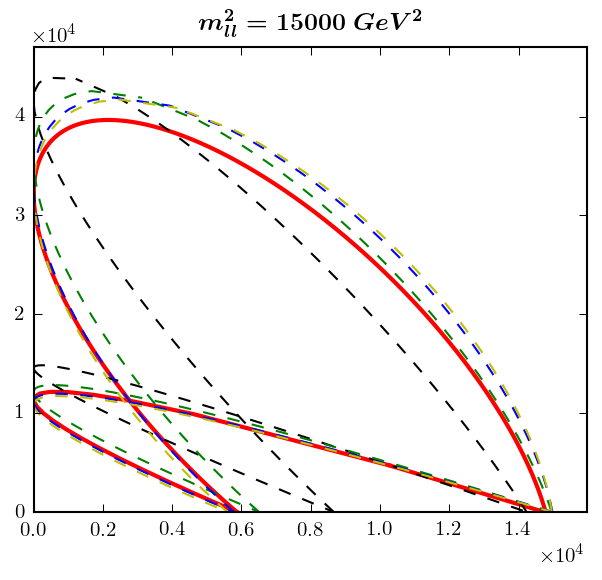}\\
\includegraphics[width=0.342\textwidth]{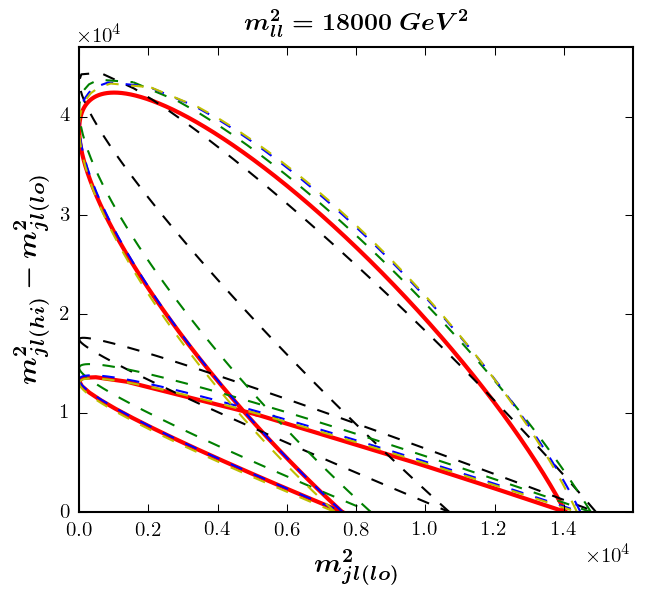}
\includegraphics[width=0.32\textwidth]{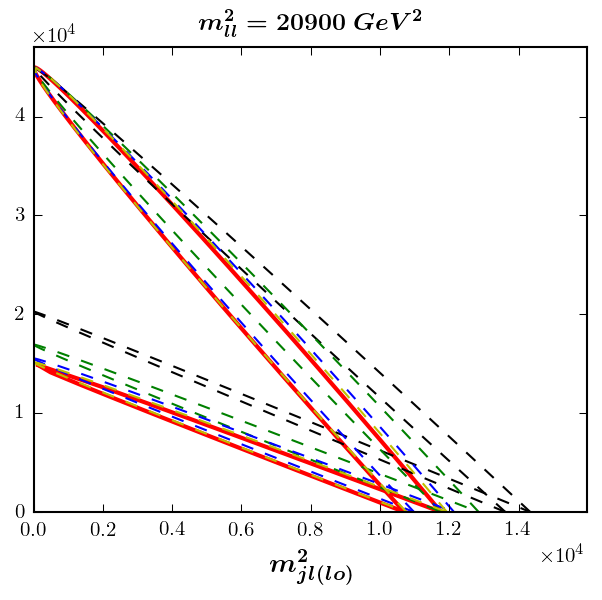}  
\includegraphics[width=0.32\textwidth]{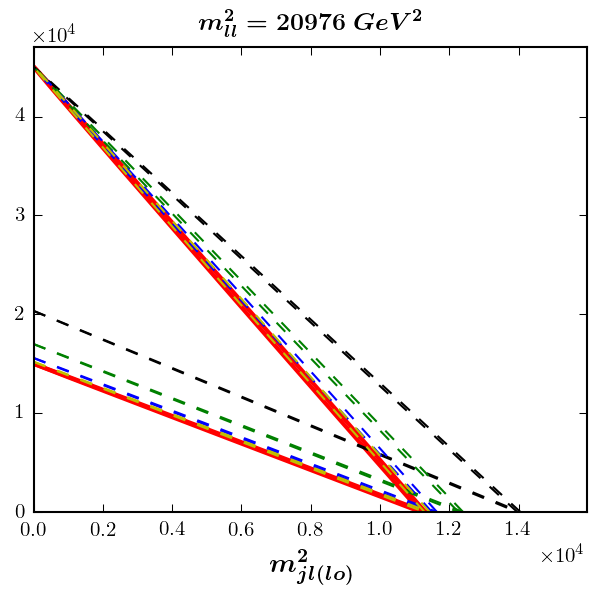}
\caption{ The same as Fig.~\ref{fig:contours_31_true}, but for the auxiliary branch going through regions $(2,3)$ and $(4,3)$.
The dashed lines represent points with 
$\tilde m_A=100$ GeV (black),
$\tilde m_A=500$ GeV (green),
$\tilde m_A=2000$ GeV (blue) and
$\tilde m_A=6000$ GeV (yellow).
%black, green boundary lines are for mass spectrum with $m_A=100, 500$ GeV in (2,3) and the blue and yellow is for $m_A=2000, 6000$ GeV in (4,3).  The 
For reference, we also show the case of the true mass spectrum for point $P_{31}$ (red solid lines), although $P_{31}$ 
does not belong to the auxiliary branch. 
\label{fig:contoursfake_31} 
}
\end{figure}

As a proof of principle, we first illustrate the change in the shape of the surface (\ref{eq:samosa2}) as we move along the 
flat direction. Our results are shown in Fig.~\ref{fig:contours_31_true} (for the true branch) and in
Fig.~\ref{fig:contoursfake_31} (for the auxiliary branch).
Following \cite{Debnath:2016mwb}, we visualize the surface (\ref{eq:samosa2}) by showing a series of 
two-dimensional slices in the $(m_{j\ell(lo)}^2, m_{j\ell(hi)}^2-m_{j\ell(lo)}^2)$ plane, where the slight
modification of the ``$y$-axis" was done in order to avoid wasted space on the plots due to the unphysical areas with 
$m_{j\ell(lo)}> m_{j\ell(hi)}$. Each slice is taken at a fixed value of $m_{\ell\ell}^2$, starting from a very low value ($10\ {\rm GeV}^2$) and going 
up all the way until the kinematic endpoint $\left(m_{\ell\ell}^{max}\right)^2=20,976\ {\rm GeV}^2$.
The red solid lines in Fig.~\ref{fig:contours_31_true} correspond to the nominal case of the study point $P_{31}$.
In each panel, the signal events will be populating the areas delineated by these red solid lines.  
As pointed out in \cite{Agrawal:2013uka}, the density of signal events is enhanced near the phase space boundary,
i.e. signal events will cluster close to the solid red lines; this property can be incorporated into the algorithm
for detecting the surface boundary \cite{Debnath:2016mwb}. It is worth noting that in general, each panel 
contains two signal populations, which arise from the reordering (\ref{mjllodef}-\ref{mjlhidef}) \cite{Burns:2009zi}. 
As we vary the value of $m_{\ell\ell}^2$, the shape of the red solid lines changes in accordance with eq.~(\ref{eq:samosa2}),
which follows from simple phase space considerations. However, the main purpose of Figs.~\ref{fig:contours_31_true} 
and \ref{fig:contoursfake_31} is to check how much the shape is modified relative to the nominal case of $P_{31}$ 
when we vary the value of $\tilde m_A$ along the flat direction (\ref{massfamily}). The dashed lines in 
Fig.~\ref{fig:contours_31_true} show results for several representative values of $\tilde m_A$ along the true branch:
$\tilde m_A=0$ (black), $\tilde m_A=100$ GeV (gray), $\tilde m_A=500$ GeV (green),
$\tilde m_A=1000$ GeV (blue), $\tilde m_A=2000$ GeV (yellow) and $\tilde m_A=5000$ GeV (magenta).
We observe noticeable shape variations, especially at low to intermediate values of $m_{\ell\ell}^2$, which bodes well
for our intended purpose of measuring the value of $m_A$. Fig.~\ref{fig:contours_31_true} aids in visualizing why
sensitivity is lost when performing one-dimensional projections. Consider, for example the variable $m_{j\ell(lo)}$.
The top two rows of Fig.~\ref{fig:contours_31_true} show that as $\tilde m_A$ is varied along the flat direction, 
the boundary contours are being stretched vertically, which does not have any effect on the $m_{j\ell(lo)}$ endpoint.
Later on, when the events are projected vertically on the $m_{j\ell(lo)}$ axis to obtain the $m_{j\ell(lo)}$ distribution 
seen in the middle panel of Fig.~\ref{fig:3dist_31}, the effects from this vertical stretching tend to be washed out and 
the resulting $m_{j\ell(lo)}$ distributions have very similar shapes.

Fig.~\ref{fig:contoursfake_31} shows the analogous results for the auxiliary branch. Once again, the red solid lines 
represent the study point $P_{31}$, while the dashed lines correspond to four values of $\tilde m_A$:
$\tilde m_A=100$ GeV (black), $\tilde m_A=500$ GeV (green), $\tilde m_A=2000$ GeV (blue) and
$\tilde m_A=6000$ GeV (yellow). This time the shape variation along the flat direction is much more significant compared to what 
we saw in Fig.~\ref{fig:contours_31_true}. This observation agrees with our expectation based on Fig.~\ref{fig:e_31}
that points on the auxiliary branch behave quite differently from our nominal study point $P_{31}$, especially at low $\tilde m_A$.

\subsection{A toy study with uniformly distributed background}
\label{sec:uni31}

In the remainder of this section we shall illustrate our proposed method for mass measurement with two exercises. 
In each case, we shall assume that the standard set of one-dimensional kinematic endpoints (\ref{4endpoints})
has already been well measured and used to reduce the relevant mass parameter space (\ref{massparspace})
to the flat direction (\ref{massfamily}) parametrized by the test mass $\tilde m_A$ for the lightest new particle $A$. 
This is done only for simplicity --- in principle, our method would also work without any prior information from endpoint measurements,
but by using those, we are reducing the 4-dimensional optimization problem in (\ref{hypothesis}) to the much simpler 
one-dimensional optimization problem 
\beq
\max_{\tilde m_A} \bar\Sigma\left(\tilde m_A,\tilde m_B(\tilde m_A),\tilde m_C(\tilde m_A),\tilde m_D(\tilde m_A)\right) \simeq
\bar\Sigma(m_A,m_B,m_C,m_D),
\label{hypothesisflatdirection}
\eeq
where $\tilde m_B(\tilde m_A)$, $\tilde m_C(\tilde m_A)$, and $\tilde m_D(\tilde m_A)$ are the masses of particles $B$, $C$ and $D$ 
along the flat direction. Our main emphasis here is on demonstrating the advantages of our method relative to the method of kinematic endpoints. 
In Section~\ref{sec:flat31} we already showed that while the method of kinematic endpoints does a good job in reducing the 
unknown mass parameter space (\ref{massparspace}) to the flat direction (\ref{massfamily}), it does a poor job of 
lifting the degeneracy along the flat direction. Thus, if we can show that our method can perform the remaining mass 
measurement along the flat direction, we will have accomplished our goal.

In order to make contact with our previous studies in \cite{Debnath:2016mwb}, we begin with a simple toy exercise where 
in addition to the signal events from the cascade decay in Fig.~\ref{fig:decay}, we also consider a certain number of background events, 
which we take to be uniformly distributed in the mass squared space of observables (\ref{3diminvmassspace}).
While the assumption of uniform background density is unrealistic, such an exercise is nevertheless worth studying 
for several reasons. First, our method is completely general and applies in any situation where we have a decay of the type
shown in Fig.~\ref{fig:decay}, while to correctly identify the relevant backgrounds, we must be a lot more specific ---  
we need to fix the signature, the type of production mechanism (which determines what else is in the event), the cuts, etc.
In order to retain generality, we choose to avoid specifying those details and instead we generate background events 
by pure Monte Carlo according to a flat hypothesis. Second, as shown in \cite{Debnath:2016mwb}, a uniform background distribution
is actually a pretty good approximation to more realistic backgrounds resulting, e.g., from dilepton $t\bar{t}$ events (compare 
to the results in Section~\ref{sec:tt31} below). Finally, our method is attempting to detect a discontinuity in the measured event density
caused by a signal kinematic boundary, so the exact shape of the background distribution is not that important, as long as 
it is smooth and without any sharp kinematic features.

In order to detect the exact location of the kinematic boundary, we shall be computing the quantity $\bar\Sigma$ defined in (\ref{Sigmabardef})
along the flat direction (\ref{massfamily}), i.e.
\beq
\bar\Sigma(\tilde m_A) \equiv \bar\Sigma\left(\tilde m_A,\tilde m_B(\tilde m_A),\tilde m_C(\tilde m_A),\tilde m_D(\tilde m_A)\right).
\label{Sigmabarflat}
\eeq
We shall perform several versions of the exercise, with varying levels of signal-to-background. For this purpose, we
vary the ratio of signal to background events inside the true ``samosa" surface ${\cal S}(m_A,m_B,m_C,m_D)$:
\beq
S/B \equiv \frac{\int_{V_{\cal S}} \rho_s\, dV} {\int_{V_{\cal S}} \rho_b\, dV},
\label{SoB}
\eeq
where $V_{\cal S}$ is the volume inside the samosa ${\cal S}(m_A,m_B,m_C,m_D)$, while
$\rho_s$ and $\rho_b$ are the signal and background event densities from Section~\ref{sec:introduction}, respectively.
In this exercise, we shall fix the overall normalization by choosing $N_B=1000$ background events inside ${\cal S}$.

\begin{figure}[t]
\centering
\includegraphics[width=0.4\textwidth]{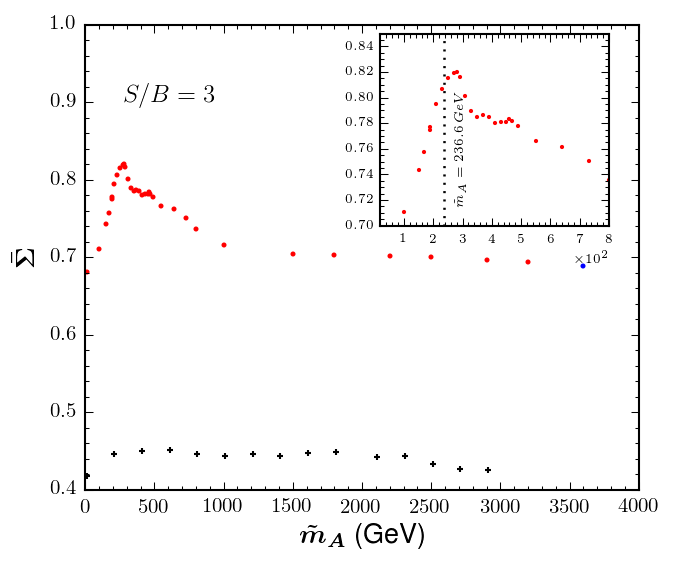}
\includegraphics[width=0.4\textwidth]{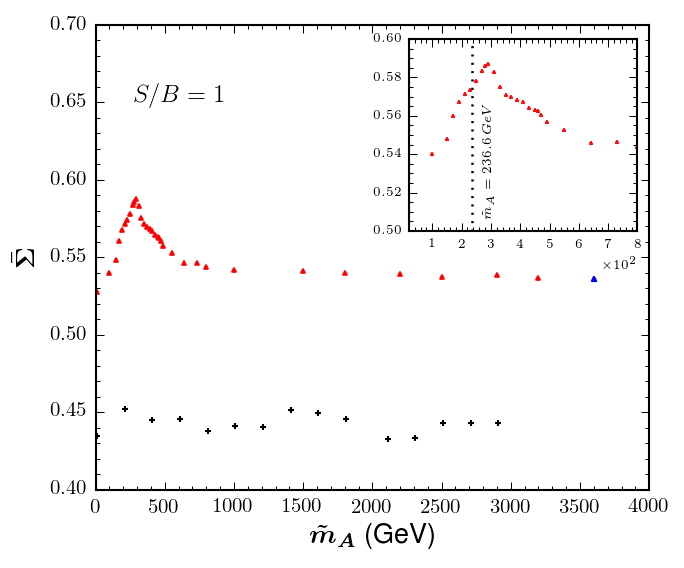}\\
\includegraphics[width=0.4\textwidth]{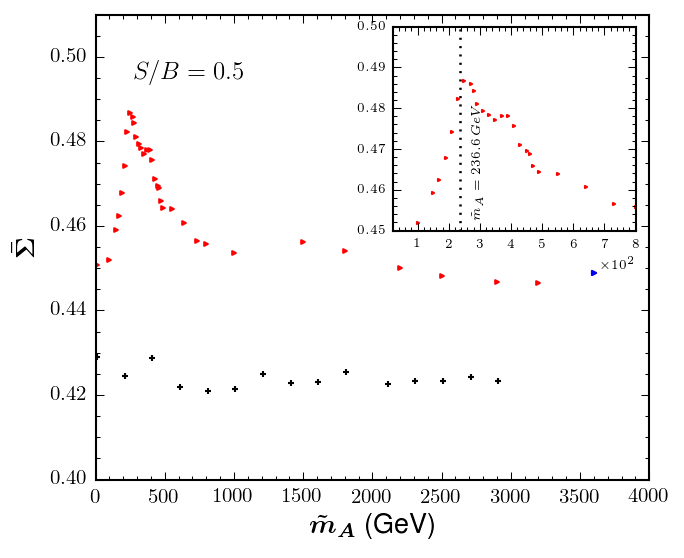}
\includegraphics[width=0.4\textwidth]{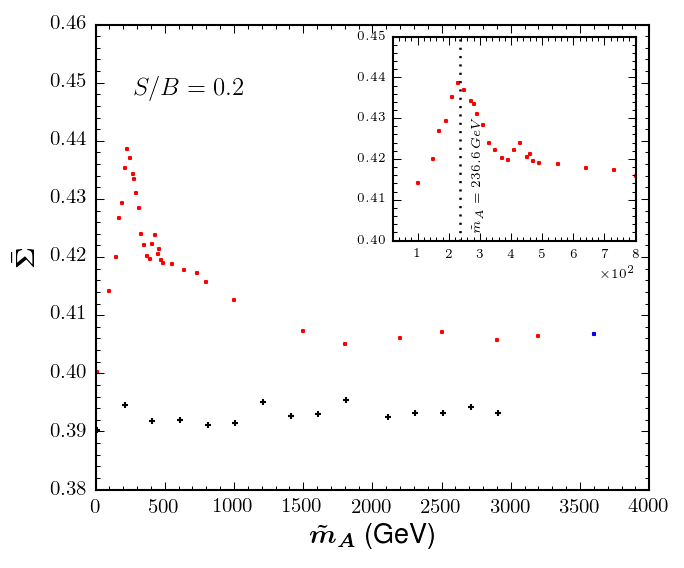}
\caption{ The quantity $\bar\Sigma(\tilde m_A)$ defined in (\ref{Sigmabarflat}) as a function of $\tilde m_A$ for 
different values of the signal to background ratio $S/B$ defined in (\ref{SoB}):
$S/B=3$ (upper left panel), $S/B=1$ (upper right panel), $S/B=0.5$ (lower left panel) and $S/B=0.2$ (lower right panel). 
The colored symbols correspond to the true branch with the color conventions from Fig.~\ref{fig:regions}, 
while the black crosses indicate points on the auxiliary branch.
The insert on each panel zooms in on the region near the peak value for $\bar\Sigma(\tilde m_A)$.
\label{fig:sigma_uniform_31} 
}
\end{figure}

Our main result is shown in Fig.~\ref{fig:sigma_uniform_31}, which plots the quantity $\bar\Sigma(\tilde m_A)$ 
along the flat direction, for several different choices of $S/B$: 
$S/B=3$ (upper left panel), $S/B=1$ (upper right panel), $S/B=0.5$ (lower left panel) and $S/B=0.2$ (lower right panel). 
Each panel contains two sets of points: the colored symbols represent points on the true branch, while the black
crosses indicate points on the auxiliary branch\footnote{Recall from Fig.~\ref{fig:mass_diff_31} that for any given choice 
of $\tilde m_A$, there is one point on the true branch and a corresponding point on the auxiliary branch.}. 

There are several important lessons from Fig.~\ref{fig:sigma_uniform_31}:
\begin{itemize}
\item {\em Viability of the method.} We see that in each panel, the maximum of $\bar\Sigma$ is obtained for 
a value of $\tilde m_A$ which is close to the true value $m_A=236.6$ GeV. This validates our 
conjecture\footnote{Strictly speaking, Fig.~\ref{fig:sigma_uniform_31} tests only the one-dimensional 
version (\ref{hypothesisflatdirection}).}, eq.~(\ref{hypothesis}), and proves the viability of our method.
\item {\em Precision of the method.} Of course, we did not recover {\em exactly} the input value for $m_A$,
but in each case, came relatively close. Each panel of Fig.~\ref{fig:sigma_uniform_31} contains an insert which zooms in on the 
region near the peak, which is sampled more finely. 
For the different values of $S/B=\{3.0, 1.0, 0.5, 0.2\}$, the $\bar\Sigma$ maxima are obtained 
at $\tilde m_A=\{280, 290, 250, 230\}$ GeV, correspondingly. Since the measurement is not perfect, it may 
be instructive to compare the theoretical boundary for the input study point $P_{31}$ to the boundary surface 
found by the fit. This is illustrated in Fig.~\ref{fig:slice_uniform_31}, where in analogy to 
Figs.~\ref{fig:contours_31_true} and \ref{fig:contoursfake_31} we show two-dimensional slices
at fixed $m_{\ell\ell}^2$ of the Voronoi tessellation of the data for the case of $S/B=3$.
The Voronoi cells are color coded by their value of $\bar\sigma_i$ defined in (\ref{defvar}).
As in Figs.~\ref{fig:contours_31_true} and \ref{fig:contoursfake_31}, the red solid line in each panel is the 
expected signal boundary for the nominal case of point $P_{31}$. We notice that the cells with the 
highest values of $\bar\sigma_i$ are indeed clustered near the nominal boundary, in agreement with the results from
Refs.~\cite{Debnath:2016mwb,Debnath:2015wra}. On the other hand, the boundary delineated by the black dashed lines in 
Fig.~\ref{fig:slice_uniform_31} corresponds to the best fit value of $\tilde m_A=280$ GeV,
which was found in the upper left panel in Fig.~\ref{fig:sigma_uniform_31}. The difference between the 
solid red and black dashed contours in Fig.~\ref{fig:slice_uniform_31} is essentially a measure of the resolution of our method.
\begin{figure}[t]
\centering
\includegraphics[width=0.342\textwidth]{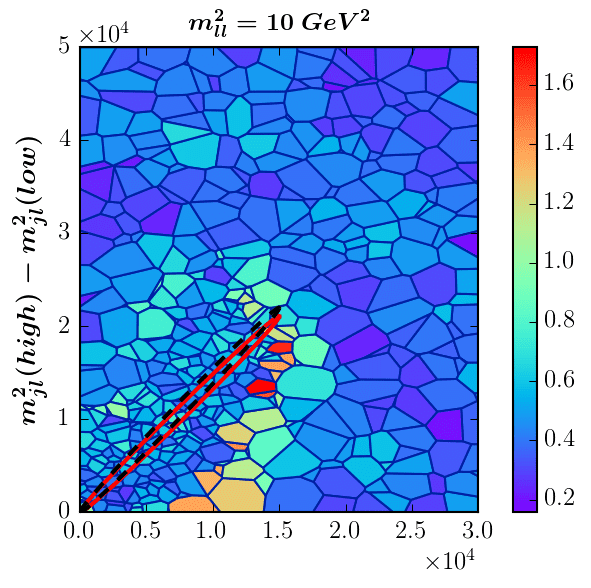}
\includegraphics[width=0.32\textwidth]{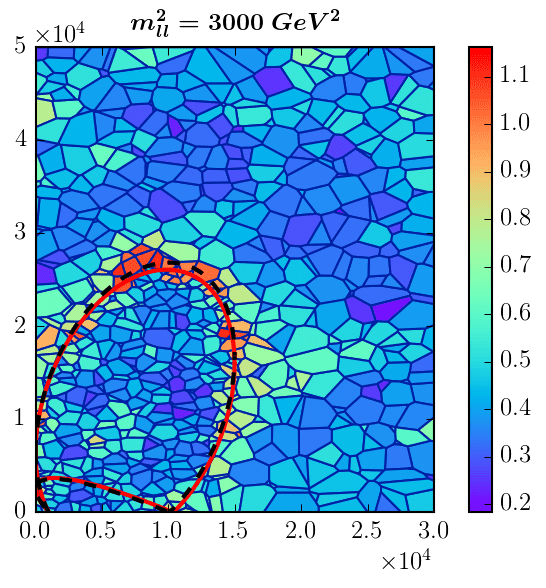}  
\includegraphics[width=0.32\textwidth]{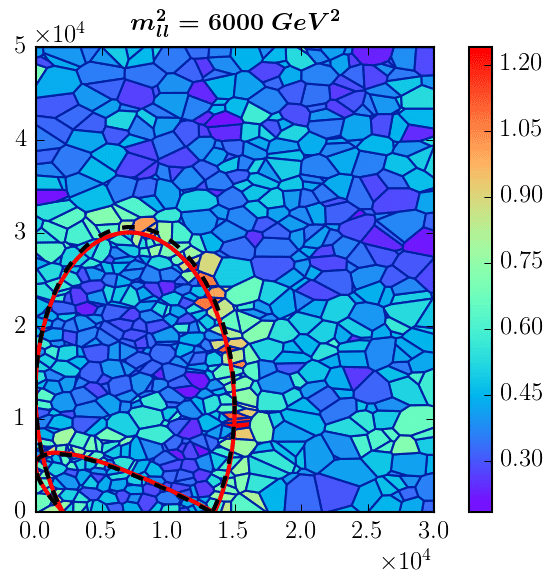}\\
\includegraphics[width=0.342\textwidth]{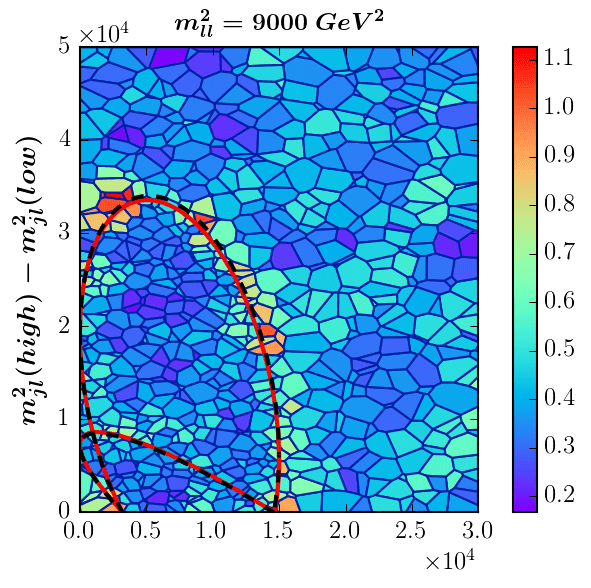}
\includegraphics[width=0.32\textwidth]{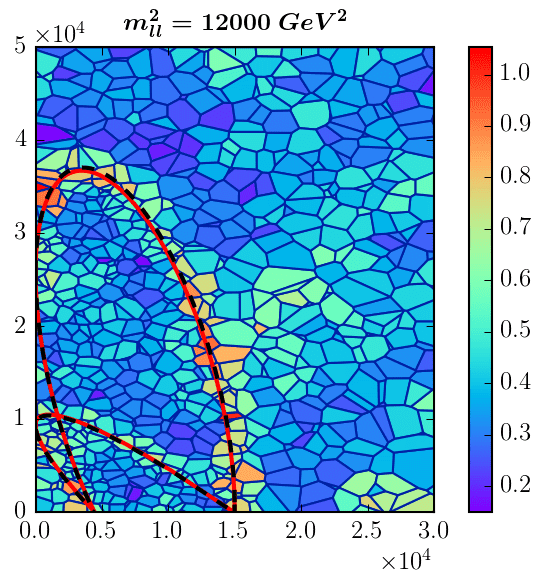}  
\includegraphics[width=0.32\textwidth]{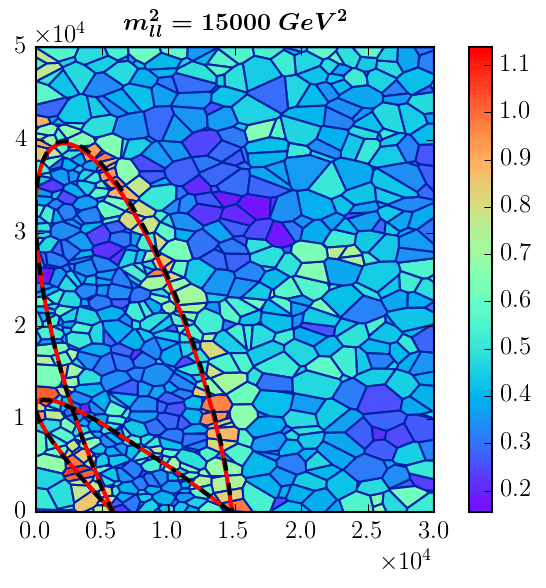}\\
\includegraphics[width=0.342\textwidth]{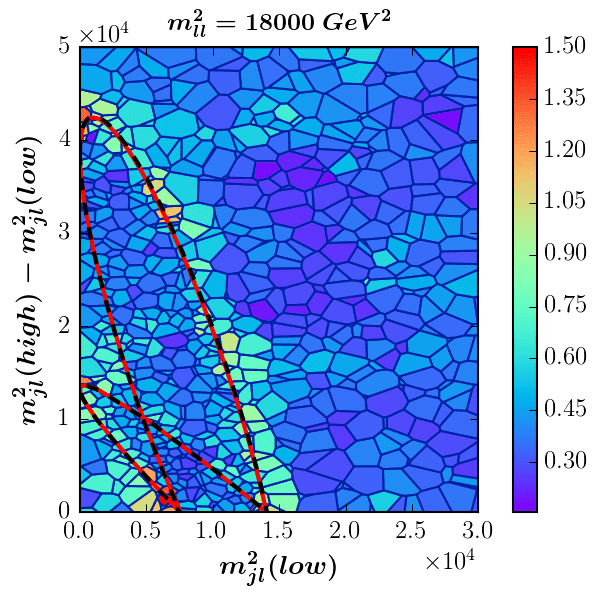}
\includegraphics[width=0.32\textwidth]{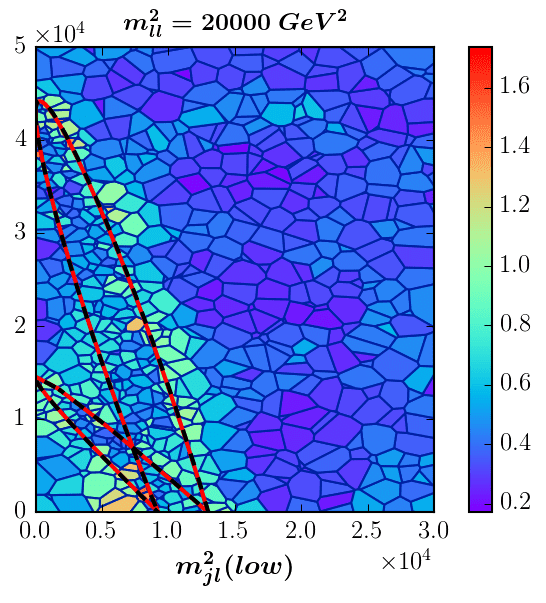}  
\includegraphics[width=0.32\textwidth]{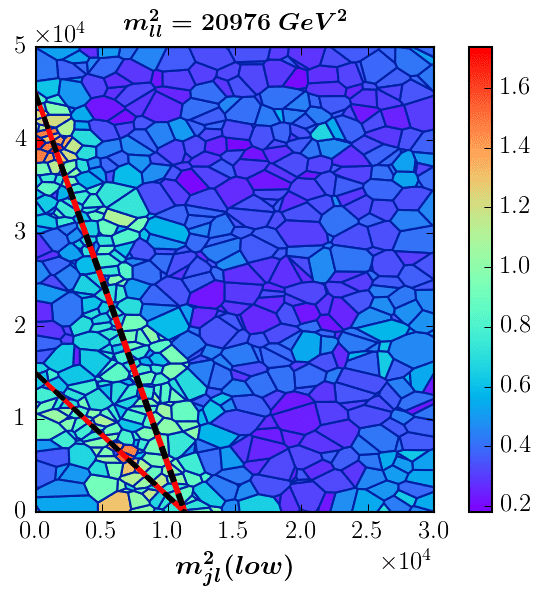}
\caption{Two-dimensional views at fixed $m_{\ell\ell}^2$ of the Voronoi tessellation of the data for the case of $S/B=3$. 
The red solid line is the expected signal boundary for the nominal case of point $P_{31}$, i.e., with the true value 
$\tilde m_A=m_A=236.6$ GeV. The black dashed line corresponds to the mass spectrum with $\tilde m_A=280$ GeV,
which was found to maximize the quantity $\bar\Sigma$ in the top left panel of Fig.~\ref{fig:sigma_uniform_31}.
\label{fig:slice_uniform_31} 
}
\end{figure}
\item {\em Elimination of the auxiliary branch and the large $\tilde m_A$ tail of the true branch.} One very positive piece of news from 
Fig.~\ref{fig:sigma_uniform_31} is that the whole auxiliary branch has very low values for $\bar\Sigma$ which makes it
easy to rule it out --- one can see that no point on the auxiliary branch was ever in contention for the top spot.
Similar comments, albeit to a lesser extent, also apply to the long tail along the true branch at large $\tilde m_A$. 
In particular, region $(4,1)$ seems to be ruled out, as well as the large $\tilde m_A$ portion of region $(3,1)$. 
In effect, the range of possible values for $\tilde m_A$ along the flat direction has been significantly narrowed down to 
a small interval within a few tens of GeV of the true value $m_A$. 
\item {\em The adverse effect of the background.} Comparing the different panels in Fig.~\ref{fig:sigma_uniform_31}, we see that 
as we make $S/B$ smaller, the difference between the true and auxiliary branch is reduced, but the auxiliary branch 
is still disfavored. As for the true branch, the peak near $m_A$ still persists, even in the case when the data is dominated 
by background events. This is not surprising, since the background distribution is relatively smooth, so that in the background-dominated 
regions of phase space there aren't too many Voronoi cells with large values of $\bar\sigma_i$, which could adversely affect the fit.
\end{itemize}

\subsection{A study with $t\bar{t}$ dilepton background events}
\label{sec:tt31}

We are now in position to repeat the exercise from Section~\ref{sec:uni31}, with signal events from $D+A$ associated production and
background taken from dilepton $t\bar{t}$ events, which represent the main background to the signature from Fig.~\ref{fig:decay}
of a jet plus two opposite sign, same flavor leptons (the electroweak backgrounds involving leptonic $Z$ decays can 
be suppressed with a $Z$ mass veto). Events were generated at parton level for LHC at 14 TeV with {\sc MadGraph5} \cite{Alwall:2014hca}
version 2.1.1 with the default PDF set {\tt cteq6l1}. For signal we used the SUSY version of the cascade decay in Fig.~\ref{fig:decay},
and considered the associate production of a squark $\tilde q$ with the lightest neutralino $\tilde \chi^0_1$, 
namely $pp\to \tilde q \tilde \chi^0_1$ \cite{Reuter:2010nx,Allanach:2010pp}. 
Since each $t\bar{t}$ background event contains two jets, there is a two-fold ambiguity in 
the jet selection. We will use both possible pairings, so that each background event will contribute two entries to our data.
Of course, we do not know {\em a priori} how many of those entries will end up inside the nominal boundary surface  
${\cal S}(m_A,m_B,m_C,m_D)$, which is why we have to use a slightly different normalization from Sec~\ref{sec:uni31}.
We shall fix the number of signal events to $N_S=3000$, and then we shall consider several values\footnote{The 
anticipated signal-to-background ratio is model-dependent. In this sense, SUSY may not be the best case for discovery,
since other scenarios, e.g., UED \cite{Macesanu:2002db,ElKacimi:2009zj,Beuria:2017jez}, have higher signal cross-sections.} for the number 
of dilepton $t\bar{t}$ events: $N_B=\{3000, 4000, 5000, 6000\}$. From Monte Carlo we then find that these choices
correspond to $S/B=\{1.52, 1.14, 0.91, 0.76\}$ inside the ${\cal S}$ boundary, see (\ref{SoB}).

\begin{figure}[t]
\centering
\includegraphics[width=0.4\textwidth]{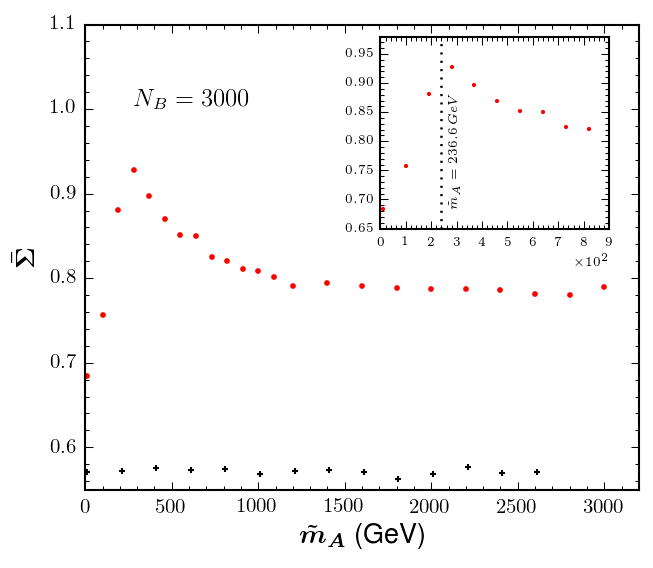}
\includegraphics[width=0.4\textwidth]{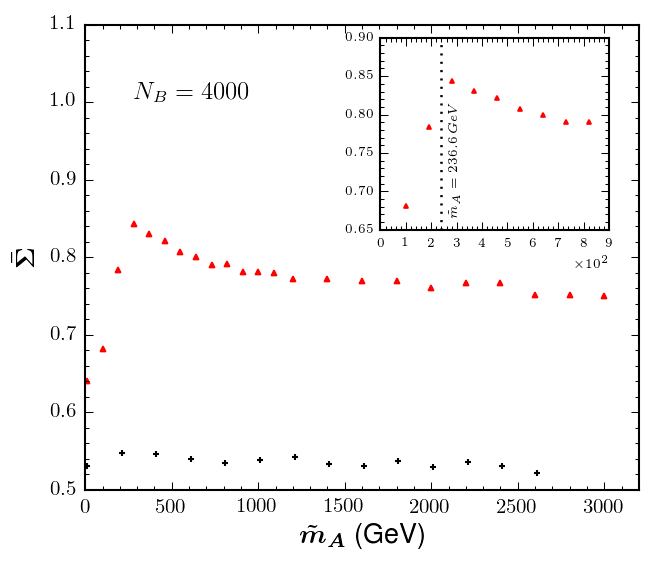}\\
\includegraphics[width=0.4\textwidth]{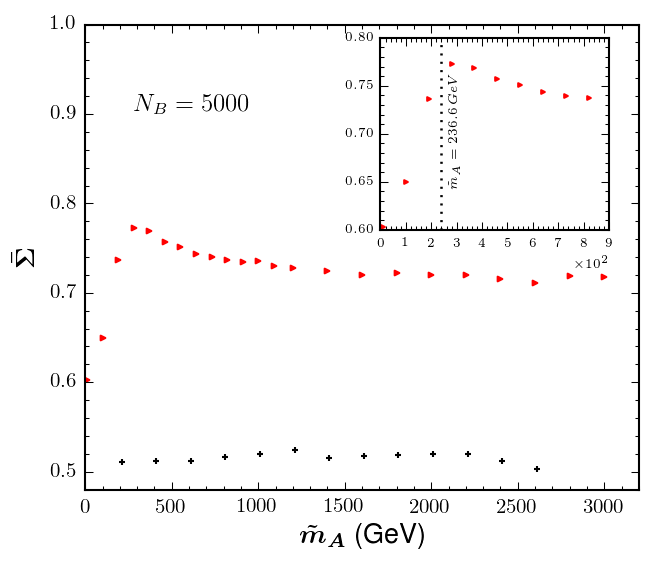}
\includegraphics[width=0.4\textwidth]{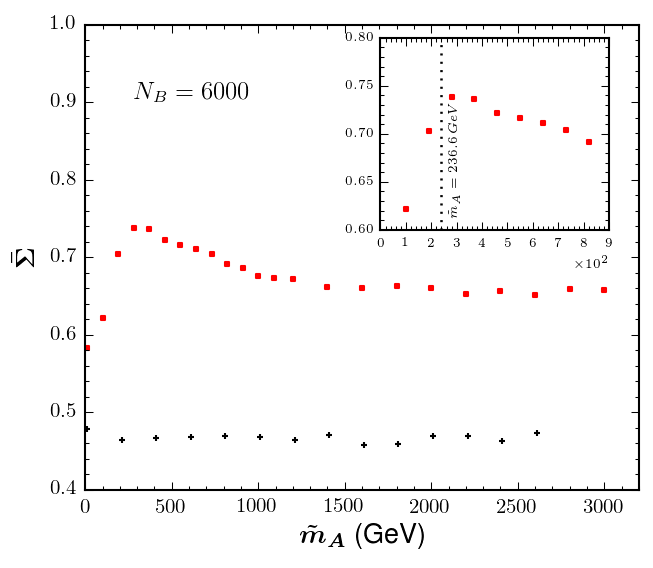}
\caption{ The analogue of Fig.~\ref{fig:sigma_uniform_31} for the exercise with $t\bar{t}$ background events considered in Section~\ref{sec:tt31}.
Results are shown for $N_S=3000$ signal events and several choices for the number of background events:
$N_B=3000$ (upper left panel), $N_B=4000$ (upper right panel), $N_B=5000$ (lower left panel) and $N_B=6000$ (lower right panel). 
\label{fig:sigma_ttbar_31} 
}
\end{figure}
Our main result is shown in Fig.~\ref{fig:sigma_ttbar_31}, which is the analogue of Fig.~\ref{fig:sigma_uniform_31} for this case.
Once again, we find that the function $\bar\Sigma(\tilde m_A)$ is maximized in the vicinity of $\tilde m_A=m_A=236.6$ GeV. 
\begin{figure}[t]
\centering
\includegraphics[height=0.37\textwidth]{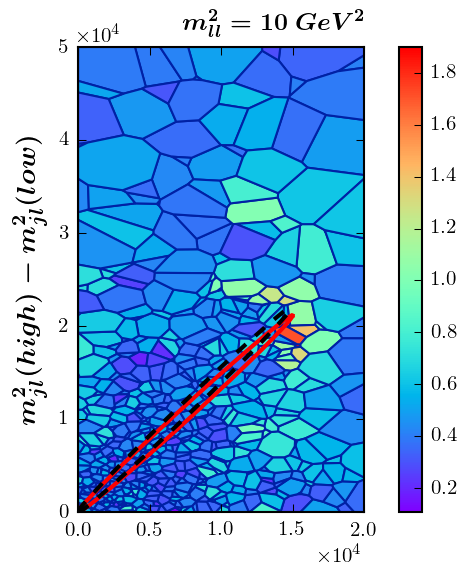}
\includegraphics[height=0.37\textwidth]{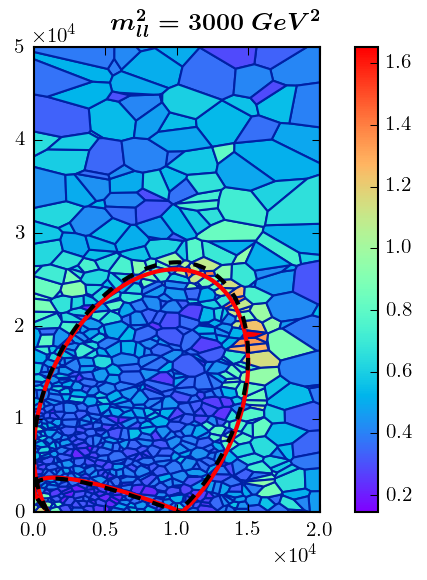}  
\includegraphics[height=0.37\textwidth]{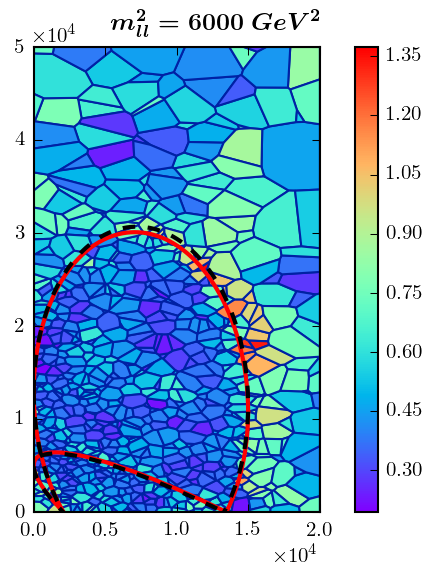}\\
\includegraphics[height=0.37\textwidth]{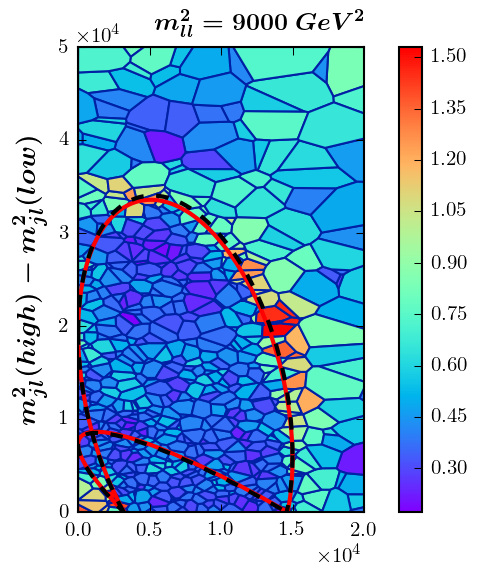}
\includegraphics[height=0.37\textwidth]{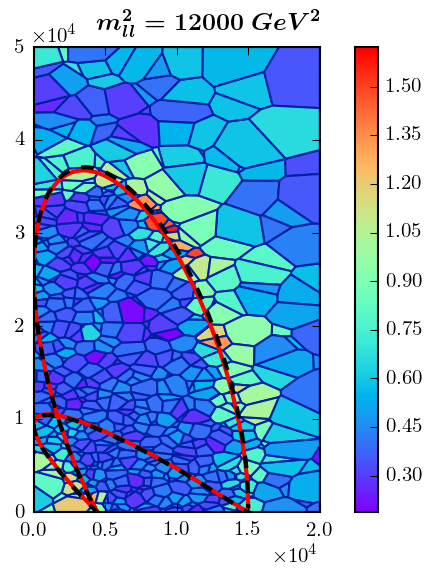}  
\includegraphics[height=0.37\textwidth]{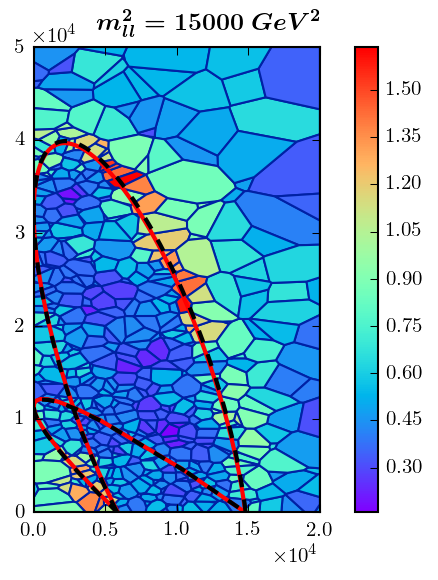}\\
\includegraphics[height=0.37\textwidth]{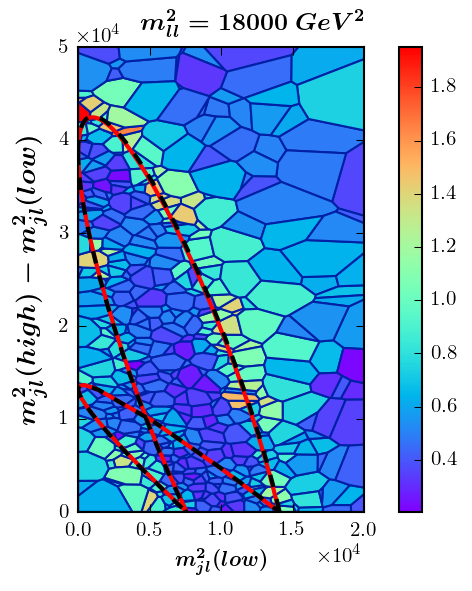}
\includegraphics[height=0.37\textwidth]{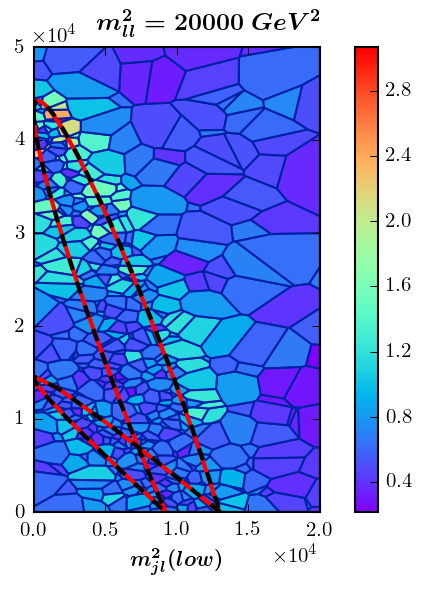}  
\includegraphics[height=0.37\textwidth]{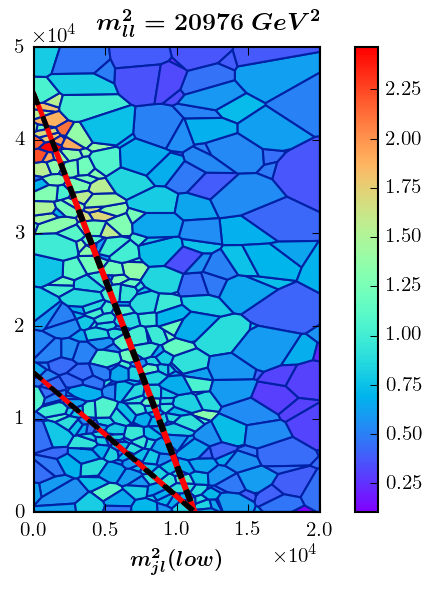}
\caption{ The analogue of Fig.~\ref{fig:slice_uniform_31} for the exercise with $t\bar{t}$ background events considered in Section~\ref{sec:tt31}.
The Voronoi tessellation was done for the case of $N_B=3000$.
The red solid line is the phase space boundary for the nominal value $m_A=236.6$ GeV, while 
the black dashed line corresponds to the best fit value $\tilde m_A=280.0$ GeV found in the 
the top left panel of Fig.~\ref{fig:sigma_ttbar_31}. 
\label{fig:slice_ttbar_31} 
}
\end{figure}
The fitting procedure is illustrated in Fig.~\ref{fig:slice_ttbar_31}, which is the analogue of Fig.~\ref{fig:slice_uniform_31}.
The red solid lines show the boundary contours for the nominal value of $m_A=236.6$ GeV, while the black dashed lines
are for the best fit value of $\tilde m_A=280.0$ GeV, which was found in the top left panel of Fig.~\ref{fig:sigma_ttbar_31}. 
Fig.~\ref{fig:sigma_ttbar_31} shows that once again, our procedure has disfavored the whole auxiliary branch and narrowed down the 
range of viable values of $\tilde m_A$ to a few tens of GeV around the nominal value $m_A$.

\section{A case study in region $(3,2)$}
\label{sec:case32}

In this section, we shall repeat the analysis from Section~\ref{sec:case31}, only this time our nominal study point, from now on 
labelled as $P_{32}$, will be chosen within the cyan region $(3,2)$ of Fig.~\ref{fig:regions}. Recall that the problematic relation (\ref{bad}) was satisfied 
in three of the colored regions in Fig.~\ref{fig:regions}, namely $(2,3)$, $(3,1)$ and $(3,2)$. The former two regions, $(2,3)$ and $(3,1)$, were already
visited by the mass trajectory studied in Section~\ref{sec:case31}, thus here for completeness we will also illustrate the
case of region $(3,2)$. 

\begin{table}[t]
\centering
\begin{tabular}{|l|c||r|r||r|r|r||}
%\hline
\cline{3-7}
\multicolumn{2}{ c||}{}
         &  \multicolumn{2}{c||}{true branch}
         &  \multicolumn{3}{c||} {mirror branch}    \\ [0.5mm]
%\cline{3-7}
\hline
\multicolumn{2}{|c||}{Region}
         & \cellcolor{cy}$(3,2)$ & \cellcolor{red}$(3,1)$   & \cellcolor{yellow}$(4,2)$  &  \cellcolor{mag}$(4,3)$  &  \cellcolor{green}$(2,3)$   \\  [0.5mm] 
%\cline{3-7}
\hline
\multicolumn{2}{|c||}{Study point}
         &\cellcolor{cy}$P_{32}$& \cellcolor{red}$P_{31}$ & \cellcolor{yellow}$P_{42}$ & \cellcolor{mag}$P_{43}$ & \cellcolor{green} $P_{23}$    \\  [0.5mm] \hline \hline
\multicolumn{2}{|c||}{$m_A$ (GeV)}
         & \cellcolor{cy}126.49  & \cellcolor{red}5000.00  & \cellcolor{yellow}90.00 & \cellcolor{mag} 150.00 & \cellcolor{green} 500.00  \\ [0.5mm] \hline 
\multicolumn{2}{|c||}{$m_B$ (GeV)}
         & \cellcolor{cy}282.84 & \cellcolor{red}5207.42  & \cellcolor{yellow}194.61 & \cellcolor{mag}250.96 & \cellcolor{green}609.04    \\ [0.5mm] \hline 
\multicolumn{2}{|c||}{$m_C$ (GeV)}
         & \cellcolor{cy}447.21 & \cellcolor{red}5324.17  & \cellcolor{yellow}399.99 &\cellcolor{mag} 460.80 &\cellcolor{green} 815.72  \\ [0.5mm] \hline 
\multicolumn{2}{|c||}{$m_D$ (GeV)}
         & \cellcolor{cy}500.00 & \cellcolor{red}5372.07  & \cellcolor{yellow}458.78  & \cellcolor{mag}518.78  &\cellcolor{green} 869.36 \\ [0.5mm] \hline 
\multicolumn{2}{|c||}{$R_{AB}$}
         & \cellcolor{cy} 0.200 &  \cellcolor{red}0.922   & \cellcolor{yellow}  0.214 & \cellcolor{mag} 0.357  & \cellcolor{green}0.674  \\ [0.5mm] \hline 
\multicolumn{2}{|c||}{$R_{BC}$}
         & \cellcolor{cy} 0.400 &  \cellcolor{red}0.957   &  \cellcolor{yellow} 0.237 & \cellcolor{mag} 0.297   &  \cellcolor{green}0.557\\ [0.5mm] \hline 
\multicolumn{2}{|c||}{$R_{CD}$}
         & \cellcolor{cy} 0.800 & \cellcolor{red} 0.982   & \cellcolor{yellow}  0.760 & \cellcolor{mag} 0.789   & \cellcolor{green}0.880 \\ [0.5mm] \hline 
\hline
$m_{ll}^{max}$ (GeV)  & $\sqrt{a}$
         & \multicolumn{5}{c||}{309.84}  \\ [0.5mm] \hline
$m_{jll}^{max}$ (GeV) & $\sqrt{b}$
         & \multicolumn{5}{c||}{368.78}   \\ [0.5mm] \hline
$m_{jl(lo)}^{max}$ (GeV)  & $\sqrt{c}$
         & \multicolumn{5}{c||}{149.07 }  \\ [0.5mm] \hline
$m_{jl(hi)}^{max}$ (GeV)  & $\sqrt{d}$
         &\cellcolor{cy} 200.00  & \cellcolor{red} 200.00    &\cellcolor{yellow} 198.23    & \cellcolor{mag}199.87 & \cellcolor{green}200.00 \\ [0.5mm] \hline
$m_{jll(\theta>\frac{\pi}{2})}^{min}$ (GeV)  & $\sqrt{e}$
         & \cellcolor{cy}247.94  & \cellcolor{red} 237.47    & \cellcolor{yellow}253.72     & \cellcolor{mag} 250.99   & \cellcolor{green}243.81\\ [0.5mm] \hline
\hline
%$m_{jl_f}^{max}$ (GeV)  & $\sqrt{f}$
%         &212     & 127      & 200     & 183                   \\ [0.5mm] \hline
%$m_{jl_f}^{(p)}$ (GeV)  & $\sqrt{p}$
%         &190     & 112      & 126     & 115                   \\ [0.5mm] \hline
%$m_{jl_n}^{max}$ (GeV)  & $\sqrt{n}$
%         &122     & 212      & 173     & 200                   \\ [0.5mm] \hline
%$m_{jl(eq)}^{max}$ (GeV)& $\sqrt{q}$
%         &  NA   %167.705 no equal point in region 1 
%                  & 122      & 149     & 149                   \\ [0.5mm] \hline
%\hline
\end{tabular}
\caption{\label{tab:32}
The same as Table~\ref{tab:31}, except now the starting point is a point ($P_{32}$)
from region $(3,2)$. }
\end{table}
\begin{figure}[t]
\centering
\includegraphics[width=0.32\textwidth]{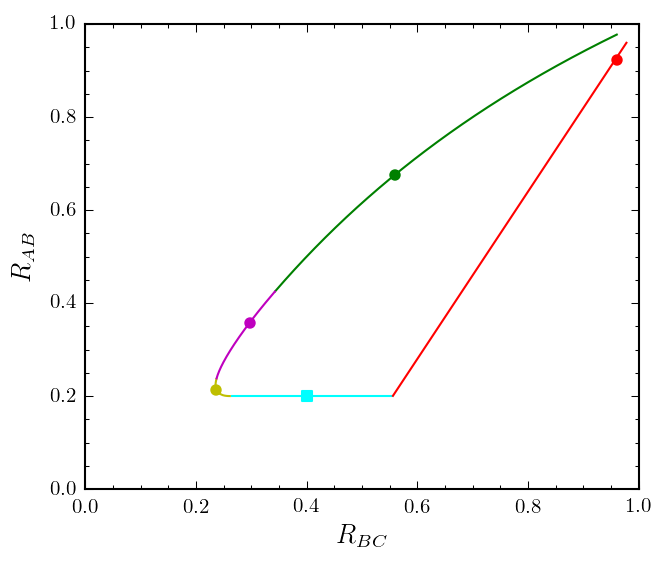}
\includegraphics[width=0.32\textwidth]{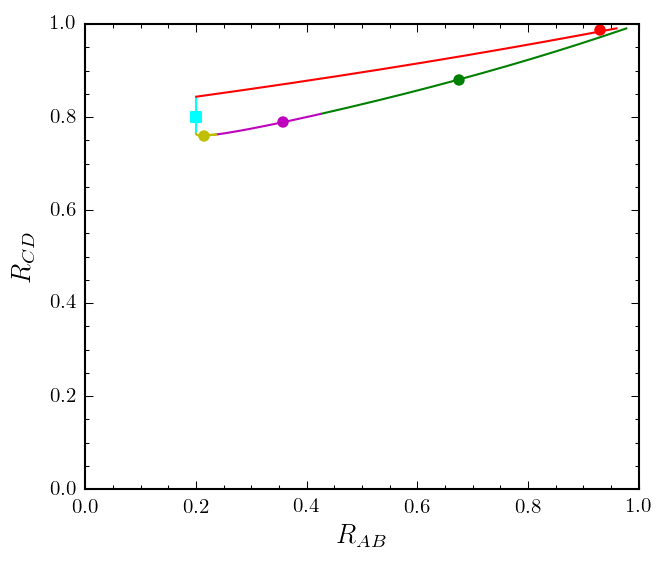}  
\includegraphics[width=0.32\textwidth]{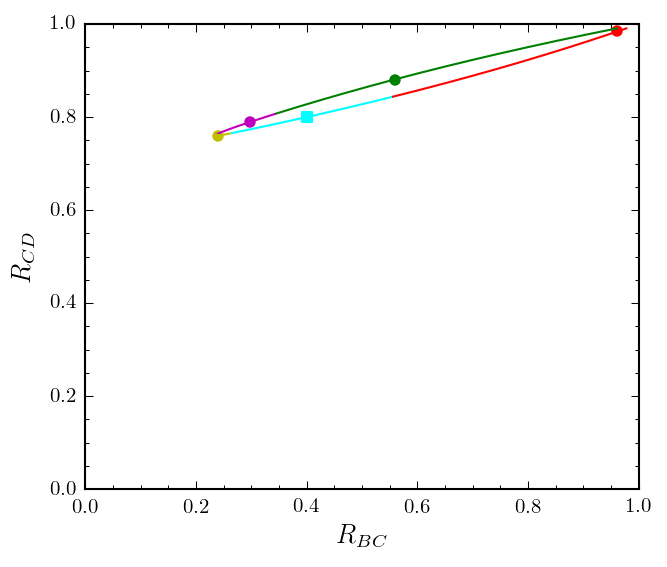}
\caption{ The same as Fig.~\ref{fig:trajectory}, but for the flat direction generated by point $P_{32}$ from Table~\ref{tab:32}.
%considered in Section~\ref{sec:case32}.
\label{fig:trajectory_32} 
}
\end{figure}
 
\subsection{Kinematical properties along the flat direction}
\label{sec:flat32}

\begin{figure}[t]
\centering
\includegraphics[width=0.5\textwidth]{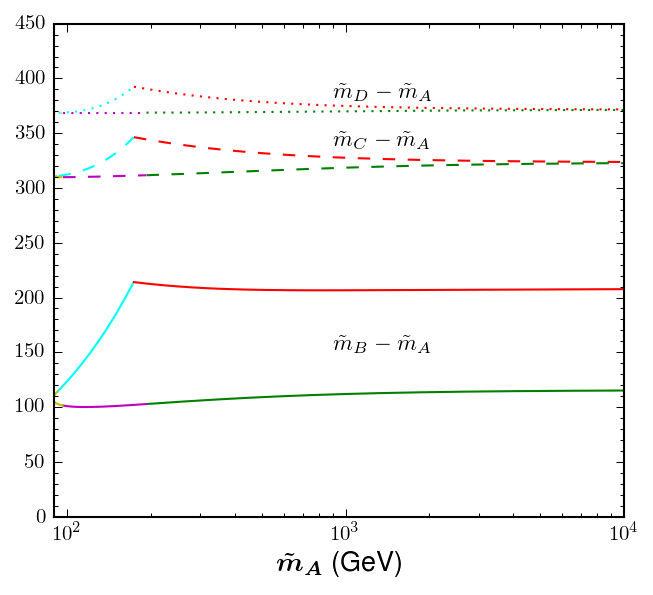}
\caption{ The analogue of Fig.~\ref{fig:mass_diff_31}, but for the flat direction defined in Fig.~\ref{fig:trajectory_32}.
\label{fig:mass_diff_32} 
}
\end{figure}

The mass spectrum for the study point $P_{32}$ and the corresponding mass squared ratios and kinematic endpoints are
shown in the cyan-shaded column of Table~\ref{tab:32}. Point $P_{32}$ was used previously in Ref.~\cite{Burns:2009zi} as an example 
of a {\em discrete} two-fold ambiguity, while here it serves to define a flat direction (\ref{massfamily}) in mass parameter space.
This flat direction is illustrated in Figs.~\ref{fig:trajectory_32} and \ref{fig:mass_diff_32}, which are the analogues of Figs.~\ref{fig:trajectory} 
and \ref{fig:mass_diff_31}, respectively.
According to Figs.~\ref{fig:trajectory_32} and \ref{fig:mass_diff_32}, the mass trajectory now goes through five of the six colored regions in 
Fig.~\ref{fig:regions}: there is a true branch through the red region $(3,1)$, the cyan region $(3,2)$ and the the yellow region $(4,2)$, as well as an auxiliary branch through 
the yellow region $(4,2)$, the magenta region $(4,3)$ and the green region $(2,3)$. As in Section~\ref{sec:case31}, we choose 
one representative study point in each of these regions. The four additional study points, $P_{31}$, $P_{42}$, $P_{43}$ and $P_{23}$, 
are also listed in Table~\ref{tab:32}, and their columns are shaded with the color of their respective regions in Fig.~\ref{fig:regions}. 

The flat direction depicted in Fig.~\ref{fig:trajectory_32} is again parametrized by the trial value $\tilde m_A$ for the mass of the 
lightest new particle $A$. However, as seen in Fig.~\ref{fig:mass_diff_32}, this time the allowed range for $\tilde m_A$ does not 
extend all the way to $\tilde m_A=0$, and instead the true and auxiliary branch meet inside the yellow region $(4,2)$ around at the lowest value $\tilde m_A\sim 89$ GeV.

\begin{figure}[t]
\centering
\includegraphics[width=0.4\textwidth]{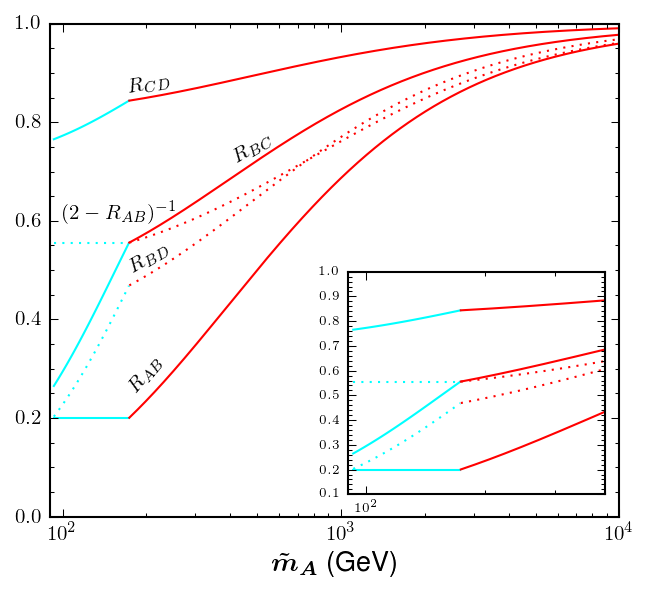}
\includegraphics[width=0.4\textwidth]{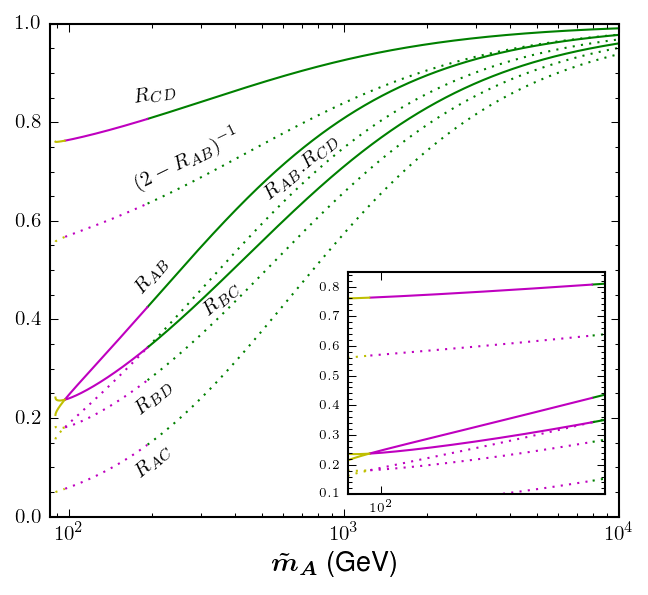}
\caption{The equivalent representation of Fig.~\ref{fig:mass_diff_32} in terms of the 
mass squared ratios $R_{AB}$, $R_{BC}$ and $R_{CD}$ (solid lines).
The dotted lines depict various quantities of interest which are used to delineate
the regions in Fig.~\ref{fig:regions}. 
The left panel shows the true branch passing through regions $(3,2)$ (cyan) and $(3,1)$ (red),
while the right panel shows the auxiliary branch through regions $(4,2)$ (yellow), $(4,3)$ (magenta) and $(2,3)$ (green). 
The left insert zooms in on the transition between regions $(3,2)$ and $(3,1)$ near $\tilde m_A=173$ GeV, while the
right insert focuses on the transitions between regions $(4,2)$ and $(4,3)$ near $\tilde m_A=97$ GeV and
between regions $(4,3)$ and $(2,3)$ near $\tilde m_A=193$ GeV.
\label{fig:mass_ratios_32} 
}
\end{figure}

The transitions between two neighboring regions along the flat direction can be understood from Fig.~\ref{fig:mass_ratios_32},
which plots the mass squared ratios $R_{AB}$, $R_{BC}$ and $R_{CD}$ (solid lines) and several other quantities (dotted lines) which
are relevant for defining the regions from Fig.~\ref{fig:regions}, as a function of the mass trajectory parameter $\tilde m_A$.
For example, relations (\ref{reg31R_BC}) and (\ref{reg32R_BC}) imply that the boundary between the cyan region $(3,2)$ 
and the red region $(3,1)$ is given by $R_{BC}=(2-R_{AB})^{-1}$. Indeed, the lines in Figs.~\ref{fig:mass_diff_32} and \ref{fig:mass_ratios_32}
change color from cyan to red when the $R_{BC}$ curve crosses the dotted line representing the function $(2-R_{AB})^{-1}$
near $\tilde m_A=173$ GeV. Similarly, it follows from (\ref{reg23R_BC}) and (\ref{reg43R_BC}) that the boundary between 
the green region $(2,3)$ and the magenta region $(4,3)$ is given by $R_{BC}=R_{AB}R_{CD}$. The right panel of Fig.~\ref{fig:mass_ratios_32}
confirms that the line color changes from magenta to green when the solid line for $R_{BC}$ is intersected by the dotted line 
for $R_{AB}R_{CD}$. Finally, according to (\ref{reg42R_BCge}) and (\ref{reg43R_AB}), the transition between the
yellow region $(4,2)$ and the magenta region $(4,3)$ occurs at $R_{AB}=R_{BC}$, and this is borne out by Fig.~\ref{fig:mass_ratios_32}
as well.

\begin{figure}[t]
\centering
\includegraphics[width=0.32\textwidth]{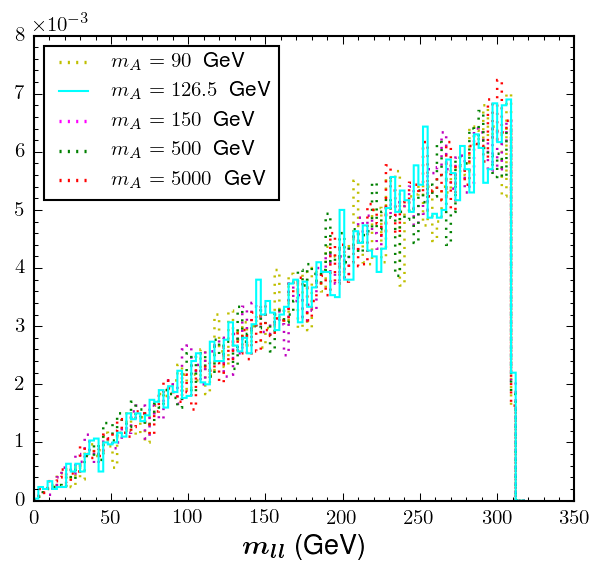}
\includegraphics[width=0.32\textwidth]{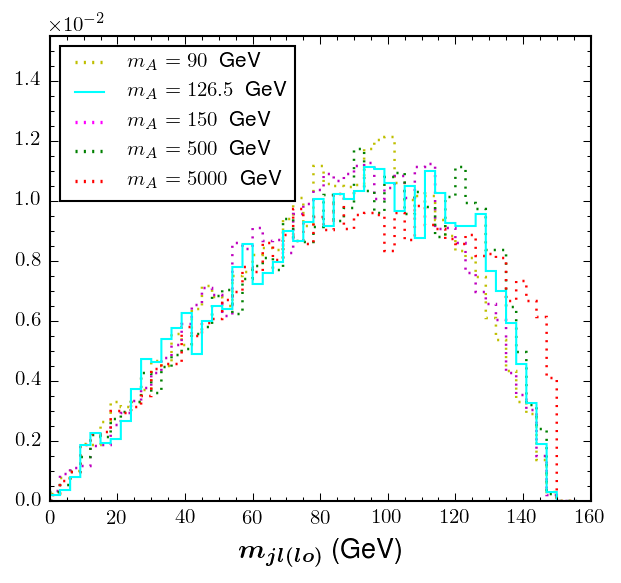}  
\includegraphics[width=0.32\textwidth]{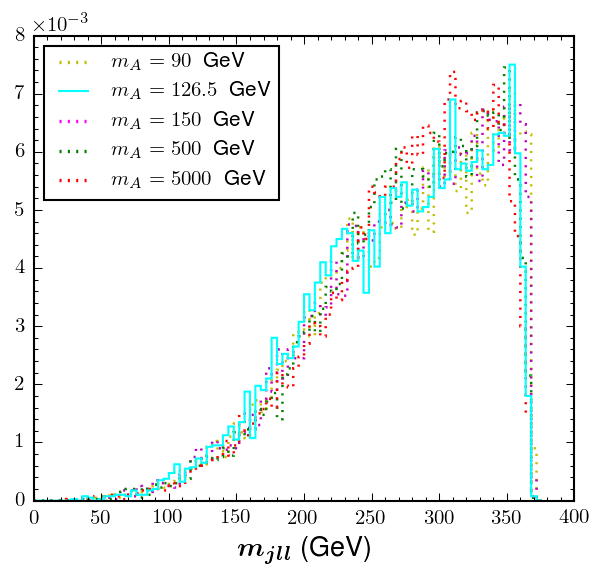}
\caption{ The analogue of Fig.~\ref{fig:3dist_31}, but for the five study points exhibited in Table~\ref{tab:32}.
The distributions are color-coded according to our color conventions for the regions in Fig.~\ref{fig:regions}.
 \label{fig:32dist} 
}
\end{figure}
\begin{figure}[t]
\centering
\includegraphics[width=0.49\textwidth]{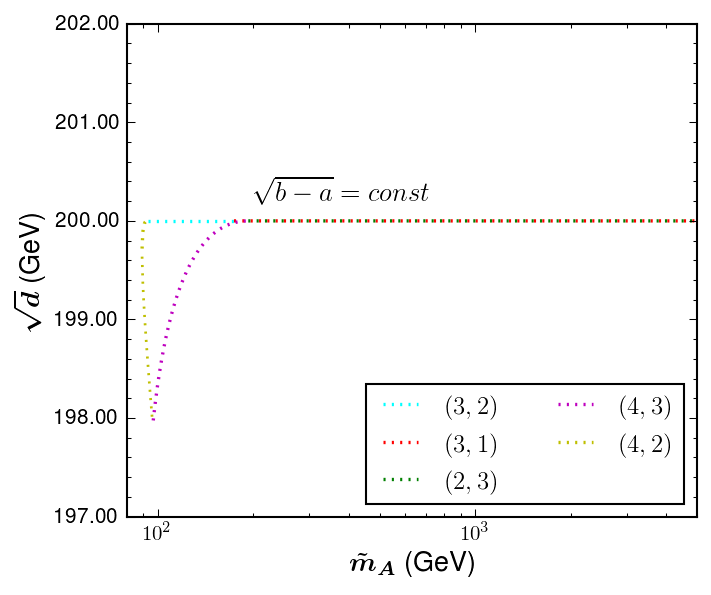}
\includegraphics[width=0.45\textwidth]{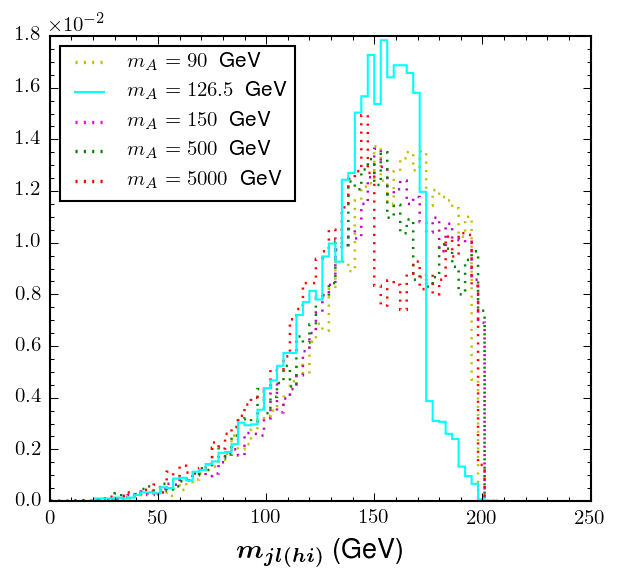}  
\caption{ The analogue of Fig.~\ref{fig:d_32}, but for the flat direction defined in Fig.~\ref{fig:trajectory_32} (left panel) and for the five study points from 
Table~\ref{tab:32} (right panel).
\label{fig:d_32} 
}
\end{figure}
\begin{figure}[t]
\centering
\includegraphics[width=0.49\textwidth]{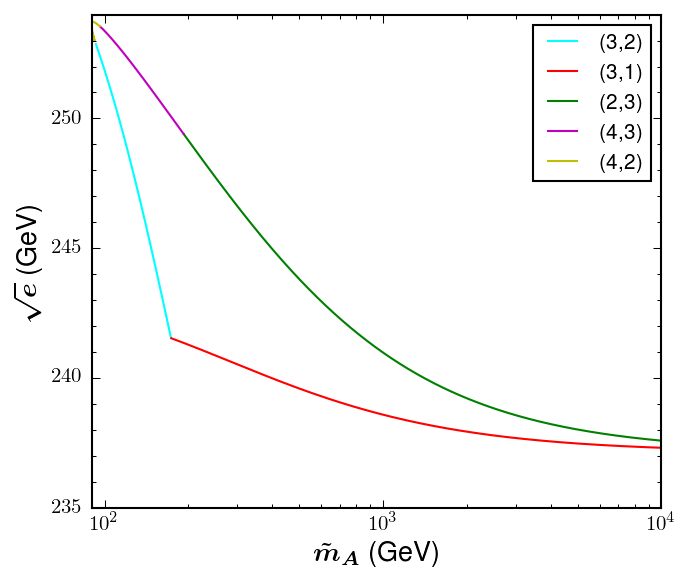}
\includegraphics[width=0.45\textwidth]{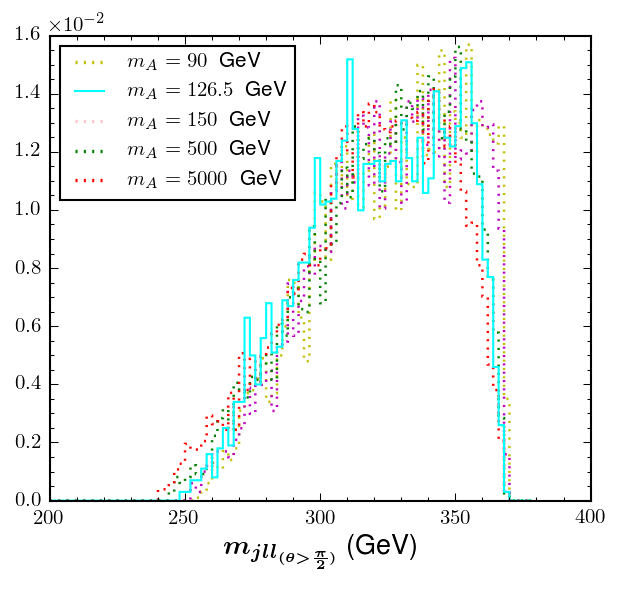}  
\caption{ The same as Fig.~\ref{fig:d_32}, but for the kinematic endpoint $\sqrt e$ and the corresponding $m_{jll(\theta>\frac{\pi}{2})}$
distribution. \label{fig:e_32} 
}
\end{figure}

By construction, all five study points from Table~\ref{tab:32} predict identical values for the 
three kinematic endpoints $a$, $b$ and $c$. This is demonstrated in Fig.~\ref{fig:32dist}, which is the analogue of Fig.~\ref{fig:3dist_31},
but for the five study points from Table~\ref{tab:32}. As before, the distributions in Fig.~\ref{fig:32dist} are color-coded 
according to our color conventions from Fig.~\ref{fig:regions}. The nominal input study point $P_{32}$ is represented by the solid line,
while the dotted lines mark the other four study points. Given that the five study points look very similar on Fig.~\ref{fig:32dist},
we now focus on the remaining two distributions, $m_{j\ell(hi)}$ and $m_{jll(\theta>\frac{\pi}{2})}$, which are investigated in 
Figs.~\ref{fig:d_32} and \ref{fig:e_32}.
The left panels show the predictions for the kinematic endpoint $\sqrt{d}=m_{j\ell(hi)}^{max}$ and the threshold
$\sqrt{e}=m_{j\ell\ell(\theta>\frac{\pi}{2})}^{min}$, respectively, while the right panels plot the corresponding
kinematic distributions for the five study points from Table~\ref{tab:32}.

By now, we should not be surprised by the extreme flatness of the curves exhibited in the left panel of Fig.~\ref{fig:d_32}.
The mass trajectory from Fig.~\ref{fig:trajectory_32} passes through all three of the regions where the endpoint $d$
is not an independent quantity, but is fixed by the relation (\ref{bad}) and is therefore strictly independent of $\tilde m_A$.
In the remaining two regions, $(4,2)$ and $(4,3)$, Fig.~\ref{fig:d_32} shows a maximal deviation of only 2 GeV from the 
prediction $\sqrt{d}=\sqrt{b-a}$ of (\ref{bad}). Taken together, the left panels of Figs.~\ref{fig:d_31} and \ref{fig:d_32} justify 
our terminology of the mass trajectory (\ref{massfamily}) as a ``flat direction" in mass parameter space.

On the other hand, the left panel of Fig.~\ref{fig:e_32} shows a much more significant variation of the kinematic threshold
variable $\sqrt{e}$ along the flat direction. The total variation is on the order of 15 GeV, which is of the same order as
our previous result in Fig.~\ref{fig:e_31}. However, as we already discussed in Section~\ref{sec:case31}, the measurement of $\sqrt{e}$
presents significant experimental challenges, as one can deduce from the very minor apparent variation of the $m_{j\ell\ell(\theta>\frac{\pi}{2})}$
distributions shown in the right panel of Fig.~\ref{fig:e_32}. This motivates searching for alternative methods for lifting the 
degeneracy along the flat direction.

\begin{figure}[t]
\centering
\includegraphics[width=0.342\textwidth]{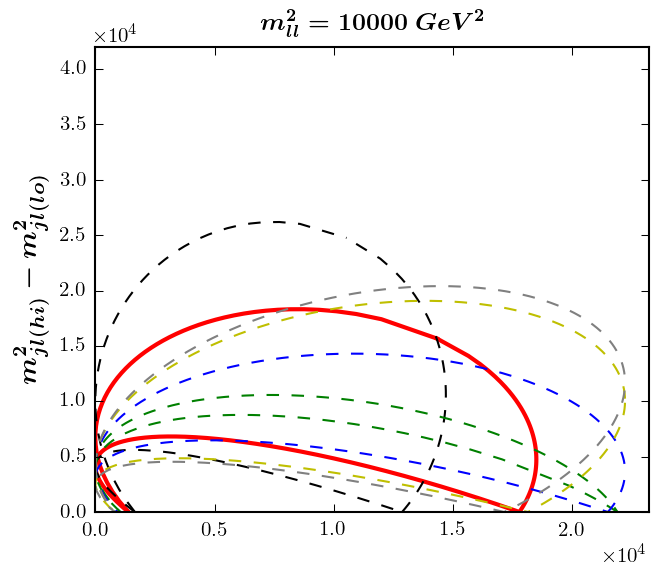}
\includegraphics[width=0.32\textwidth]{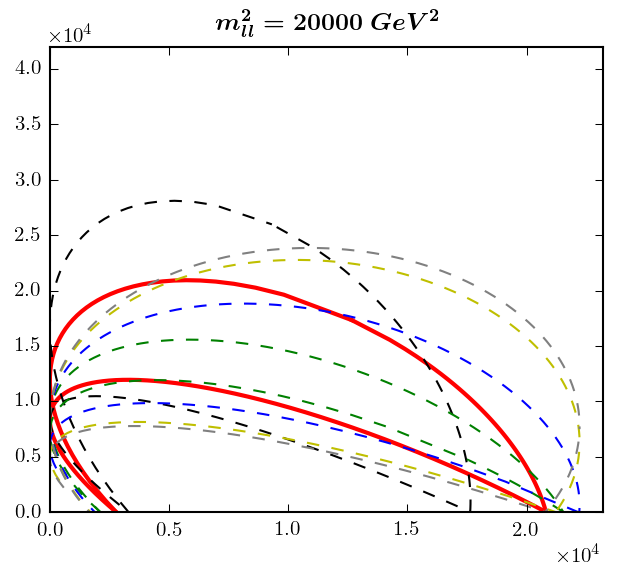}  
\includegraphics[width=0.32\textwidth]{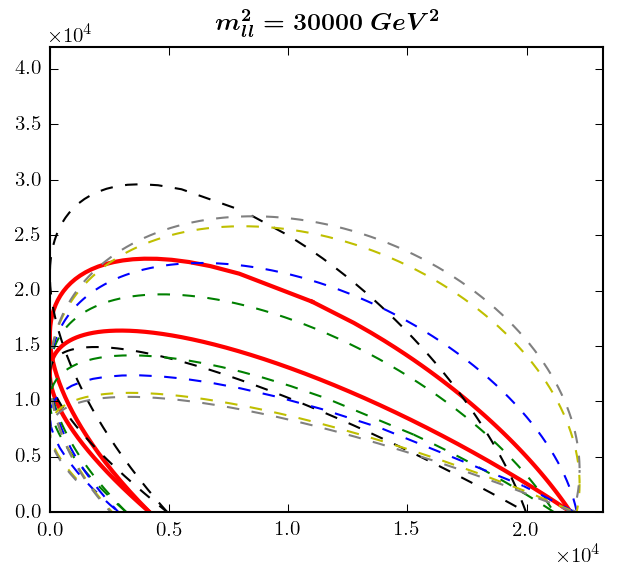}\\
\includegraphics[width=0.342\textwidth]{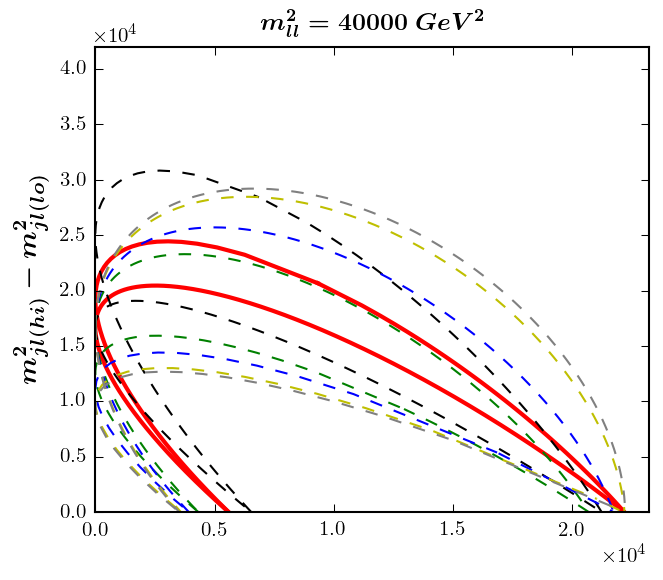}
\includegraphics[width=0.32\textwidth]{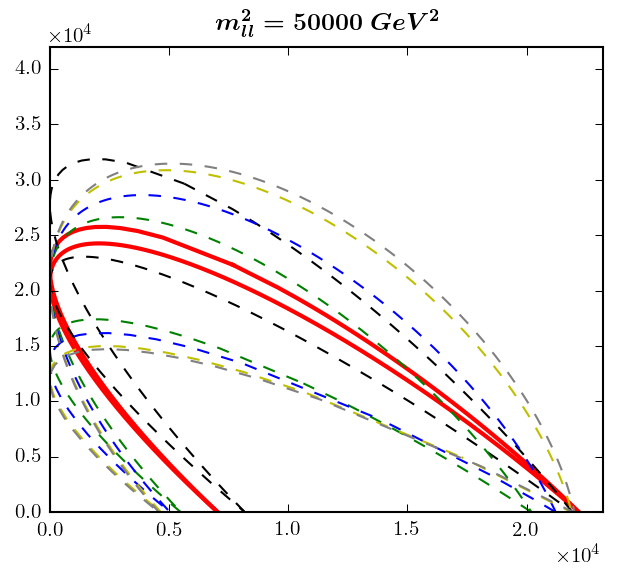}  
\includegraphics[width=0.32\textwidth]{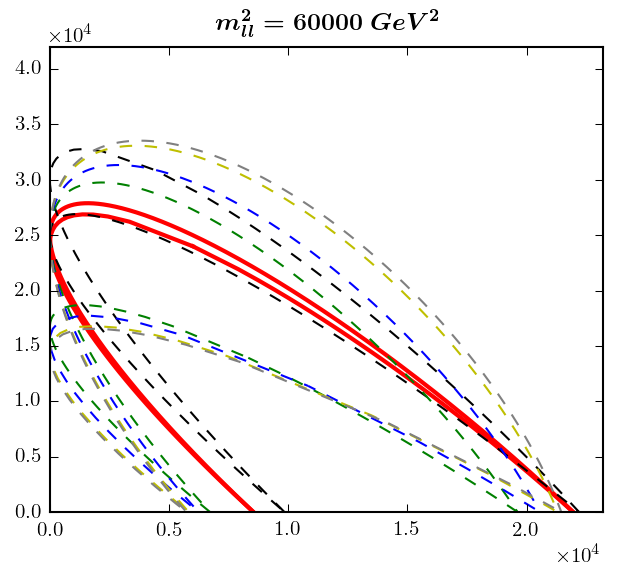}\\
\includegraphics[width=0.342\textwidth]{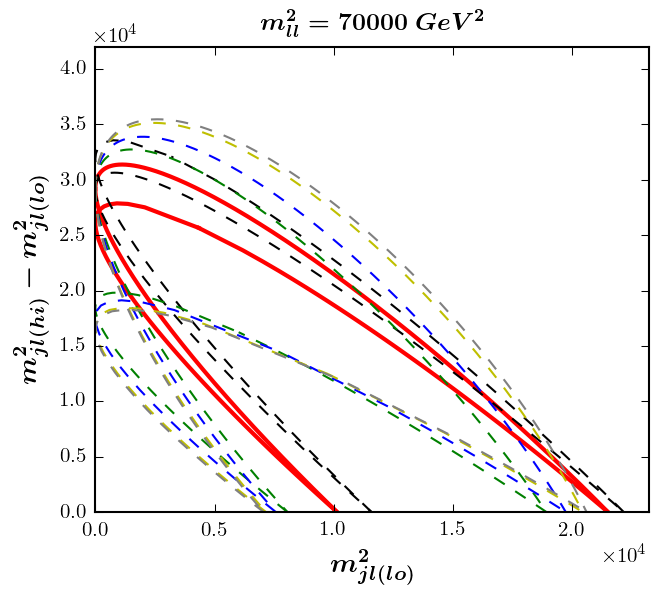}
\includegraphics[width=0.32\textwidth]{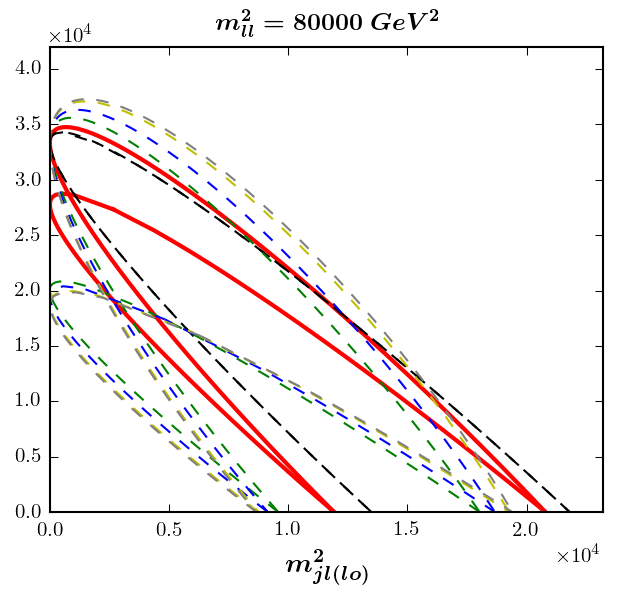}  
\includegraphics[width=0.32\textwidth]{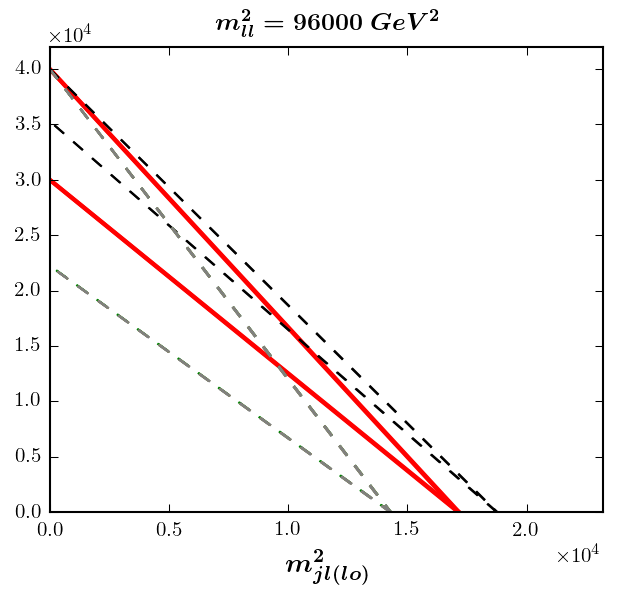}
\caption{ The same as Fig.~\ref{fig:contours_31_true}, but for the true branch in Fig.~\ref{fig:trajectory_32}.
The red solid line represents the case of the $P_{32}$ study point with $\tilde m_A=126.5$ GeV, 
while the dashed lines correspond to other values of $\tilde m_A$ along the true branch:
$\tilde m_A=100$ (black),
$\tilde m_A=173$ GeV (green),
$\tilde m_A=500$ GeV (blue),
$\tilde m_A=2000$ GeV (yellow) and
$\tilde m_A=4000$ GeV (gray). 
\label{fig:contours_32_true} 
}
\end{figure}
\begin{figure}[t]
\centering
\includegraphics[width=0.342\textwidth]{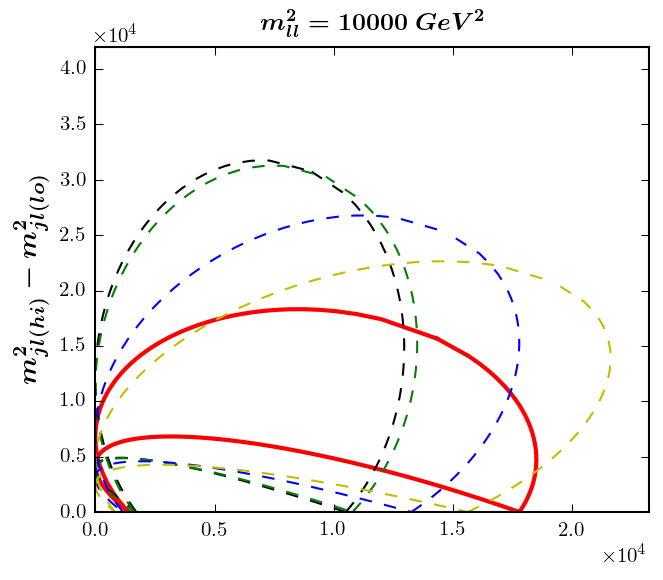}
\includegraphics[width=0.32\textwidth]{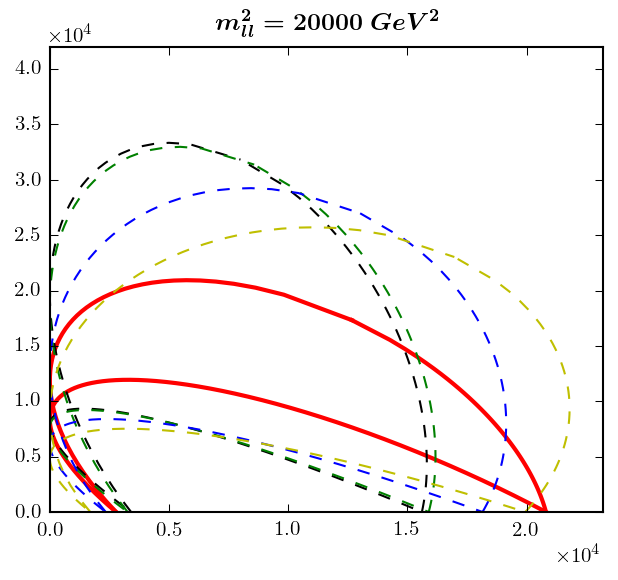}  
\includegraphics[width=0.32\textwidth]{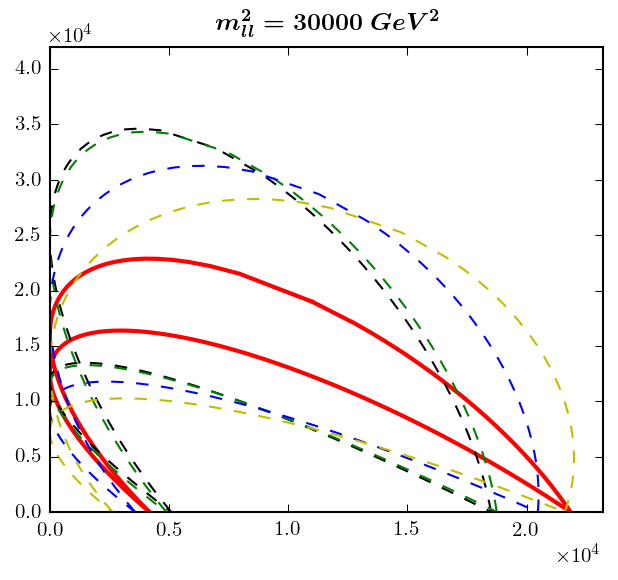}\\
\includegraphics[width=0.342\textwidth]{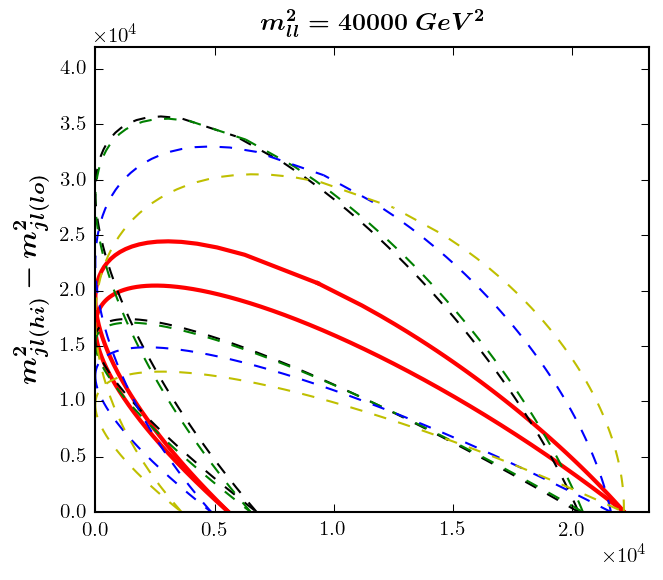}
\includegraphics[width=0.32\textwidth]{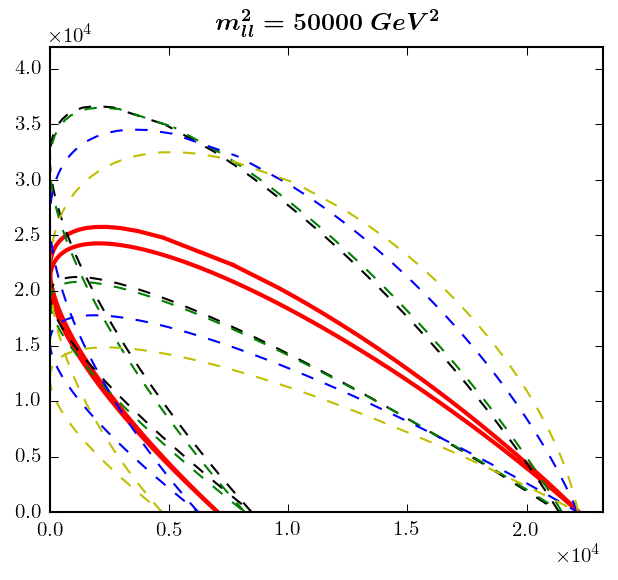}  
\includegraphics[width=0.32\textwidth]{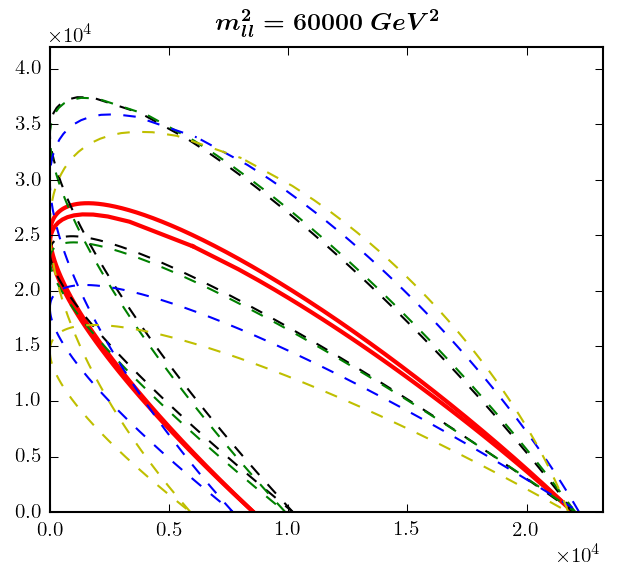}\\
\includegraphics[width=0.342\textwidth]{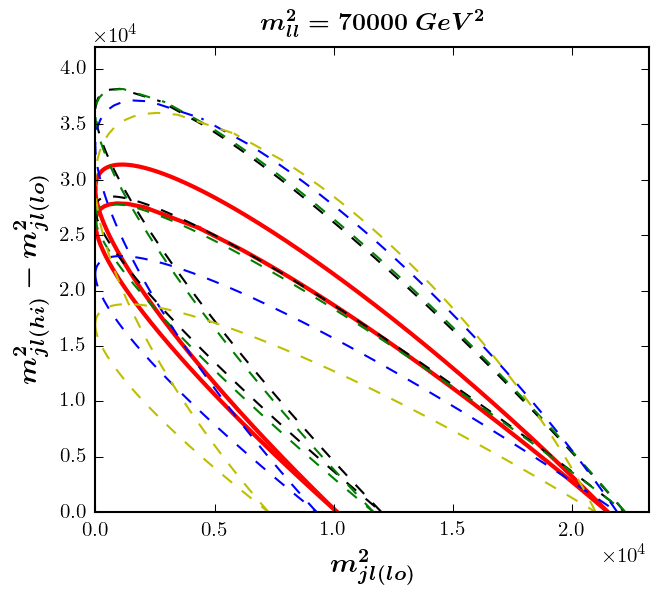}
\includegraphics[width=0.32\textwidth]{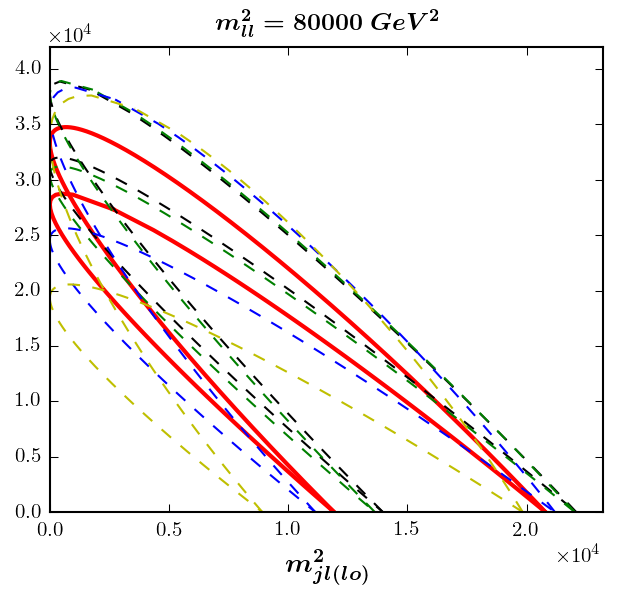}  
\includegraphics[width=0.32\textwidth]{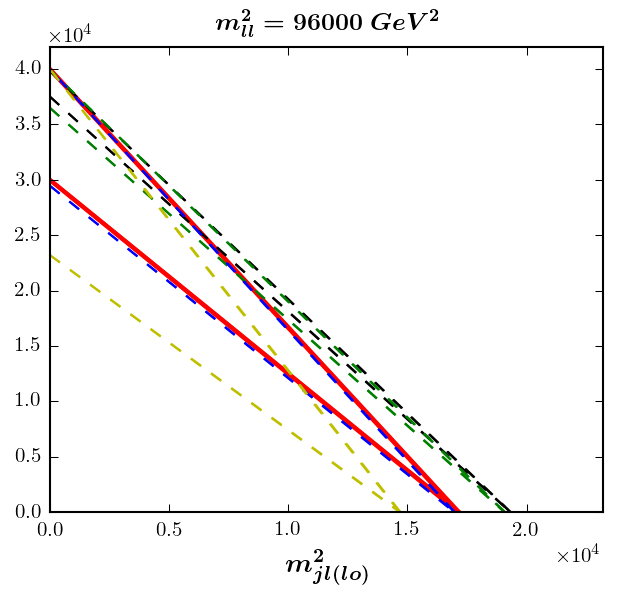}
\caption{ The same as Fig.~\ref{fig:contours_32_true}, but for the auxiliary branch in regions $(4,2)$, $(4,3)$ and $(2,3)$. 
The dashed lines represent points with 
$\tilde m_A=90$ GeV (black),
$\tilde m_A=150$ GeV (green),
$\tilde m_A=500$ GeV (blue) and
$\tilde m_A=5000$ GeV (yellow).
For reference, we also show the case of the true mass spectrum for point $P_{32}$ (red solid lines), although $P_{32}$ 
itself does not belong to the auxiliary branch. 
\label{fig:contoursfake_32} 
}
\end{figure}

As already discussed in Section~\ref{sec:case31}, one such method is to track the deformation of the shape of the 
kinematic boundary (\ref{eq:samosa2}) along the flat direction. The effect is illustrated in Figs.~\ref{fig:contours_32_true}
and \ref{fig:contoursfake_32}, which are the analogues of Figs.~\ref{fig:contours_31_true} and \ref{fig:contoursfake_31} 
for the example of a flat direction considered in this section.
%
%%%%  UNCOMMENT LATER
%
Once again, the solid red lines in each panel indicate the kinematic boundaries for the nominal study point $P_{32}$ with
$\tilde m_A=126.5$ GeV, while the dashed lines are drawn for several other values of $\tilde m_A$, chosen so that 
they illustrate the typical range of shape fluctuations. Along the true branch, in Fig.~\ref{fig:contours_32_true} we
plot contours for $\tilde m_A=100$ (black), $\tilde m_A=173$ GeV (green), $\tilde m_A=500$ GeV (blue),
$\tilde m_A=2000$ GeV (yellow) and $\tilde m_A=4000$ GeV (gray). Even though we are confined to the true branch only,
when we compare Fig.~\ref{fig:contours_32_true} to its analogue, Fig.~\ref{fig:contours_31_true}, we observe a much 
larger variation in the shape of the kinematic boundary in the present case, which promises good prospects for the 
mass measurement exercise to follow. The results shown in Fig.~\ref{fig:contoursfake_32} for the auxiliary branch 
are also quite good. This should not come as a surprise, since the exercise in Section~\ref{sec:case31} already
indicated that the auxiliary branch has a different kinematic behavior, as reflected in the shape of the phase space boundary.

\subsection{A toy study with uniformly distributed background}
\label{sec:uni32}

In the remainder of Section~\ref{sec:case32} we shall repeat the two exercises from Sections~\ref{sec:uni32} and
\ref{sec:tt32}, only this time we shall use $P_{32}$ as our input study point, and perform the measurement along the 
corresponding flat direction described in Figs.~\ref{fig:trajectory_32} and \ref{fig:mass_diff_32}.

\begin{figure}[t]
\centering
\includegraphics[width=0.4\textwidth]{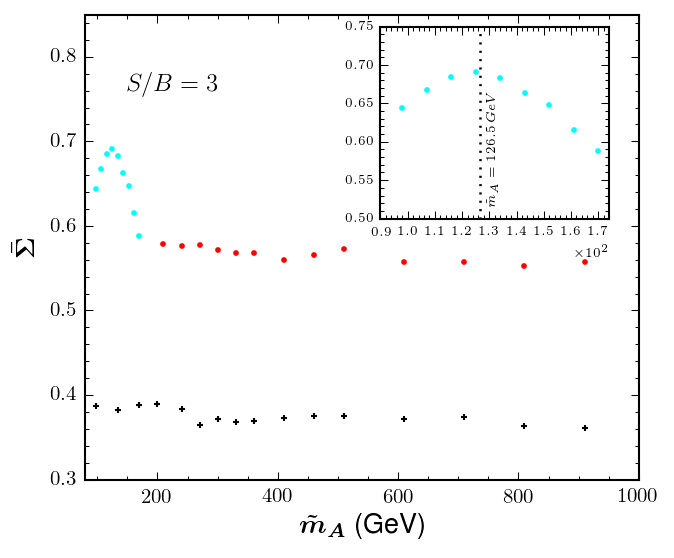}
\includegraphics[width=0.4\textwidth]{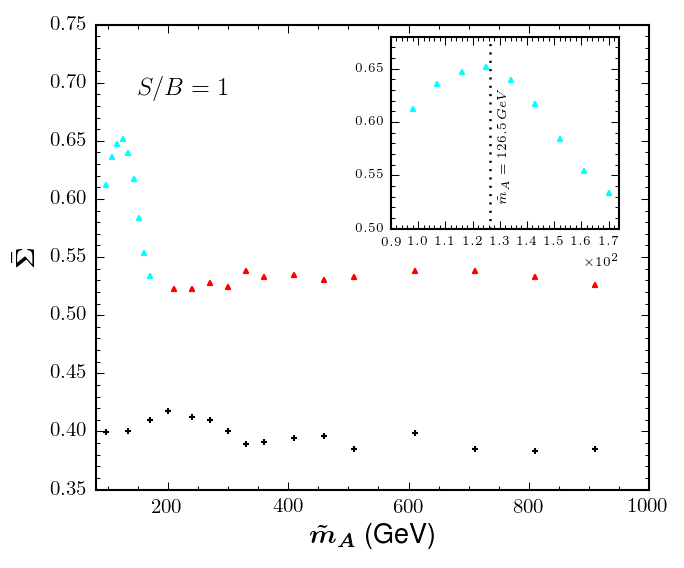}\\
\includegraphics[width=0.4\textwidth]{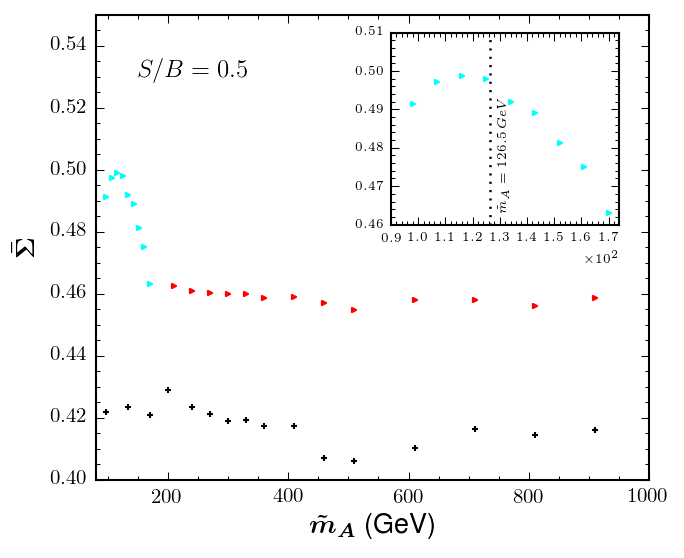}
\includegraphics[width=0.4\textwidth]{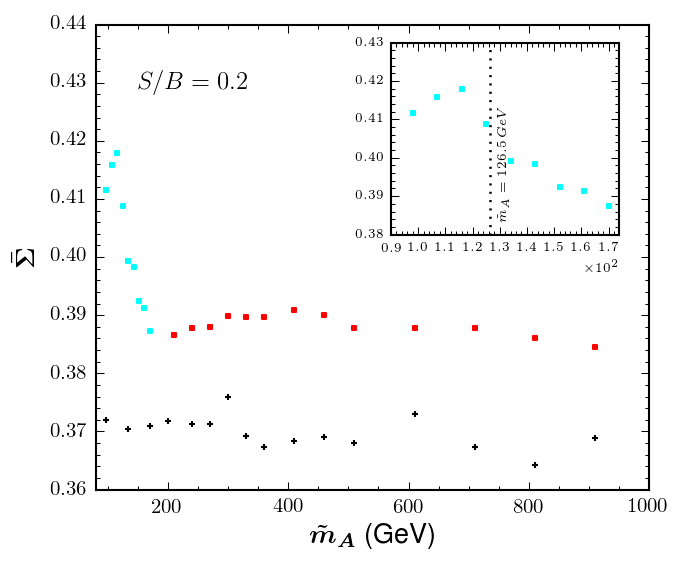}
\caption{ The same as Fig.~\ref{fig:sigma_uniform_31}, but now taking point $P_{32}$ as input and 
measuring along the flat direction depicted in Fig.~\ref{fig:trajectory_32}.
\label{fig:sigma_uniform_32} 
}
\end{figure}
First we consider the case of uniformly distributed (in mass squared) background events, and proceed to evaluate the 
quantity $\bar\Sigma$ along the flat direction. As before, we fix $N_B=1000$ and then vary the signal-to-background ratio
inside the boundary surface ${\cal S}$. Fig.~\ref{fig:sigma_uniform_32} shows our results for
the same choices of $S/B$ as in Fig.~\ref{fig:sigma_uniform_31}: 
$S/B=3$ (upper left panel), $S/B=1$ (upper right panel), $S/B=0.5$ (lower left panel) and $S/B=0.2$ (lower right panel). 
We find that the function $\bar\Sigma(\tilde m_A)$ once again peaks in the vicinity of the true value $m_A=126.5$ GeV.
Specifically, for $S/B=\{3.0, 1.0, 0.5, 0.2\}$, the maxima are found at $\tilde m_A=\{125, 125, 116, 116\}$ GeV, to be
contrasted with the true value of $m_A=126.5$ GeV. In all four cases, the auxiliary branch is disfavored, as it 
always gives low values for $\bar\Sigma$, while the true branch is restricted to a very narrow region near the true 
mass spectrum.

\begin{figure}[t]
\centering
\includegraphics[width=0.342\textwidth]{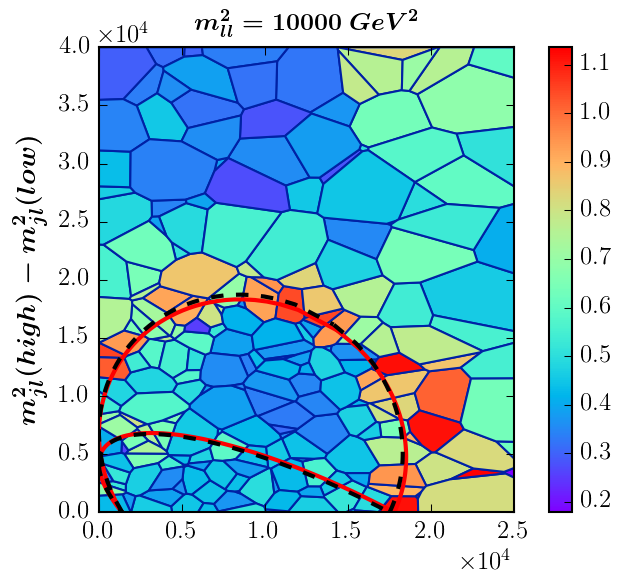}
\includegraphics[width=0.32\textwidth]{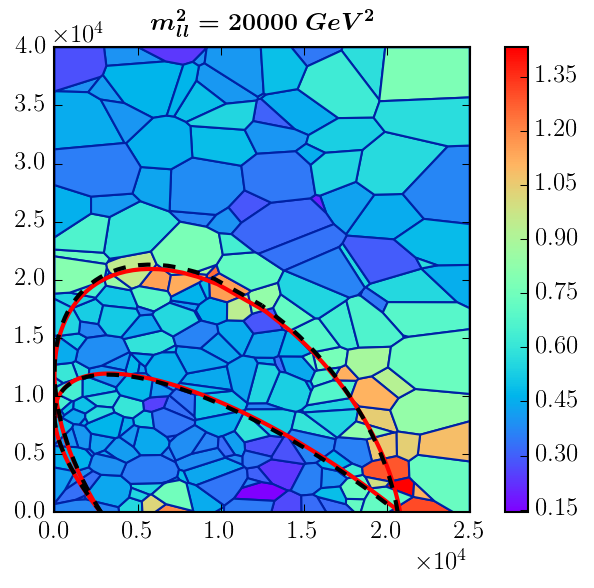}  
\includegraphics[width=0.32\textwidth]{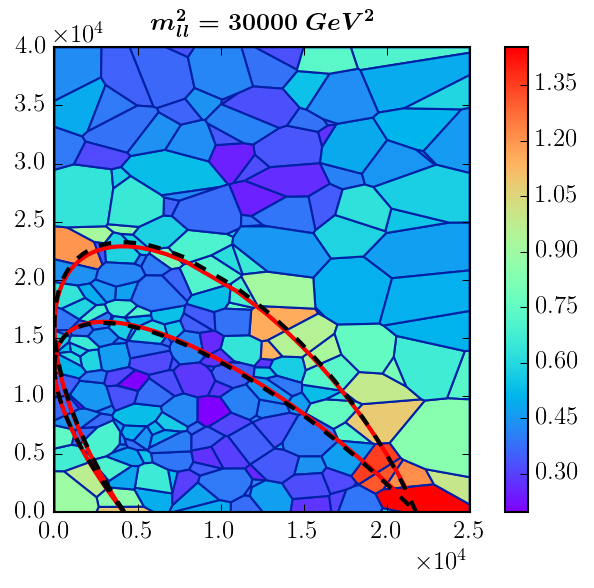}\\
\includegraphics[width=0.342\textwidth]{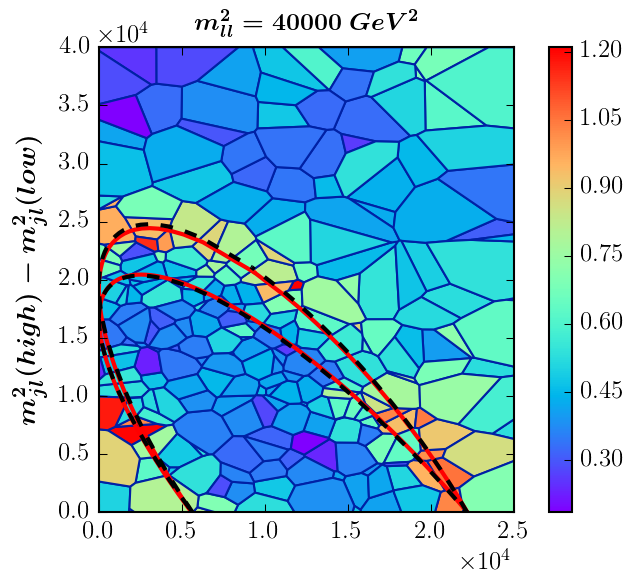}
\includegraphics[width=0.32\textwidth]{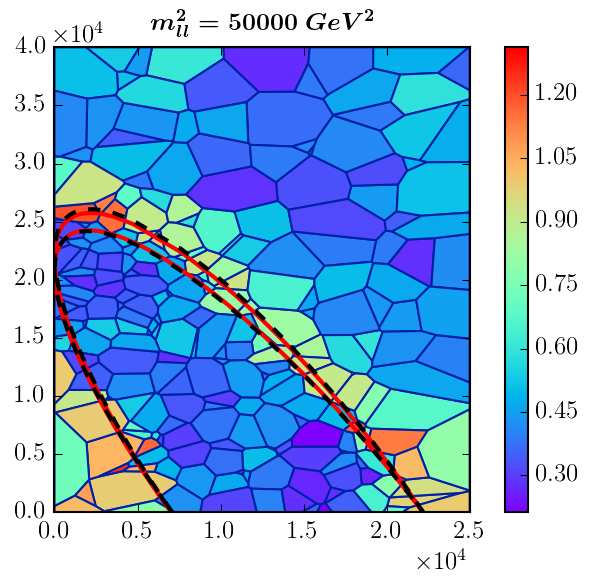}  
\includegraphics[width=0.32\textwidth]{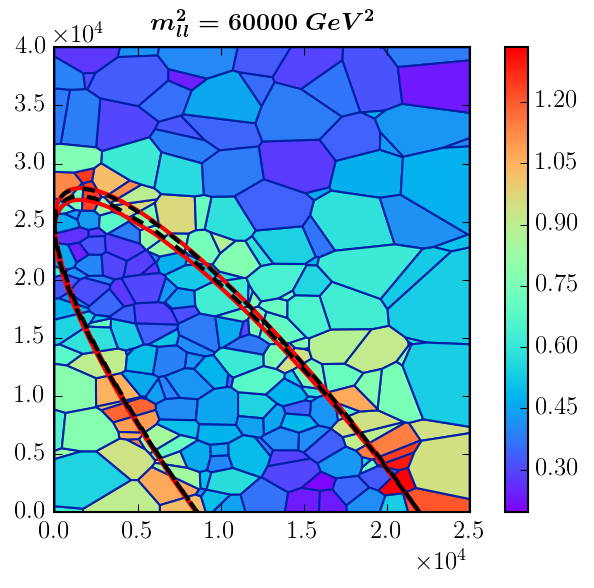}\\
\includegraphics[width=0.342\textwidth]{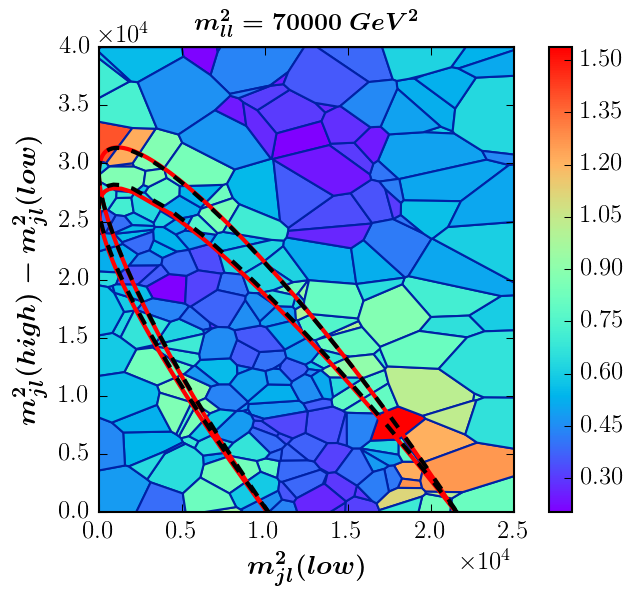}
\includegraphics[width=0.32\textwidth]{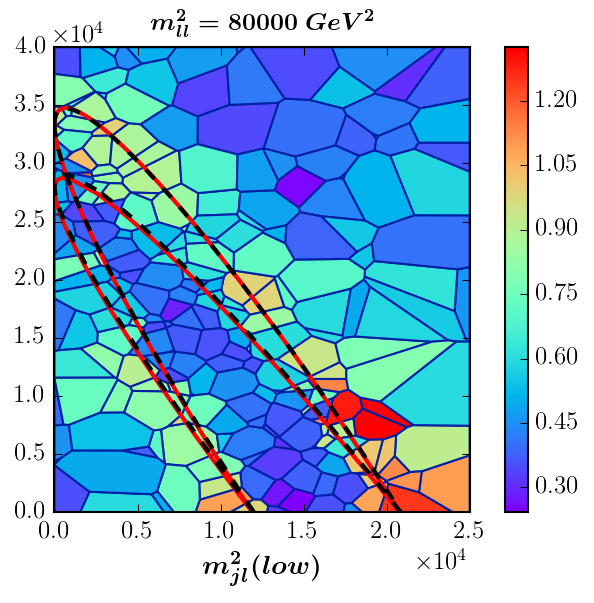}  
\includegraphics[width=0.32\textwidth]{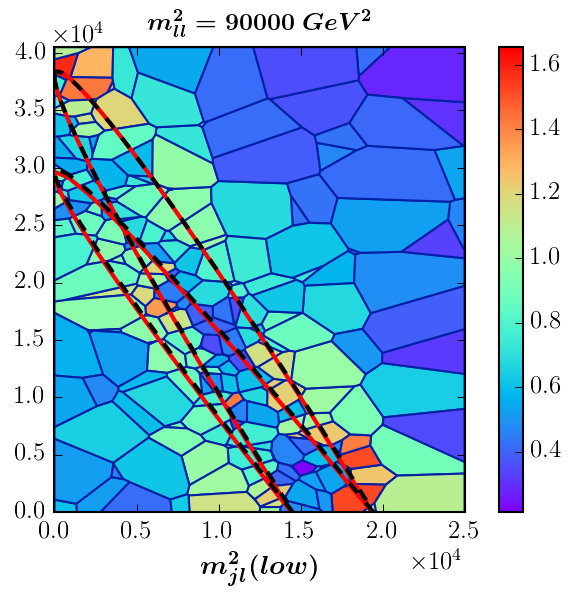}
\caption{ The same as Fig.~\ref{fig:slice_uniform_31}, but for the exercise performed in Section~\ref{sec:uni32},
with point $P_{32}$ as input (solid red lines). The black dashed line corresponds to the mass spectrum with $\tilde m_A=125$ GeV,
which was found to maximize the quantity $\bar\Sigma$ in the top left panel of Fig.~\ref{fig:sigma_uniform_32}.
\label{fig:slice_uniform_32}
}
\end{figure}
Fig.~\ref{fig:slice_uniform_32} provides a consistency check on our fitting procedure, similarly to Fig.~\ref{fig:slice_uniform_31}.
We show two-dimensional views at fixed $m_{\ell\ell}^2$ of the Voronoi tessellation of the data for the case of $S/B=3$. 
The red solid line is the expected signal boundary for the nominal case of point $P_{32}$, i.e., with the true value 
$\tilde m_A=m_A=126.5$ GeV. The black dashed line then corresponds to the best fit, i.e., a mass spectrum with $\tilde m_A=126$ GeV,
which was found to maximize the quantity $\bar\Sigma$ in the top left panel of Fig.~\ref{fig:sigma_uniform_32}.

\subsection{A study with $t\bar{t}$ dilepton background events}
\label{sec:tt32}

Our final task will be to repeat the $P_{32}$ exercise with dilepton $t\bar{t}$ events as was done in Section~\ref{sec:tt31}.
As before, we fix the number of signal events $N_S=3000$ and then consider several values for the number of background events:
$N_B=\{3000,4000,5000,6000\}$. In each case, we compute the function $\bar\Sigma(\tilde m_A)$ along the flat direction of Fig.~\ref{fig:trajectory_32}.
\begin{figure}[t]
\centering
\includegraphics[width=0.4\textwidth]{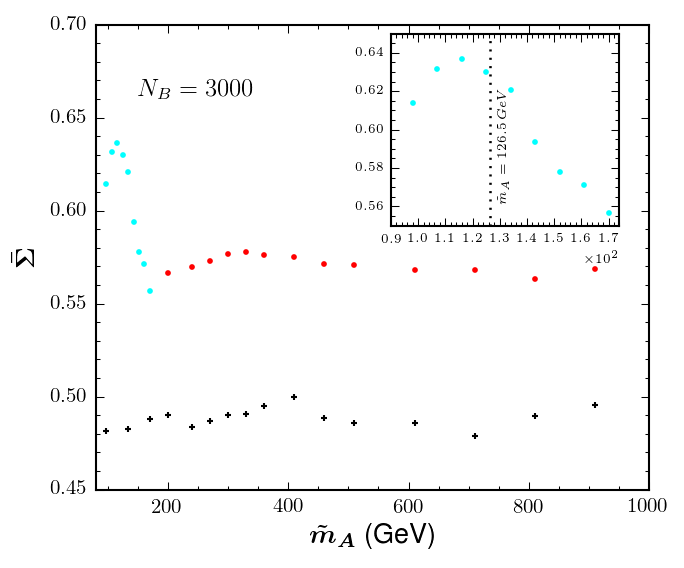}
\includegraphics[width=0.4\textwidth]{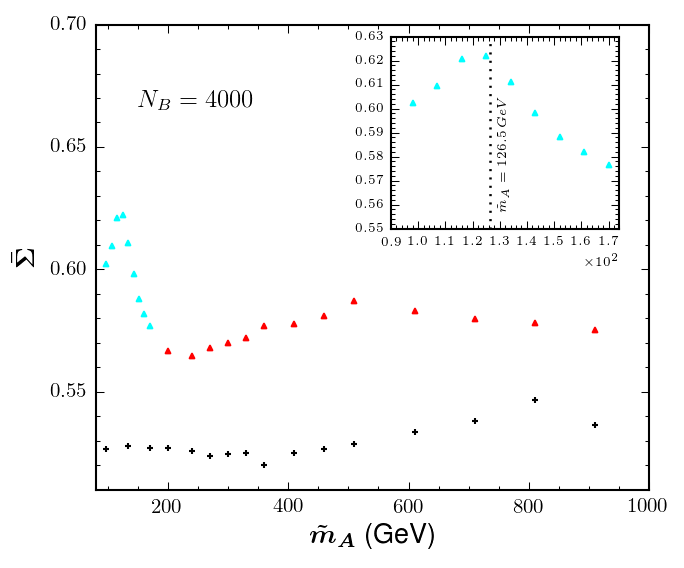}\\
\includegraphics[width=0.4\textwidth]{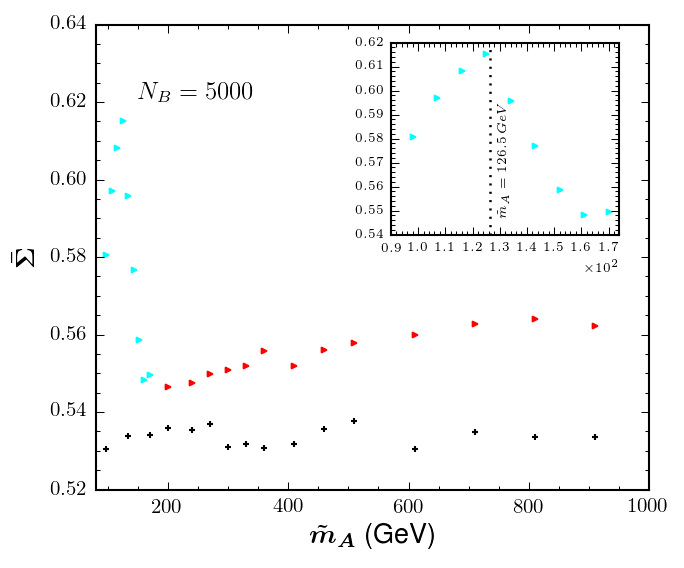}
\includegraphics[width=0.4\textwidth]{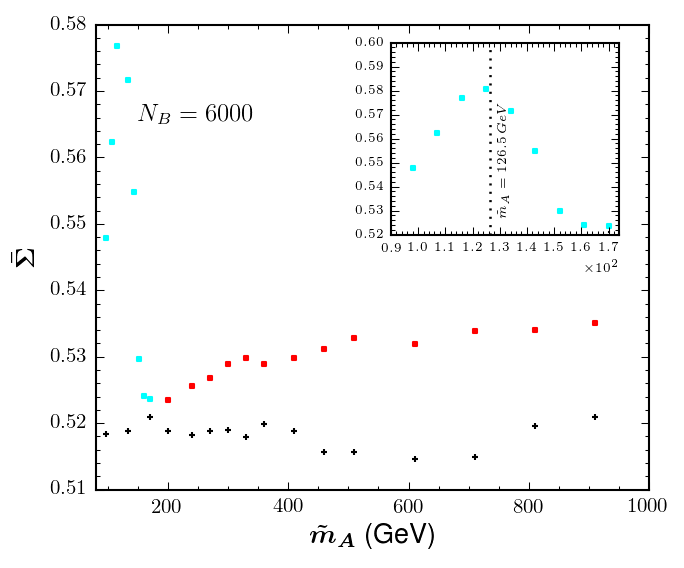}
\caption{ The same as Fig.~\ref{fig:sigma_ttbar_31}, but using study point $P_{32}$ as input.
% Sigma as a function of $m_A$ with $N_S=3000$ and $N_B=$ 3000 (red dots), 4000 (gren triangles) , 
% 5000 (magenta triangles), 6000 (blue squares). The black $``+"$ indicates the sigma value for the fake brunches. 
\label{fig:sigma_ttbar_32} 
}
\end{figure}
The results are shown in Fig.~\ref{fig:sigma_ttbar_32}, which has the same qualitative behavior as Fig.~\ref{fig:sigma_uniform_32}.
The $\bar\Sigma$ values for the auxiliary branch tend to be low, and the branch is disfavored. The global peak of 
$\bar\Sigma(\tilde m_A)$ is again found in the vicinity of the right answer (for $N_B=\{3000,4000,5000,6000\}$, the peak is at
$\tilde m_A=\{116,125,125,125\}$ GeV), and the large $\tilde m_A$ tail of the true branch 
is also disfavored. 
\begin{figure}[t]
\centering
\includegraphics[width=0.342\textwidth]{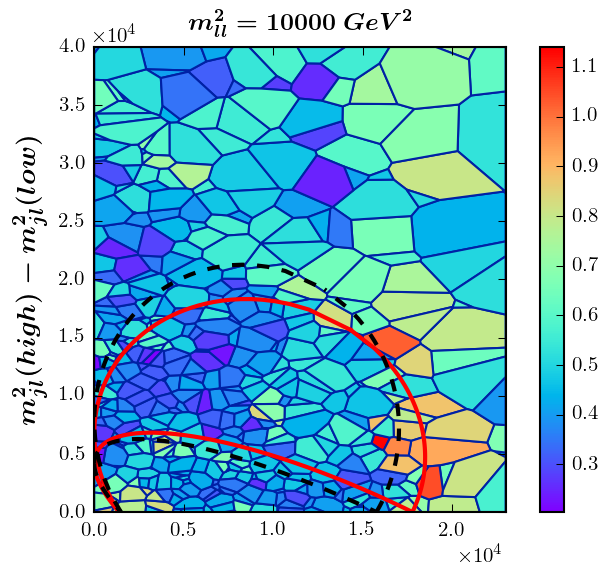}
\includegraphics[width=0.32\textwidth]{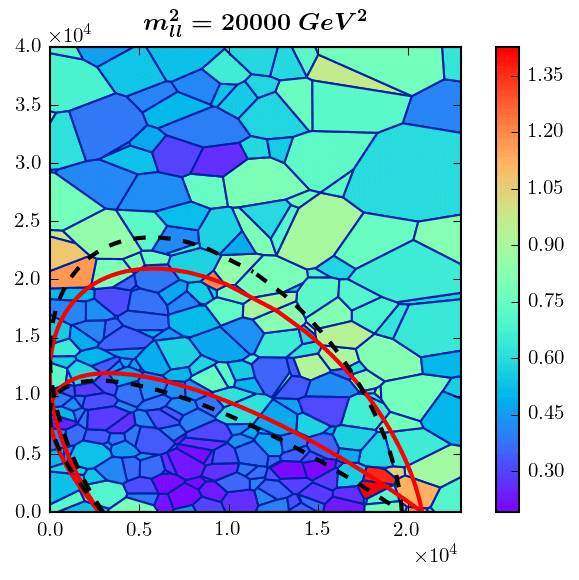}  
\includegraphics[width=0.32\textwidth]{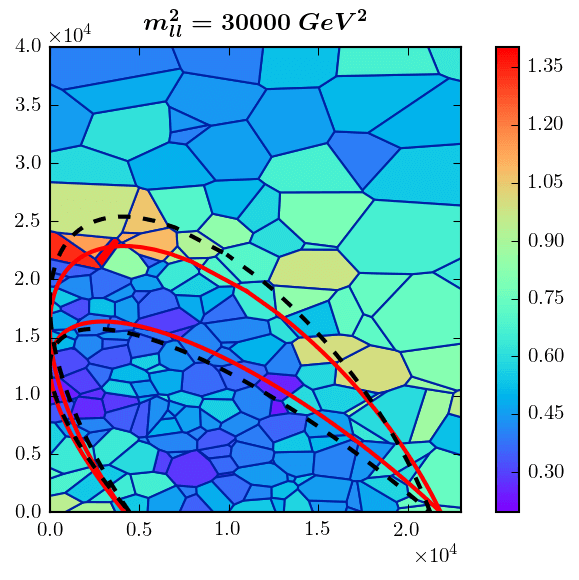}\\
\includegraphics[width=0.342\textwidth]{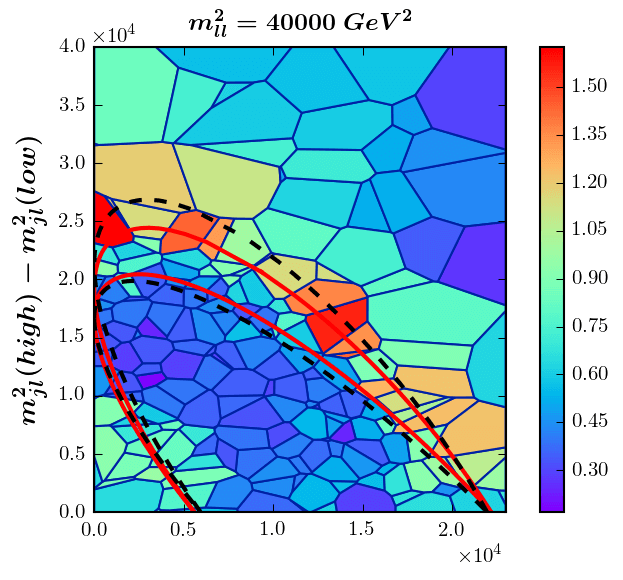}
\includegraphics[width=0.32\textwidth]{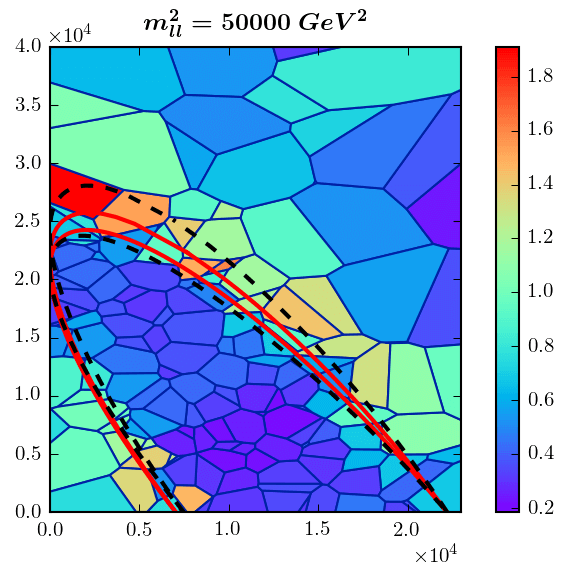}  
\includegraphics[width=0.32\textwidth]{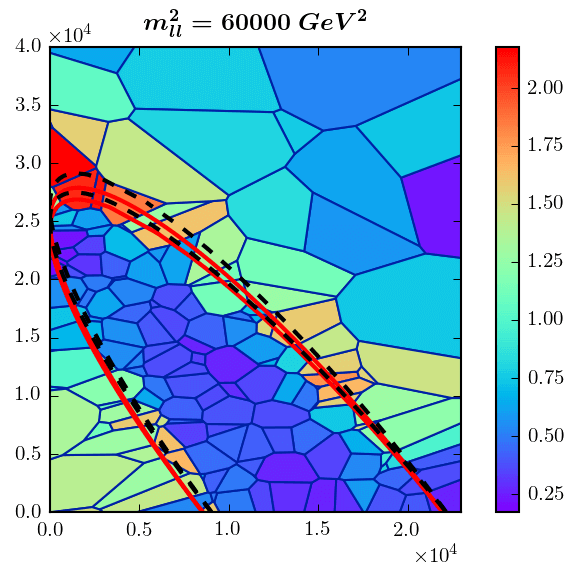}\\
\includegraphics[width=0.342\textwidth]{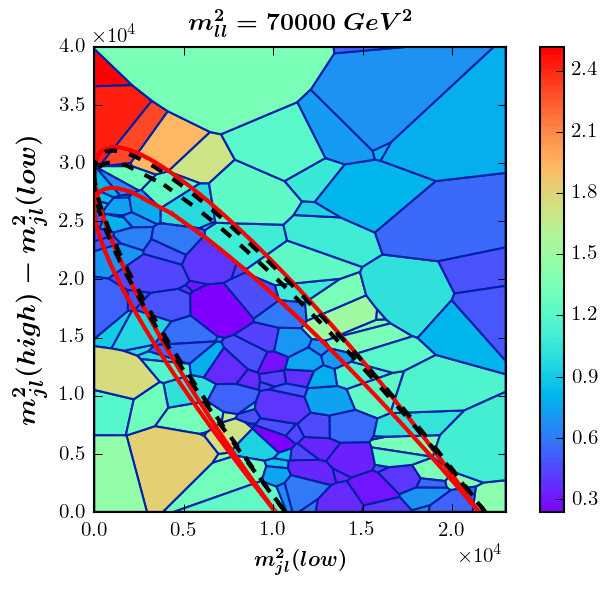}
\includegraphics[width=0.32\textwidth]{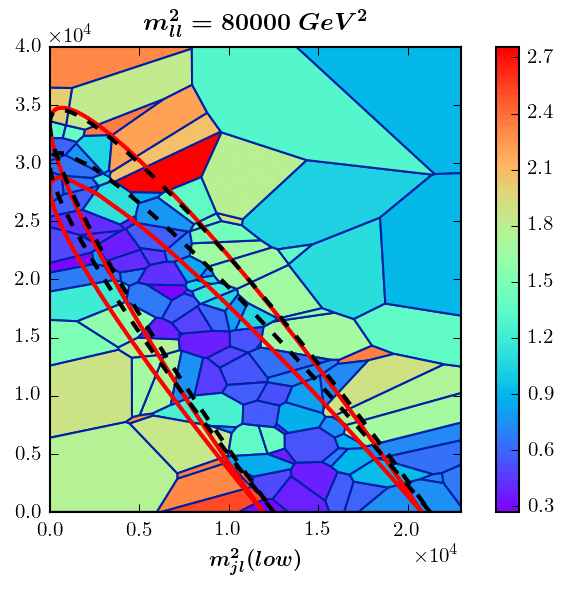}  
\includegraphics[width=0.32\textwidth]{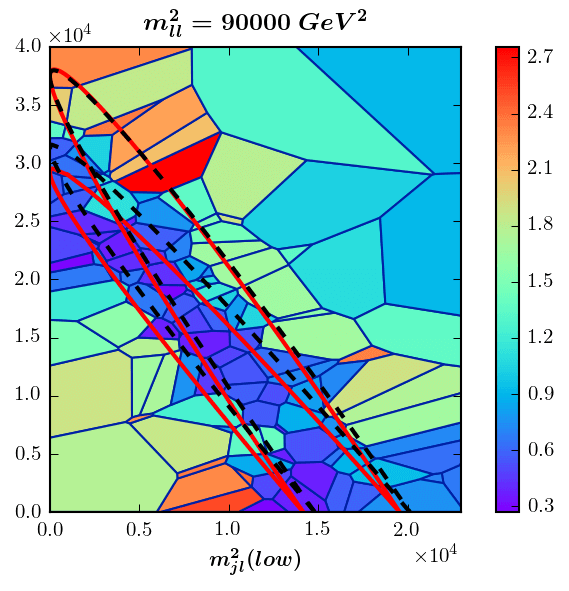}
\caption{ The same as Fig.~\ref{fig:slice_ttbar_31}, but for the exercise performed in Section~\ref{sec:tt32}, 
using study point $P_{32}$ as input.
%The Voronoi tessellation for signal $N_S=3000$ and background $N_B=3000$ (top quark pair production). 
The red solid line is the phase space boundary for the nominal value $m_A=126.5$ GeV, while 
the black dashed line corresponds to the best fit value $\tilde m_A=116$ GeV found in the 
the top left panel of Fig.~\ref{fig:sigma_ttbar_32}. 
\label{fig:slice_ttbar_32} 
}
\end{figure}
One final consistency check is provided by Fig.~\ref{fig:slice_ttbar_32}, which shows a comparison of the 
kinematic boundaries for the nominal study point $P_{32}$ with $m_A=126.5$ GeV (red solid lines), and the boundaries 
for the best fit value $\tilde m_A=125$ GeV (black dashed lines).

\subsection{A detector level study}
\label{sec:det}

\begin{figure}[t]
\centering
\includegraphics[width=0.4\textwidth]{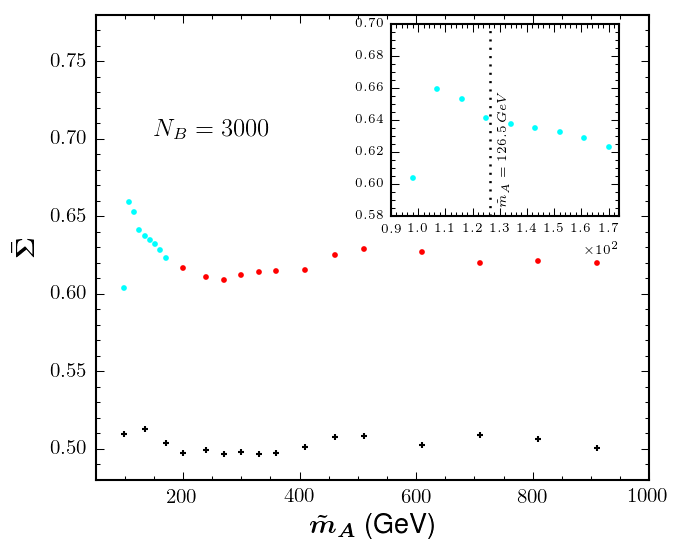}
\includegraphics[width=0.4\textwidth]{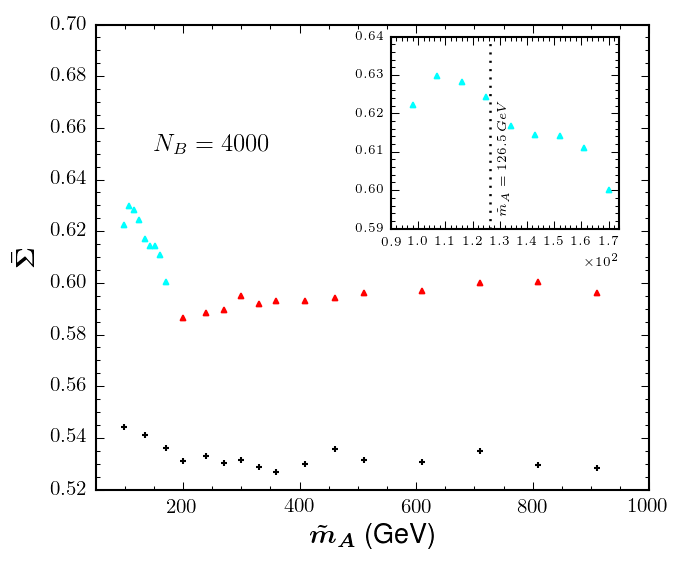}\\
\includegraphics[width=0.4\textwidth]{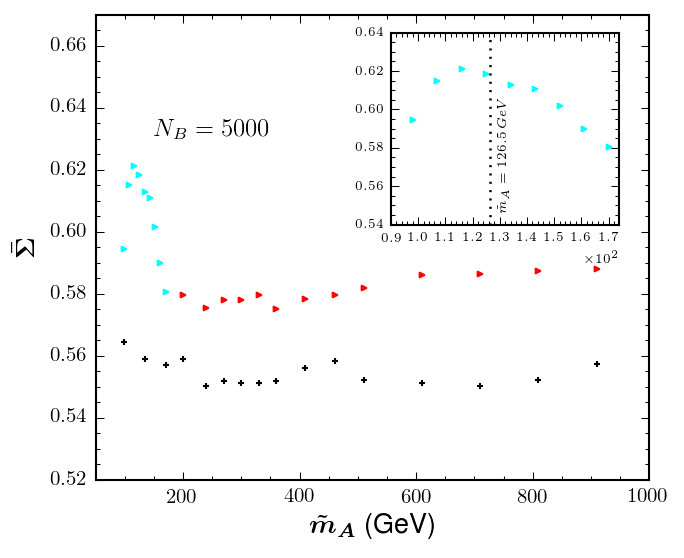}
\includegraphics[width=0.4\textwidth]{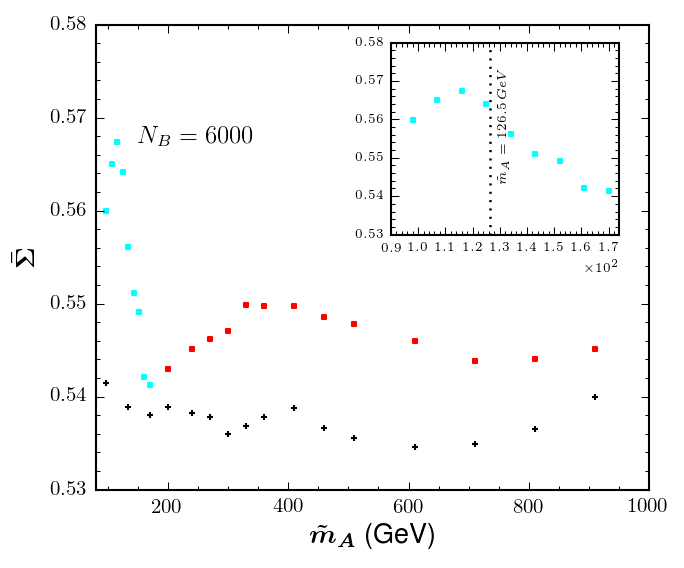}
\caption{ The same as Fig.~\ref{fig:sigma_ttbar_32}, but accounting for the detector resolution as described in the text.
% Sigma as a function of $m_A$ with $N_S=3000$ and $N_B=$ 3000 (red dots), 4000 (gren triangles) , 
% 5000 (magenta triangles), 6000 (blue squares). The black $``+"$ indicates the sigma value for the fake brunches. 
\label{fig:sigma_ttbar_smear_32} 
}
\end{figure}

In this paper, we introduced the new Voronoi-based method for mass measurement as a proof of principle, 
and showed that at the parton level it does reasonably well in the two examples considered so far in 
Sections~\ref{sec:case31} and \ref{sec:case32}. Before concluding, we would like to also test the method in the presence
of detector effects (this subsection) and combinatorics (see Sec.~\ref{sec:combinatorics} below). 
For this purpose, we first repeat the exercise from Section~\ref{sec:tt32}, only this time we
account for the finite detector resolution by smearing the jet energies with the typical hadronic calorimeter resolution
\beq
\frac{\sigma}{E} = \left(\frac{1}{\sqrt{E}}\right) 
\label{hcal}
\eeq
and electromagnetic calorimeter resolution 
\beq
\left(\frac{\sigma}{E}\right)^2= \left( \frac{0.0363}{\sqrt{E}} \right)^2
+ \left( \frac{0.124}{E} \right)^2 + 0.0026^2
\label{ecal}
\eeq
in CMS \cite{Bayatian:2006zz}, with the energy measured in GeV. Smearing of the muon momenta is done according to the
``Full System" values in Fig.~1.5 of \cite{Bayatian:2006zz}. 
The result of the fitting exercise is shown in Fig.~\ref{fig:sigma_ttbar_smear_32}.
We see that the peak structure in the vicinity of the correct mass value ($126.5$ GeV) is preserved, 
but somewhat degraded due to the detector resolution effects.

\subsection{$D$-pair production and combinatorics effects}
\label{sec:combinatorics}

Our proposed mass measurement method uses a single decay chain like the one depicted in Fig.~\ref{fig:decay}.
In that sense, the method is inclusive and model-independent, since it does not depend on what else is going on in the event.
In particular, the method is equally applicable when particle $D$ is produced singly, in pairs, or in association with another object.
Nevertheless, a well-motivated and widely studied class of models are the SUSY-like dark matter scenarios in which all particles 
$A$, $B$, $C$ and $D$ carry negative parity under the additional $Z_2$ symmetry. In that case, $D$ has to be produced 
in association with another negative parity object. If $A$ is a neutral dark matter candidate, then $D$ must carry color, therefore
$D$ pair-production is strong and may dominate the inclusive cross-section for $D$ production. 

\begin{figure}[t]
\centering
\includegraphics[width=0.4\textwidth]{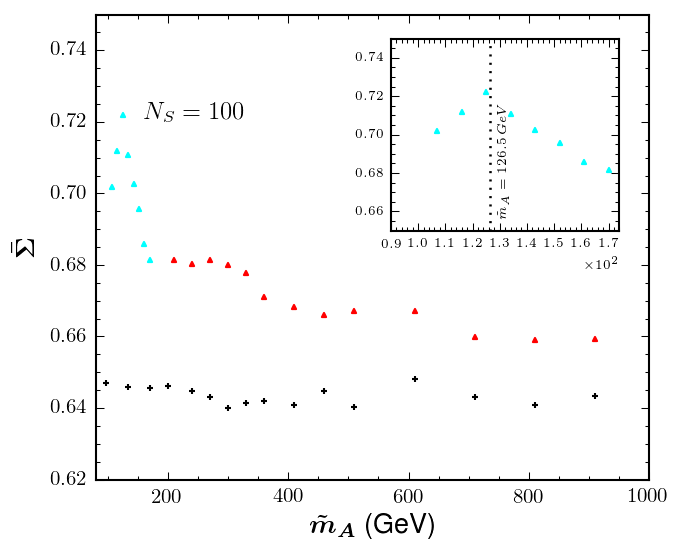}
\includegraphics[width=0.4\textwidth]{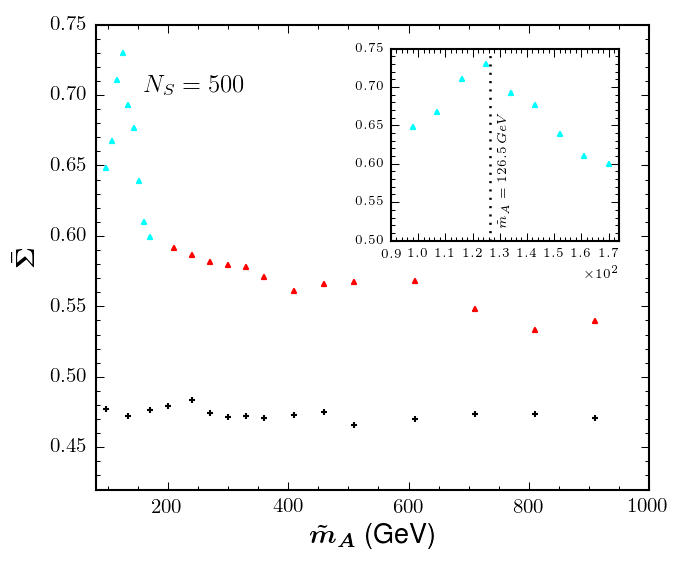}
\caption{ The same as Fig.~\ref{fig:sigma_ttbar_32}, but for signal events where particles $D$ are pair-produced and decay as 
in Fig.~\ref{fig:decay}. The left (right) panel is made with 100 (500) signal events.
\label{fig:sigma_pairpro_noback} 
}
\end{figure}

The presence of a second $D$ decay chain in the event can either be a blessing or a curse. If both $D$ particles decay 
the same way, as in Fig.~\ref{fig:decay}, we can attempt to double our statistics by considering the second decay chain as well.
However, this comes at a cost, since now we have to face the combinatorial problem of associating the different reconstructed 
final state objects to one of the two decay chains. First, there is a two-fold ambiguity of associating each jet to the correct side,
and furthermore, there is an additional two-fold ambiguity in the case when all leptons are the same flavor. The simplest approach 
would be to consider all possible combinations and use the resulting data set for building the Voronoi tessellation, then proceeding with
the fitting of the boundary surface as before. The result from this exercise is shown in Fig.~\ref{fig:sigma_pairpro_noback}
for the case of 100 (left panel) and 500 (right panel) signal events. We see that despite the pollution from wrong combinatorics, 
the peak in $\bar\Sigma$ is still very well visible, and found in the right location.

\begin{figure}[t]
\centering
\includegraphics[width=0.4\textwidth]{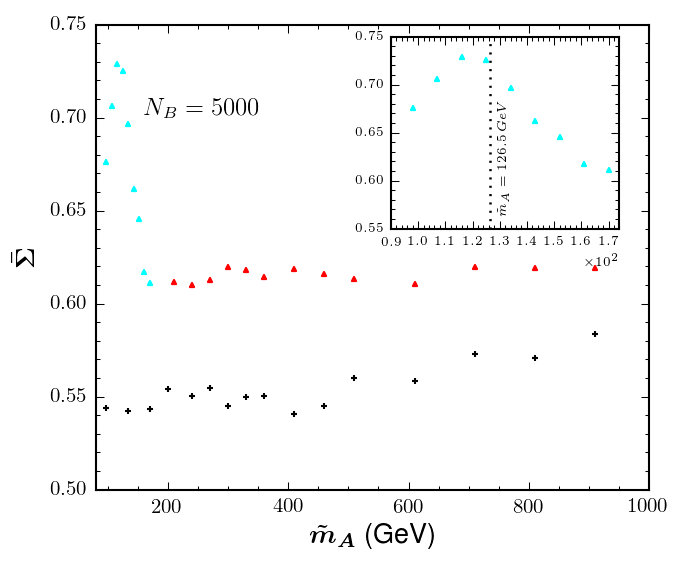}
\includegraphics[width=0.4\textwidth]{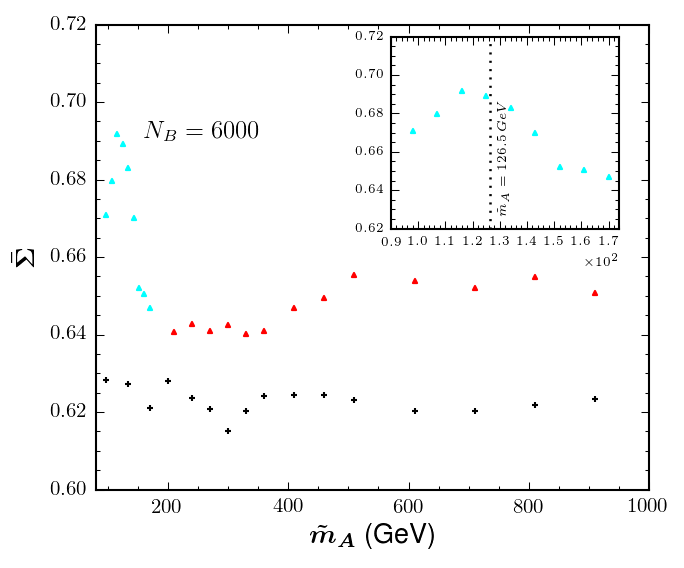}
\caption{ The same as the lower two plots in Fig.~\ref{fig:sigma_ttbar_32}, but for signal events where particles $D$ are pair-produced and one of them decays as 
in Fig.~\ref{fig:decay}, while the other decays to 3 jets and particle A. 
\label{fig:sigma_pairpro_ttbar} 
}
\end{figure}

Our final study is reserved for the most challenging example so far --- a case with a severe combinatorics problem,
in the presence of SM ($t\bar{t}$) background events. For signal, let us again consider $D$-pair production, only this time let
{\em all three} of the decay products of the second $D$ particle be QCD jets. Thus, each one of our signal events has
2 leptons and 4 jets, and picking the correct jet becomes a difficult task. Now, instead of using all possible combinations, we 
design a preselection cut in order to improve our chances of capturing the correct jet pairing. For this purpose, we consider 
the four possible jet-lepton-lepton combinations and compute the corresponding three-body invariant masses. Next, 
we rank-order these four values \cite{Kim:2015bnd,Klimek:2016axq} and eliminate from further consideration the two jets which 
correspond to the two largest jet-lepton-lepton invariant masses, since those jets are very likely to come from the decay chain 
opposite the two leptons. The remaining two jets still cause a two-fold ambiguity, which we handle as in Fig.~\ref{fig:sigma_pairpro_noback}:
by simply plotting both combinations. The end result from the analysis is shown in Fig.~\ref{fig:sigma_pairpro_ttbar}.
As in the example from Sec.~\ref{sec:tt32}, here we also include a certain number of $t\bar{t}$ background events:
5000 in the left panel and 6000 in the right panel. The fact that the $\bar\Sigma$ peak is again obtained in the correct location
indicates that our method can be viable in the presence of combinatorial background due to pair production of particles $D$.

\section{Conclusions}
\label{sec:conclusions}

In this paper, we reconsidered the classic endpoint method for particle mass determination
in SUSY-like decay chains like the one shown in Fig.~\ref{fig:decay}. Our main points are:
\begin{itemize}
\item We have identified a ``flat direction" in the mass parameter space (\ref{massparspace}), 
along which mass differences can be measured relatively well, but  the overall mass scale 
remains poorly constrained. (The analytical formulas parametrizing this flat direction can be found in Appendix~\ref{sec:inversion}.)
We quantified the problem with examples of specific study points, $P_{31}$ and $P_{32}$, 
considered in Sections~\ref{sec:case31} and \ref{sec:case32}, respectively.
\item We then proposed a new method for mass measurements in general,
and for extracting the mass scale along the flat direction, in particular.
The method takes advantage of the changes in the {\em shape} of the two-dimensional kinematic boundary surface within the 
fully differential three-dimensional space of observables, as one moves along the flat direction. We have tested 
our Voronoi-based algorithm \cite{Debnath:2016mwb,Debnath:2015wra} for detecting the boundary surface
and demonstrated that it can be usefully applied in order to lift the degeneracy along the flat direction. 
This approach represents the natural extension of the 
one-dimensional kinematic endpoint method to the relevant three dimensions of invariant mass phase space.
\item We introduced a new variable, $\bar\Sigma$, which is the average RSD per unit area, calculated over the hypothesized 
kinematic boundary. We showed that the location of the $\bar\Sigma$ maximum correlates very well with the true 
values of the new particle masses, see Figs.~\ref{fig:sigma_uniform_31}, \ref{fig:sigma_ttbar_31}, \ref{fig:sigma_uniform_32}, \ref{fig:sigma_ttbar_32}, and \ref{fig:sigma_ttbar_smear_32}. 
\end{itemize}

The work reported here can be extended in several directions. First of all, the method can be readily generalized to longer decay chains with 
more visible particles, where the boundary enhancement is even more pronounced \cite{Altunkaynak:2016bqe}, and therefore, the detection of the boundary surface 
should be in principle easier. One could also try to apply Voronoi-based boundary detection algorithms for the {\em discovery} of new physics.
It is also interesting to develop a general and universal method for estimating the statistical significance of the local
 peaks found in Figs.~\ref{fig:sigma_uniform_31}, \ref{fig:sigma_ttbar_31}, \ref{fig:sigma_uniform_32} and \ref{fig:sigma_ttbar_32},
 and hence the statistical precision of our mass measurement. These, along with many other interesting questions, 
 will be investigated in future studies \cite{del4short}.

\acknowledgments
The research of the authors is supported by the 
National Science Foundation Grant Number PHY-1620610 and
by the Department of Energy under Grants DE-SC0010296 and DE-SC0010504. 
DK is presently supported in part by the Korean Research Foundation (KRF) through
the CERN-Korea Fellowship program, and also acknowledges support by the LHC-TI postdoctoral fellowship under grant
NSF-PHY-0969510.

\appendix

\section{Inverse formulas}
\label{sec:inversion}

In this appendix we derive the inverse relations which define $m_B$, $m_C$ and $m_D$ in terms of the three measured endpoints
\bea
a \equiv \left(m_{ll}^{max}\right)^2, \qquad
b \equiv \left(m_{j\ell\ell}^{max}\right)^2, \qquad
c \equiv \left(m_{jl(low)}^{max}\right)^2,
%d \equiv \left(m_{jl(high)}^{max}\right)^2.
\eea
and the remaining mass parameter $m_{A}$. For simplicity of notation, in this appendix we shall omit the 
tildes on the trial mass parameters $m_{A}$, $m_B$, $m_C$ and $m_D$.

\subsection*{The case of region $(3,1)$}

\textcolor{red}{Region $(3,1)$} is defined by the following conditions
\bea
R_{AB} &\le& R_{BC}R_{CD} = R_{BD},  \label{reg31R_AB}\\
R_{BC} &\ge& \frac{1}{2-R_{AB}}. \label{reg31R_BC}
\eea
The kinematic endpoints are given by the following formulas:
\bea
a &=& m_D^2\, R_{CD}\,(1-R_{AB})(1-R_{BC}), \label{reg31a} \\ [2mm]
b &=& m_D^2\,(1-R_{BD})(1-R_{AB}), \label{reg31b} \\ [2mm]
c &=& m_D^2\,(1-R_{BC})(1-R_{CD}), \label{reg31c} \\ [2mm]
d &=& m_D^2\,(1-R_{CD})(1-R_{AB}). \label{reg31d}
\label{fsin31}
\eea
The masses of $B$, $C$ and $D$ are given by
\bea
m_B^2&=&\frac{1}{2c}\left\{
2m_A^2c+a(b-a-c)+\left[4m_A^2 ac(b-a) +a^2(b-a-c)^2\right]^{1/2}
\right\}, \label{reg31mB} \\[2mm]
m_C^2&=&m_B^2 \left( 1+ \frac{a}{m_B^2-m_A^2}\right),      \label{reg31mC}\\[2mm]
m_D^2&=&m_B^2 \left( 1+ \frac{b}{m_B^2-m_A^2}\right),    \label{reg31mD}
\eea
where in the right hand sides of the last two equations $m_A$ is an input, while $m_B$ is calculated from (\ref{reg31mB}).
Using (\ref{reg31R_AB}) it is easy to show that in this region, we always have 
$$b-a-c=m_D^2\, (1-R_{CD})(R_{BC}-R_{AB}) \ge 0,$$ 
so that (\ref{reg31mB}) always gives a non-negative result for $m_B^2$.
Substituting (\ref{reg31mB}-\ref{reg31mD}) into (\ref{reg31d}), one can explicitly 
check that the $m_A$ dependence drops out and we recover the ``bad" relation (\ref{bad}) in the form
\beq
d=b-a. \label{reg31dformula}
\eeq
%Finally, by substituting (\ref{reg31mB}-\ref{reg31mD}) into (\ref{jllthetadef}),
%we find 
%\beq
%e = f(a,b,c,m_A^2).\  {\bf Derive\ me\ please!}\label{reg31eformula}
%\eeq

\subsection*{The case of region $(3,2)$}

\textcolor{cyan}{Region $(3,2)$} is defined by the following two conditions:
\bea
R_{AB} &\le& R_{BC}R_{CD} = R_{BD},  \label{reg32R_AB}\\[2mm]
R_{BC} &\le& \frac{1}{2-R_{AB}}. \label{reg32R_BC}
\eea
The kinematic endpoints are given by the following formulas:
\bea
a &=& m_C^2(1-R_{AB})(1-R_{BC}), \label{reg32a} \\ [2mm]
b &=& m_D^2(1-R_{AB})(1-R_{BD}), \label{reg32b} \\ [2mm]
c &=& m_D^2(1-R_{CD})(1-R_{AB})(2-R_{AB})^{-1},  \label{reg32c} \\ [2mm]
d &=& m_D^2(1-R_{CD})(1-R_{AB}).\label{reg32d} 
\label{fsin32}
\eea
The masses of $B$, $C$ and $D$ are given by \cite{Gjelsten:2006tg}
\bea
m_B^2&=&\frac{cm_A^2}{2c-b+a}, \label{reg32mB}\\[2mm]
m_C^2&=&m_B^2 \left( 1+ \frac{a}{m_B^2-m_A^2}\right),      \label{reg32mC}\\[2mm]
m_D^2&=&m_B^2 \left( 1+ \frac{b}{m_B^2-m_A^2}\right).    \label{reg32mD}
\eea
In this region, we always have 
$$2c-b+a=m_D^2\, (1-R_{CD})\left(\frac{2}{2-R_{AB}}-1\right) \ge 0,$$ 
so that (\ref{reg32mB}) always gives a non-negative result for $m_B^2$.
The ``bad" relation (\ref{bad}) is again satisfied in this region, so that along
the flat direction the endpoint $d$ is again constant and given by (\ref{reg31dformula}),
providing a useful cross-check on the obtained solution (\ref{reg32mB}-\ref{reg32mD}).
%Finally, the $e$ endpoint along the mass family (\ref{massfamily}) is given by
%\beq
%e = f(a,b,c,m_A^2).\  {\bf Derive\ me\ please!}\label{reg32eformula}
%\eeq

\subsection*{The case of region $(2,3)$.}

\textcolor{green}{Region $(2,3)$} is defined by the following condition:
\bea
R_{BC} &\le& R_{AB}R_{CD}, \label{reg23R_BC}
%R_{AB} &\ge& R_{BC}. \label{reg23R_BC}
\eea
The kinematic endpoints are given by the following formulas:
\bea
a &=& m_C^2(1-R_{AB})(1-R_{BC}), \label{reg23a} \\ [2mm]
b &=& m_D^2(1-R_{BC})(1-R_{AB}R_{CD}), \label{reg23b} \\ [2mm]
c &=& m_D^2(1-R_{CD})(1-R_{AB})(2-R_{AB})^{-1}, \label{reg23c} \\ [2mm]
d &=& m_D^2(1-R_{CD})(1-R_{BC}). \label{reg23d} 
\label{fsin23}
\eea
The masses of $B$, $C$ and $D$ are given by 
\bea
m_B^2&=&\frac{2m_A^2(b-a)+a(2c-b+a)+\left[4m_A^2 ac(b-a) +a^2(2c-b+a)^2\right]^{1/2}}{2(b-a)}, 
\label{reg23mB} \\ [2mm]
m_C^2&=&m_B^2 \left( 1+ \frac{a}{m_B^2-m_A^2}\right),      \label{reg23mC}\\ [2mm]
m_D^2&=&\left( 1+ \frac{a}{m_B^2-m_A^2}\right)\left[\frac{b}{a}\left(m_B^2-m_A^2\right)+m_A^2\right]. \label{reg23mD}
\eea
Using (\ref{reg23a}), (\ref{reg23b}) and (\ref{reg23d}) it is easy to show that in this region, we always have 
$$b-a=d \ge 0,$$
and while the term $(2c-b+a)$ can have either sign, the discriminant
(i.e., the term inside the square root in (\ref{reg23mB})) is always larger than $a^2(2c-b+a)^2$, 
which guarantees a non-negative result for $m_B^2$. In this region, the relation (\ref{bad}) is again satisfied, 
so that $d$ is again given by (\ref{reg31dformula}), which can be used to cross-check the result (\ref{reg23mB}-\ref{reg23mD}).

\subsection*{The case of region $(4,1)$.}

\textcolor{blue}{Region $(4,1)$} is defined by the following conditions:
\bea
R_{AB} &\ge& R_{BC}R_{CD} = R_{BD},  \label{reg41R_AB}\\[2mm]
R_{BC} &\ge& \frac{1}{2-R_{AB}},             \label{reg41R_BC} \\ [2mm]
R_{CD} &\ge& R_{AB}R_{BC} = R_{AC}.  \label{reg41R_CD} 
\eea
For the case (4,1) the endpoints are given by the following formulas:
\bea
a &=& m_C^2(1-R_{AB})(1-R_{BC}), \label{reg41a} \\[2mm]
b &=& m_D^2(1-\sqrt{R_{AD}})^2,     \label{reg41b}  \\[2mm]
c &=& m_D^2(1-R_{CD})(1-R_{BC}), \label{reg41c}  \\[2mm]
d &=& m_D^2(1-R_{CD})(1-R_{AB}).  \label{reg41d} 
\label{fsin41}
\eea
The masses of $B$, $C$ and $D$ in terms of  $a$, $b$, $c$ and $m_{A}$ are given by
\bea
m_B^2&=&\frac{am_D^2+cm_A^2+(m_A^2-a)(a+c)+a\left[(a+c-m_A^2-m_D^2)^2-4m_A^2m_D^2\right]^{1/2}}{2(a+c)},~~~~ 
\label{reg41mB} \\ [2mm]
%m_B^2&=&\frac{-B+\sqrt{B^2-4AC}}{2A}   \label{reg41mB}  \\
m_C^2&=&m_B^2 \left( 1+ \frac{a}{m_B^2-m_A^2}\right),      \label{reg41mC}\\ [2mm]
m_D&=&m_A+\sqrt{b},                             \label{reg41mD}
\eea
where in (\ref{reg41mB}) $m_D$ should be taken from (\ref{reg41mD}), and the obtained result for $m_B$ should be used in (\ref{reg41mC}).
%The other solution of $m_B^2$ with negative sign is ignored as it violates the constraint of the region (4,1): $R_{BC}>(2-R_{AB})^{-1}$.
The $d$ endpoint is given by
\begin{subequations}
\bea
d &=& b - a - m_D^2 \left( \sqrt{R_{AB}} - \sqrt{R_{BD}}\right)^2  \label{reg41drelation} \\[2mm]
&=&  \frac{ac\left\{(b-a-c)\left[b(1+\sqrt{r})+2m_A\sqrt{b}\sqrt{r} \right] +2(2b-a-c)(m_A\sqrt{b}+m_A^2) \right\}}{(a+c)\left[ a(b-a-c)(1+\sqrt{r})+2am_A\sqrt{b}+2(a+c)m_A^2\right]},~~~~~~~
\label{reg41dformula}
\eea
\end{subequations}
where
\beq
r\equiv 1+\frac{4m_A(m_A+\sqrt{b})}{b-a-c}.
\label{r41}
\eeq
%The $e$ endpoint along the mass family (\ref{massfamily}) is given by
%\beq
%e = f(a,b,c,m_A^2).\  {\bf Derive\ me\ please!}\label{reg41eformula}
%\eeq

\subsection*{The case of region $(4,2)$.}

\textcolor{yellow}{Region $(4,2)$} is defined by the following conditions:
\bea
R_{AB} &\ge& R_{BC}R_{CD} = R_{BD}, \label{reg42R_AB}\\[2mm]
R_{BC} &\ge& R_{AB},                              \label{reg42R_BCge} \\ [2mm]
R_{BC} &\le& \frac{1}{2-R_{AB}},              \label{reg42R_BCle} \\ [2mm]
R_{CD} &\ge& R_{AB}R_{BC} = R_{AC}.  \label{reg42R_CD} 
\eea
The kinematic endpoints are given by the following formulas:
\bea
a &=& m_C^2(1-R_{AB})(1-R_{BC}), \label{reg42a} \\[2mm]
b &=& m_D^2(1-\sqrt{R_{AD}})^2,  \label{reg42b} \\[2mm]
c &=& m_D^2(1-R_{CD})(1-R_{AB})(2-R_{AB})^{-1},   \label{reg42c}\\[2mm]
d &=& m_D^2(1-R_{CD})(1-R_{AB}).  \label{reg42d}
\label{fsin42}
\eea
The masses of $B$, $C$, $D$ in terms of  $a,b,c, m_{A}$ are the following
\bea
m_B^2&=&\frac{1}{2}\left[m_D^2+m_A^2-a-2c\pm\sqrt{(m_D^2+m_A^2-a-2c)^2-4m_A^2(m_D^2-c)}
\right],        \label{reg42mB} \\[2mm]
m_C^2&=&m_B^2 \left( 1+ \frac{a}{m_B^2-m_A^2}\right),        \label{reg42mC} \\[2mm]
m_D&=&m_A+\sqrt{b}.                              \label{reg42mD} 
\eea
Here, just as in (\ref{reg41mB}-\ref{reg41mD}), one should first find $m_D$ from (\ref{reg42mD}),
then use the result in (\ref{reg42mB}) to obtain $m_B$, which will be needed in (\ref{reg42mC}).
The $d$ endpoint is then given by
\begin{subequations}
\bea
d &=& b - a - m_D^2 \left( \sqrt{R_{AB}} - \sqrt{R_{BD}}\right)^2 \label{reg42drelation} \\[2mm]
&=& 
\frac{c\left[ a+3b-2c+6m_A\sqrt{b}+2m_A^2-\sqrt{(b-a-2c+2m_A\sqrt{b})^2-4(a+c)m_A^2}\right]}{2(b-c+2m_A\sqrt{b}+m_A^2)}. ~~~~~~~
\label{reg42dformula}
\eea
\end{subequations}
%where
%\beq
%r_{42}\equiv 1+\frac{4m_A(m_A+\sqrt{b})}{b-a-c}.
%\label{r42}
%\eeq
%The $e$ endpoint along the mass family (\ref{massfamily}) is given by
%\beq
%e = f(a,b,c,m_A^2).\  {\bf Derive\ me\ please!}\label{reg42eformula}
%\eeq

\subsection*{The case of region $(4,3)$.}

\textcolor{magenta}{Region $(4,3)$} is defined by the following conditions:
\bea
R_{AB} &\ge& R_{BC},                              \label{reg43R_AB}\\[2mm]
R_{BC} &\ge& R_{AB}R_{CD},                  \label{reg43R_BC} \\ [2mm]
R_{CD} &\ge& R_{AB}R_{BC} = R_{AC}.  \label{reg43R_CD} 
\eea
For the case (4,3) the endpoints are given by the following formulas:
\bea
a &=& m_C^2(1-R_{AB})(1-R_{BC}),     \label{reg43a} \\[2mm]
b &=& m_D^2(1-\sqrt{R_{AD}})^2,        \label{reg43b} \\[2mm]
c &=& m_D^2(1-R_{CD})(1-R_{AB})(2-R_{AB})^{-1},   \label{reg43c}   \\[2mm]
d &=& m_D^2(1-R_{CD})(1-R_{BC}).    \label{reg43d}
\label{fsin43}
\eea
The masses of $B$, $C$, $D$ in terms of  $a,b,c, m_{A}$ are the following
\bea
m_B^2&=&\frac{1}{2}\left[m_D^2+m_A^2-a-2c-\sqrt{(m_D^2+m_A^2-a-2c)^2-4m_A^2(m_D^2-c)}
\right],        \label{reg43mB} \\[2mm]
m_C^2&=&m_B^2 \left( 1+ \frac{a}{m_B^2-m_A^2}\right),        \label{reg43mC} \\[2mm]
m_D&=&m_A+\sqrt{b},                              \label{reg43mD} 
\eea
where again the masses are calculated in the order $m_D$, $m_B$ and then $m_C$.
The $d$ endpoint is given by
\begin{subequations}
\bea
d &=& b - a - m_D^2 \left( \sqrt{R_{BC}} - \sqrt{R_{AB}R_{CD}}\right)^2 \label{reg43drelation} \\ [2mm]
&=& 
\frac{4ac\left[
b-a-2c+2m_A\sqrt{b} +m_A^2 - \sqrt{(b-a-2c+2m_A\sqrt{b})^2-4(a+c)m_A^2}
\right]}{\left(b-2c+2m_A\sqrt{b}-\sqrt{(b-a-2c+2m_A\sqrt{b})^2-4(a+c)m_A^2}\right)^2-a^2}.~~~~~~~~
\label{reg43dformula}
\eea
\end{subequations}
%The $e$ endpoint along the mass family (\ref{massfamily}) is given by
%\beq
%e = f(a,b,c,m_A^2).\  {\bf Derive\ me\ please!}\label{reg43eformula}
%\eeq

\end{document}